\documentclass[11pt,tightenlines,eqsecnum,floats,aps,amsmath,amssymb,nofootinbib,prd,shownopacs,floatfix]{revtex4}
\usepackage[english]{babel}
\usepackage[utf8x]{inputenc}
\usepackage{amsmath,amsfonts,amssymb, extarrows,mathrsfs}
\usepackage{xcolor}
\usepackage{graphicx}
\usepackage{empheq}
\usepackage{hyperref}
\usepackage{bbm}
\usepackage{pdfpages}
\usepackage{makecell}

\newcommand{\be}{\begin{equation}}
\newcommand{\ee}{\end{equation}}
\newcommand{\ba}{\begin{eqnarray}}
\newcommand{\ea}{\end{eqnarray}}

\DeclareMathAlphabet{\mathpzc}{OT1}{pzc}{m}{it}
\newcommand*\bigcdot{\mathpalette\bigcdot@{.5}}
\begin{document}

\title{{Gauge invariant variables for cosmological perturbation theory using geometrical clocks}}

\author{Kristina Giesel$^{1}$}
\thanks{kristina.giesel@gravity.fau.de}
\author{Adrian Herzog$^1$}
\thanks{adri.a.herzog@studium.uni-erlangen.de}
\author{Parampreet Singh$^2$}
\thanks{psingh@lsu.edu}
\affiliation{$^1$ Institute for Quantum Gravity, Department of Physics,  \\ FAU Erlangen -- N\"urnberg,
Staudtstr. 7, 91058 Erlangen, Germany}
\affiliation{$^2$ Department of Physics and Astronomy,\\
Louisiana State University, Baton Rouge, LA 70803, U.S.A.}

\begin{abstract}
Using the extended ADM-phase space formulation in the canonical framework we analyze the relationship between various gauge choices  made in cosmological perturbation theory and the choice of geometrical clocks in the relational formalism. We show that various gauge invariant variables obtained in the conventional analysis of cosmological perturbation theory correspond to Dirac observables tied to a specific choice of geometrical clocks. As examples, we show that the Bardeen potentials and the Mukhanov-Sasaki variable emerge naturally in our analysis as observables when gauge fixing conditions are determined via clocks in the Hamiltonian framework. Similarly other gauge invariant variables for various gauges  can  be  systematically  obtained. We demonstrate this by analyzing five common gauge choices:  longitudinal, spatially flat, uniform field, synchronous and comoving gauge. For all these, we apply the observable map in the context of the
relational formalism and obtain the corresponding Dirac observables associated with these choices of clocks. At the linear order, our analysis generalizes the existing results in canonical cosmological perturbation theory twofold. On the one hand we can include also gauges that can only be analyzed in the context of the extended ADM-phase  space and furthermore, we obtain a set of natural gauge invariant variables, namely the Dirac observables, for each considered choice of gauge conditions.  Our analysis provides insights on which clocks should be used to extract the relevant natural physical observables both at the classical and quantum level. We also discuss how to generalize our analysis in a straightforward way to higher orders in the perturbation theory to understand gauge conditions and the construction of gauge invariant quantities beyond linear order.
\end{abstract}

\maketitle

\section{Introduction}
\label{Sec:Intro}
In general relativity, only those quantities are physically observable which are invariant under spacetime diffeomorphisms. In the  formulation of general relativity as a constraint theory, namely the ADM-phase space formulation, these invariant quantities are the Dirac observables. These observables commute with the Hamiltonian in the canonical description of general relativity. If these observables do not depend explicitly on time, then they are constants of motion. But if there is an explicit time dependence, these observables are evolving, but, then are not invariant under spacetime coordinate transformations. To permit evolving observables one needs to go to a relational formalism where the evolution of observables is studied with respect to some other fields \cite{Bergmann1,Bergmann2,Komar,RovelliPartial,RovelliObservable}. The latter are the clocks or reference fields which capture the dynamics of other fields in the spacetime without using spacetime coordinates.  These clocks can be matter fields, such as scalar fields or dust, or can be chosen from metric variables. Given a choice of clocks, 
the relational formalism can be used to systematically construct Dirac observables in the canonical description of gravity \cite{Vytheeswaran, Dittrich, Dittrich2, Thiemann2, Thiemann3}. For instance this has been used to derive a reduced phase space, that is the phase space of the gauge invariant quantities, for general relativity in \cite{Giesel:2007wn,Giesel:2016gxq, Giesel:2017mfc} that was taking as the starting point for quantization. At the classical level such a reduced phase space approach was considered in order to formulate general relativistic perturbation theory in \cite{Dust1}. An application of this to linearized cosmological perturbation theory in the ADM-phase space can be found in \cite{Dust2}, whereas a generalization to scalar-tensor theories has been presented in \cite{Giesel-Ma}. The reduced phase space of LTB spacetimes has been derived in \cite{Giesel-LTB}. The work in  \cite{Dittrich-Tambornino1,Dittrich-Tambornino2} follows more the conventional approach in cosmological perturbation theory, that is first considering linearized perturbations and afterwards constructing quantities invariant under linearized gauge transformations, but based on Ashtekar-variables.  

In the canonical framework, there have been earlier works on cosmological perturbation theory, most notably by Langlois \cite{Langlois}. Using the ADM-phase space, where the lapse function and the shift vector are treated as Lagrange multipliers, Langlois found a phase space formulation of the Mukhanov-Sasaki variable and its evolution in the canonical setting. However, in the usual ADM-phase space formulation, the lapse function and the shift vector are not treated as phase space variables. Often this is called the reduced ADM-phase space.
As a result, it is not possible to understand various gauges and gauge invariant variables such as the Bardeen potentials in this setting. In other words, one can not generalize Langlois' work to various other gauge invariant quantities in the ADM-phase space based on gauge fixing conditions involving lapse and shift variables. To accomplish this, which is one of the main objectives of our work, we have to go beyond the (reduced) ADM-phase space and use the extended ADM-phase space studied in pioneering works by Pons, Salisbury, Sundermeyer and others \cite{Pons1,Pons2,Pons3,Pons4}. An advantage of this phase space, as discussed in  \cite{Giesel:2017roz} and demonstrated here, is that it is well suited to capture the full scope of the relational formalism. While in the companion work \cite{Giesel:2017roz}, we formulated cosmological perturbation theory in the setting of the extended phase space, in this work we will apply the relational formalism to it. In particular, we will aim to understand the relationship between gauge fixing conditions and gauge invariant quantities, used in conventional cosmological perturbation theory, in the setting of the relational formalism in the extended phase space. In comparison with Langlois' earlier results,  the novelty with the current approach based on the extended phase space is that we are able to provide  a phase space formulation of various gauges and find their associated gauge invariant variables using reference clock fields even for those cases where lapse and shift variables are involved in the gauge fixing conditions. Furthermore, next to the results of Langlois \cite{Langlois} there is more recent work in the canonical formalism \cite{Domenech:2017ems} that also considers second order perturbations. Canonical cosmological perturbation theory has been also formulated in Ashtekar variables in \cite{Bojowald:2006tm}, but also here by treating lapse and shift as Lagrange multipliers. As a result, also in this work understanding of the issue of gauge invariant variables for various gauge choices is restricted to the reduced ADM-phase space. 

The issue of gauge fixing and gauge invariant quantities is a subtle one in cosmological perturbation theory. To relate the goals of our analysis in comparison to the conventional treatment, let us revisit it briefly.  
In the conventional analysis of cosmological perturbations, in order to extract any meaningful predictions, one must handle the problem of gauge freedom in perturbations. 
This arbitrariness arising due to the freedom in performing the space-time coordinate transformations, called  diffeomorphisms, affects the scalar and vector metric perturbations, as well as the matter perturbations. 
There are two ways to deal with this this gauge freedom. One can either fix a gauge from the very onset and work entirely in it, or one can construct gauge invariant variables.  In the  method of gauge fixing, one imposes conditions on metric and/or matter variables to eliminate the underlying freedom due to spacetime diffeomorphisms. An example where this freedom can be completely eliminated is the longitudinal or the Newtonian gauge in which the longitudinal part of the spatial metric perturbations is put to zero, and as a result the physically relevant scalar metric perturbations are the ones in the lapse and the trace of the spatial metric. However, in general some residual freedom can remain even after gauge fixing. This happens for example for  the synchronous gauge where the perturbations in the shift vector vanish, yet the freedom under diffeomorphisms  can not be fully eliminated.  

In the method of gauge invariant variables, one finds the right combination of metric and/or matter perturbations which are invariant under diffeomorphisms up to corrections that are of higher order than the considered order in perturbation theory. There are numerous ways to construct such gauge invariant variables. As examples, the Bardeen potentials $\Phi$ and $\Psi$ are obtained using the gauge transformation properties of linear perturbations in the lapse and 
the trace of the spatial metric perturbation respectively. Similarly, the Mukhanov-Sasaki variable $v$ is constructed using the transformation properties of the linear perturbations in the scalar field under gauge transformation. The Bardeen potentials are in fact linear 
gauge invariant extensions of the lapse and the trace of the spatial metric perturbations. Similarly, the Mukhanov-Sasaki variable $v$  is a linear gauge invariant extension of the scalar field perturbation.

Though, in conventional treatments there is 
an emphasis on the usage of gauge invariant variables, it is to be noted, and as is stressed by Bardeen himself \cite{bardeen}, that the latter approach is not  
 more beneficial than the former to extract physical predictions about the relevant metric and matter perturbations. As is well known, this is because the physical relevance of gauge invariant variables is tied to the gauge fixing conditions. As an example, it is only in the 
longitudinal gauge that the Bardeen potentials can be naturally related to the metric perturbations, namely the perturbations in the lapse and the trace of the spatial metric. In any other gauge, the physical relevance of Bardeen potentials is rather unnatural. This can easily seen for the case of the spatially flat gauge in which the gauge invariant extension of the trace of the perturbed metric vanishes. Similarly, it is only in the spatially flat gauge that the Mukhanov-Sasaki variable can be naturally interpreted as the gauge invariant variable corresponding to the perturbations in scalar field. 

Thus, there is a subtle relationship between the gauge fixing conditions and the gauge invariant variables. While working with the latter, the gauge fixing conditions may not be apparent but they are critical to extract any physically relevant prediction for cosmological perturbations. For every gauge fixing condition, there is a natural set of physically relevant gauge invariant variables. Here it is important to note that such a correspondence is further tied to the order of perturbation theory in which the gauge invariant variables are constructed. 

In the relational formalism, the choice of gauge fixing is tied to a particular choice of clocks, see for instance also the early work in \cite{DeWitt:1967yk,DeWitt:1967ub} for a discussion on clocks in the context of general relativity and its quantization. As discussed earlier, these clocks can be chosen using geometrical or matter degrees of freedom.
In this manuscript, we choose components of the perturbed metric as well as combinations of perturbed metric and matter degrees of freedom as the clocks.  The linearized gauge fixing conditions $\delta G^\mu = 0$ define a hypersurface and involve the perturbations in the clock variables, $\delta T^\mu$. This allows us to express the linearized gauge fixing conditions in terms of the linearized perturbations of the clock variables. Thus, for various choices of gauge fixing conditions, we can identify corresponding clocks whose stability conditions are equivalent to the stability of the gauge fixing condition.  
Once a particular choice of clock variables is made, the relational formalism can be applied to explicitly construct  Dirac observables associated with those clocks. These Dirac observables turn out to be  the gauge invariant variables usually constructed from combinations of the metric and matter perturbations. And as discussed above, these gauge invariant variables are exactly the ones which are physically tied to the corresponding gauge fixing conditions. As an example, the gauge invariant variables such as the Bardeen potentials emerge naturally as the Dirac observables for the clocks corresponding to the longitudinal gauge fixing conditions. 

 We carry out this analysis  for five commonly used gauge conditions in cosmological perturbation theory. The first three gauge conditions are the ones where the longitudinal part of spatial metric perturbations vanishes. These are the longitudinal or the Newtonian,  the spatially flat and the uniform field gauges. The other two gauge conditions have vanishing perturbation in the shift vector. The gauges considered in this category are the synchronous and the comoving  gauges. In each of these gauges, we apply the relational formalism in the extended phase space to identify consistent clocks  and find the Dirac observables which are the gauge invariant variables naturally relevant for these particular gauges. It is to be noted that to achieve this task for various common gauges in cosmological perturbation theory, it is necessary to have an extended phase space formulation. Without it, one is 
severely restricted in the above objective and can not even understand for example the Bardeen potentials in the relational formalism. As noted earlier, in particular one faces this limitation if one works with Langlois' formulation of canonical cosmological perturbation theory.

Our analysis using the relational formalism based on clocks not only provides an alternate path to understand various gauge fixing conditions and gauge invariant variables, it further comes with some advantages not available in  the conventional approach.  The first of which distinguishes it from the conventional procedure in the classical cosmological perturbation theory is its straightforward applicability to any higher order in perturbations. In the conventional framework, going beyond the linear order requires finding new gauge invariant quantities afresh at each order. Where as in relational formalism, once a choice of non-linear clocks is made, one can compute the gauge invariant quantities for any order. Thus, providing a systematic procedure to obtain gauge invariant quantities at linear and higher orders.  The second important advantage is that this formalism can provide useful insights on how to extract predictions for the cosmological perturbations in the canonical quantization framework. Given this formalism the path to understand clocks and gauge invariant quantities in quantum cosmological spacetimes becomes much clearer. Further, the formalism can guide us on how to choose the right clocks relevant for extracting analogs of Bardeen potentials and Mukhanov-Sasaki variable in a quantum cosmological perturbation theory.

This manuscript is organized as follows. In section \ref{Sec:ReviewObs} we start with an overview of the canonical cosmological perturbation theory in the extended phase space. Here we present a brief introduction in the relational observable formalism for general relativity on the extended phase space. For details of this discussion, the reader is referred to our companion article \cite{Giesel:2017roz}.  We divide section \ref{Sec:ReviewObs} in two parts. The first part focuses on various details of the cosmological perturbation theory in the extended ADM-phase space, and the second part deals with the relational formalism and construction of observables in the extended ADM-phase space. In section \ref{Sec:ConstrObs} we apply the relational formalism to understand how to choose clocks corresponding to various gauge conditions in the cosmological perturbation theory. The formalism is then used to derive Dirac observables in each of the cases which yield the relevant gauge invariant quantities. We study the longitudinal, spatially flat, uniform field, synchronous and comoving  gauges in detail and obtain geometric clocks and gauge invariant quantities for each of them. At the end of this section for part of the gauges we also discuss possible modifications of the gauge fixing conditions that arise naturally in the relational formalism. The results from various gauges are summarized in tables \ref{tab:clockchoices1} and \ref{tab:clockchoices2}. We conclude with a summary and discussion of open issues in section \ref{Sec:Concl}.\\

For the benefit of the reader, in the following we provide a list of our main notation and conventions that will be used throughout the article. We use the Einstein sum convention. The spacetime metric signature is chosen to be $(-1,1,1,1)$. 
We use the regular Poisson bracket convention, i.e. $\{ q,p \}= 1$. And, the Legendre map is denoted by $\mathcal{LM}$ and the corresponding inverse Legendre map by $\mathcal{LM}^*$.

The following tables summarize various symbols used in the manuscript (see also table \ref{tab:symbols} in section \ref{Sec:ConstrObs}).

\vspace{1cm}

\begin{center}
\textbf{General notations}
\begin{tabular}{r|l}
Notation & Meaning \\
\hline
$a,b,c,... = 1,2,3$ & Spatial indices \\
$\mu,\nu,\rho,... = 0,...,3$ & Spacetime or temporal-spatial indices \\
$\partial_af, f_{,a}, \frac{\partial f}{\partial x^a}$ & Partial derivatives \\
$D_a f, f_{|a}$ & Spatial covariant derivatives \\
$x^\mu,y^\mu,...$ & Spacetime coordinates \\
$x^j,y^j,...$ & Spatial coordinates \\
$g_{\mu\nu}$ & Spacetime metric \\
$q_{ab}$ & Induced metric on $\Sigma$, spatial metric \\
$\Gamma^a_{bc}$ & Spatial Christoffel symbols \\
$P^{ab}$ & Conjugate momentum to $q_{ab}$ \\
$N,N^a$ & Lapse function and shift vector field \\
$\Pi,\Pi_a$ & Momenta of lapse and shift, primary constraints \\
$\lambda,\lambda^a$ & Lagrange multipliers associated to $\dot{N}$,$\dot{N}^a$ \\
$R^{(3)~d}_{abc}$ & Spatial Riemann tensor $[D_a,D_b]\omega_c = R^{(3)~d}_{abc}\omega_d$ \\
$K_{ab}$ & Extrinsic curvature \\
$\text{Tr}(T) = q^{ab}T_{ab}$ & Trace of the tensor $T$ \\
$T^T_{ab} = T_{ab} - \tfrac{1}{3}q_{ab}\text{Tr}(T)$ & Traceless part of $T$ \\
$T_{<ab>} = T_{(ab)}-\tfrac{1}{3}q_{ab}\text{Tr}(T)$ & Symmetric and traceless combination \\
$\kappa$ & $:=16\pi G$, $G$: Newton's constant \\
$\{ q_{ab}(x),P^{cd}(y) \}$ & $ := \kappa\delta_{(a}^c\delta_{b)}^d\delta(x,y)$ \\
$\varphi,\pi_\varphi$ & Scalar field and its momentum \\
$\lambda_\varphi$ & Coupling constant of the scalar field action \\
$V(\varphi)$ & Scalar field potential. Note we have a factor of $\frac{1}{2}$ in the  \\
& definition of our potential due to a global factor of $\frac{1}{2}$ in our scalar field action.
\end{tabular}
\end{center}

\newpage

\begin{center}
\textbf{Cosmological background quantities}
\begin{tabular}{r|l}
Notation & Meaning \\
\hline
$a$ & Cosmological scale factor \\
$A= a^2$ & The squared scale factor \\
$\mathcal{H} = \tfrac{\dot{a}}{a}$ & Hubble parameter \\
$\bar{N}$ & Background lapse, $\bar{N} = \sqrt{A}$: conformal time, $\bar{N} = 1$: proper time \\
$\tilde{\mathcal{H}} = \mathcal{LM}(\mathcal{H})$ & Hubble parameter in phase space \\
$\tilde{P} = - \frac{2\sqrt{A}\tilde{\mathcal{H}}}{\bar{N}}$ & $\propto$ momentum of $A$, $\bar{P}^{ab} = \tilde{P}\delta^{ab}$ \\
$\rho,p$ & Energy-density and pressure \\
$\bar{\varphi}, \bar{\pi}_\varphi$ & Background scalar field and its momentum
\end{tabular}
\end{center}

\begin{center}
\textbf{Cosmological perturbation theory}
\begin{tabular}{r|l}
Notation & Meaning \\
\hline
$T^{(n)} \equiv \tfrac{1}{n!}\delta^nT$ & $n$'th perturbation of $T$ \\
$\phi,B,\psi,E,p_\psi,p_E$ & Scalar perturbations \\
$S^a,F_b,p_F^c$ & Transversal vector perturbations \\
$h^{TT}_{ab},p_{hTT}^{cd}$ & Transversal traceless tensor perturbations \\
$\delta N, \delta N^a$ & Lapse and shift perturbations $\delta N = \bar{N}\phi$, $\delta N^a = B^{,a} + S^a$ \\
$\delta q_{ab}$ & Spatial metric perturbation $\delta q_{ab} = 2A( \psi\delta_{ab} + E_{,<ab>} + F_{(a,b)} + \tfrac{1}{2}h^{TT}_{ab} )$ \\
$\delta P^{ab}$ & Spatial momentum perturbation $\delta P^{ab} = 2\tilde{P}( p_\psi\delta^{ab} + p_E^{,<ab>} + p_F^{(a,b)} + \tfrac{1}{2}p_{h^{TT}}^{ab} )$ \\
$\delta \varphi,\delta \pi_\varphi$ & Scalar field perturbation and momentum thereof \\
$\mathcal{E}, \mathcal{P}$ & Matter part of energy/pressure perturbations \\
$\Phi,\Psi$ & Bardeen potentials \\
$\Upsilon$ & Gauge invariant perturbation related to the momentum of $\Psi$ \\
$v$ & Mukhanov-Sasaki variable \\
$\nu^a$ & Gauge invariant vector perturbation 
\end{tabular}
\end{center}

\vskip1cm

\begin{center}
\textbf{Observables in canonical general relativity} \\
\begin{tabular}{r|l}
Notation & Meaning \\
\hline
$T^\mu$ & Clock fields \\
$G^\mu = \tau^\mu-T^\mu$ & Gauge fixing constraints \\
$\mathcal{O}_{f,T}[\tau]$ & Observable of $f$ \\
$\mathcal{A}^\mu_\nu(x,y)$ & $:= \{ T^\mu(x),C_\nu(y) \}$, matrix for weak abelianization \\
$\mathcal{B}$ & $:= \mathcal{A}^{-1}$ \\
$\tilde{C}_\mu$ & weakly abelianized constraints for reduced ADM \\
$( \tilde{\Pi}_\mu,\tilde{\tilde{C}}_\nu )$ & weakly abelianized constraints for full ADM-phase space \\
$\{\cdot,\cdot\}^*$ & Dirac bracket with respect to constraint set $(G^\mu, \tilde{C}_\nu)$
\end{tabular}
\end{center}

\vskip0.5cm
\newpage
\section{Review of canonical cosmological perturbation theory in extended phase space in the relational formalism}
\label{Sec:ReviewObs}
The goal of this section is to provide an overview of cosmological perturbation theory in the canonical setting in the extended phase space. While canonical cosmological perturbation theory goes back to the work of Langlois \cite{Langlois}, its generalization to the extended phase space has been recently introduced in the companion article \cite{Giesel:2017roz}, to which the reader is referred to for various details.  A short review is presented in this section which is divided into two parts. The first one includes a discussion of canonical cosmological perturbation theory in extended phase space, whereas the second part focuses on the relational formalism and how it can be used to formulate cosmological perturbation theory in terms of so called Dirac observables.

\subsection{Canonical cosmological perturbations in extended ADM phase space}
In cosmology literature, cosmological perturbation theory is often discussed in the Lagrangian framework. In this case one chooses a given background solution such as for instance spatially flat FLRW spacetime and considers perturbations of all ten metric components including particularly the $g_{00}$ and the $g_{0a}$ components that are parametrized by the lapse function and the shift vector in the canonical framework. If in addition, we have  matter degrees of freedom as for instance a scalar field, one considers perturbations of these degrees of freedom in a similar manner.
At the level of linearized perturbation theory one chooses a certain subset among the perturbed metric and matter variables and adds to them specific combinations of the remaining metric and matter degrees of freedom such that the final resulting quantities are invariant under linearized diffeomorphisms. By this we mean that the quantities constructed as above are invariant under diffeomorphisms up to corrections that are second order and higher. At linear order, prominent examples of such gauge invariant quantities are the Bardeen potentials and the Mukhanov-Sasaki variable. Given these gauge invariant quantities, that will also be called  observables in the following, one can derive their corresponding equations of motion and hence obtain the linearized Einstein equations expressed exclusively in terms of gauge invariant objects. 

If we want to carry the framework over to the Hamiltonian formalism, the first observation we make is that for the usual ADM phase space, only the spatial metric components and their momenta are treated as elementary phase space variables whereas the lapse and shift degrees of freedom take the role of Langrage multipliers and are therefore phase space independent quantities. The reason that such a different treatment of the two sets of variables is possible is that one obtains the reduced ADM phase space by working on the constraint hypersurface that is obtained from the primary constraints. This reduces the 10 metric degrees of freedom by four to six independent degrees of freedom that are encoded in the six components of the spatial ADM metric $q_{ab}$, whereas the lapse function and the shift vector become arbitrary Lagrange multipliers. However, as discussed in detail in \cite{Giesel:2017roz} if our goal is to find the canonical 
counter-parts of for instance the Bardeen potentials then it is necessary to treat all of the metric variables on an equal footing also in the Hamiltonian framework of general relativity. Therefore, in \cite{Giesel:2017roz} as well as in this article,  we consider the full or the so called extended ADM phase space for which the primary constraints have not yet been reduced. As a consequence, the lapse and shift take the role of elementary phase space variables likewise to the spatial metric components in the extended phase space.

In the next subsection we will summarize the results of linearized canonical cosmological perturbation theory in extended ADM-phase space, that has been analyzed in the review \cite{Giesel:2017roz} to which we refer for more details. Afterwards we will briefly introduce the relational formalism and the way conventional cosmological perturbation theory can be embedded into it. For the latter step we will build on the seminal work of Pons, Salisbury, Sundermeyer, and others \cite{Pons1,Pons2,Pons3,Pons4}.

\subsubsection{Canonical cosmological perturbation theory in extended ADM-phase space}
Throughout this article we consider linear perturbations around $k=0$ FLRW cosmological spacetimes. Hence, we will start our presentation with  summarizing the necessary equations for the background solutions in the canonical framework. This will also serve to fix the notation for the background quantities.  Again the derivation of these results as well as a more elaborate discussion on this topic can be found in \cite{Giesel:2017roz}.
~\\
~\\
\paragraph*{\centerline{1.a Background FLRW cosmologies}}
~\\

For the spatially flat FLRW spacetime,  the line element is given by:
\begin{equation}
\mathrm{d}s^2 = g_{\mu\nu}dx^\mu dx^\nu=
-\bar N^2(t)\mathrm{d}t^2 + a^2(t)\delta_{ab}\mathrm{d}x^a\mathrm{d}x^b~.
\end{equation}
In our notation greek indices are spacetime indices and run from 0 to 4, while latin indices label the spatial coordinates and run from 1 to 3 only. In the following we will denote background quantities with a bar on the top. The background metric components can be written as:
\begin{align}
\bar{q}_{ab} &= A(t)\delta_{ab} & \bar{N} &= \bar{N}(t) & \bar{N}^a &= 0,
\end{align}
where we chose $A:=a^2$ as our elementary configuration variable. This choice is convenient to write various formulae in canonical perturbation theory in a simpler form. We denote the corresponding momenta of above quantities by $\overline{P}^{ab}, \overline{\Pi}$ and $\overline{\Pi}_a$ respectively.
Note, that in order to keep a certain freedom for the time parametrization of the background solutions we do not completely specify the lapse function ($\bar N$) of the background here. In case one chooses  $\bar{N}=1$, the background evolution will be measured in cosmic time, while for the choice $\bar{N}=\sqrt{A}=a$, the line element is parameterized by conformal time. Performing a Legendre transformation we realize that it is singular which is reflected in the fact that the momenta associated with lapse and shift vanish, as can be seen below:
\be
\label{eq:BgrdMomenta}
\overline{P}^{ab}=:\tilde{P}\delta^{ab}=-\frac{A^{\frac{3}{2}}}{\bar{N}}\frac{\dot{A}}{A}\delta^{ab}, 
\quad \overline{\Pi}=0,\quad\overline{\Pi}_a=0 ~.
\ee
Using $\tilde P$ we can introduce $\tilde{\cal H}$ the analogue of the Hubble parameter on phase space, that is,
\begin{equation}
\label{eq:tildeH}
\tilde{\mathcal{H}} := - \frac{\bar{N}\tilde{P}}{2\sqrt{A}} ~.
\end{equation}
The definition of $\tilde{\mathcal{H}}$ follows from the requirement that $\tilde{\mathcal{H}}$ when pulled back to the tangent bundle should agree with 
conventional Hubble parameter which in the Lagrangian picture is the relative velocity of the scale factor $\mathcal{H} = \frac{\dot{a}}{a}$. Thus we require $\mathcal{LM}^*\tilde{\mathcal{H}} = \mathcal{H}$. Using $\tfrac{\dot{a}}{a}=\tfrac{1}{2}\tfrac{\dot{A}}{A}$ and the equation of motion for $A$ in (\ref{eq:dotAdotPcosm}) we get exactly the expression in (\ref{eq:tildeH}). 
We denote the canonically conjugate momentum of $A$ as $P_A$ which is related to ${\tilde P}$ by ${P_A}(t)=3\tilde{P}(t)$, where the factor of 3 is due to the trace of $\delta^{ab}$ being equal to 3.

The momenta of lapse and shift, $\overline{\Pi}=0$ and $\quad\overline{\Pi}_a=0$, are primary constraints and are assumed to be satisfied for the background solution. A stability analysis of the primary constraints yields for flat FLRW cosmologies the following secondary constraints:
\begin{align}
\label{eq:barCbarCa}
\bar{C} &= -\frac{3}{2}\sqrt{A}\tilde{P}^2 + \kappa A^{3/2}\rho = 0 , ~~~~~~~~~~ \mathrm{and}, \nonumber \\
\bar{C}_a &= 0 ~.
\end{align}
Here $\bar{C}$ is the background Hamiltonian constraint and $\bar{C}_a$ the background spatial diffeomorphism constraints. That the latter trivially vanish for FLRW spacetimes is expected because it is linear in the  spatial derivatives of momenta and configuration variables, and these derivatives need to vanish in order to be consistent with the homogeneity and isotropy symmetries assumed for FLRW solutions.  Thus, the symmetry reduced ADM Hamiltonian for flat FLRW cosmologies reads:
\begin{equation}
H=\int d^3x \left(\bar{N}\bar{C}+\bar{N}^a\bar{C}_a+\bar{\lambda}\overline{\Pi}+\bar{\lambda}^a\overline{\Pi}_a\right)
=\int d^3x \bar{N}\bar{C} ~.
\end{equation}
Here in the last step we used the fact that for the background solution the primary constraints are satisfied and that the shift vector identically vanishes.

The Hamiltonian equations of motion can then be derived by computing the Poisson bracket of the canonically conjugate phase space variables $A$ and $3\tilde{P}$. An alternative but equivalent way to obtain these equations of motion is to specialize the full relativistic equations of motions for $q_{ab}$ and $P^{ab}$ to the case of flat FLRW cosmologies. In both cases we end up with the following equations for $\dot{A}$ and $\dot{\tilde{P}}$:
\begin{align}
\label{eq:dotAdotPcosm}
\dot{A} &= -\bar{N}\sqrt{A}\tilde{P}, & \dot{\tilde{P}} &= \bar{N}\left( \frac{1}{4}\frac{\tilde{P}^2}{\sqrt{A}} + \frac{\kappa}{2}\sqrt{A}p \right) ~.
\end{align}
Using $\tilde{\cal H}$ these equations can be written as:
\begin{align}
\dot{A} &= 2\tilde{\mathcal{H}}A,& \dot{\tilde{P}} &= -\frac{1}{2}\tilde{\mathcal{H}}\tilde{P} + \frac{\kappa}{2}\bar{N}\sqrt{A}p ~.
\end{align}
As the matter content we introduce a minimally coupled Klein-Gordon scalar field with an arbitrary potential $V(\bar{\varphi})$ whose phase space variables we denote by $\bar{\varphi}$ and $\bar{\pi}_\varphi$. In order that these variables comply with the symmetries of  FLRW spacetime, both $\bar{\varphi}$ and $\bar{\pi}_\varphi$ do only depend on the temporal coordinate. The corresponding contribution to the Hamiltonian constraint is given by:
\begin{equation}
\bar{C}_\varphi=\frac{\kappa}{2\lambda_\varphi}\left(\frac{\lambda_\varphi^2 \bar{\pi}_\varphi^2}{\sqrt{\det(\bar{q})}}+V(\varphi)\right) .
\end{equation}
Here the coupling constant $\lambda_\varphi$ should not be confused with the Lagrange multiplier $\lambda$ associated with the primary constraints. Also, $\mathrm{det}(\bar q) = A^3$ with $\bar q^{ab} = \tfrac{1}{A} \delta^{ab}$. The total Hamiltonian constraint becomes $\bar{C}^{\rm tot}=\bar{C}^{\rm geo}+\bar{C}_\varphi$. The resulting 
Hamiltonian equations for the matter variables are:
\begin{align}
\label{eq:dotbarvarphidotbarpivarphi}
\dot{\bar{\varphi}} &= \bar{N}\frac{\lambda_\varphi}{A^{3/2}}\bar{\pi}_\varphi, & \dot{\bar{\pi}}_\varphi = -\bar{N}\frac{A^{3/2}}{\lambda_\varphi}\frac{1}{2}\frac{\mathrm{d}V}{\mathrm{d}\varphi}(\bar{\varphi}) ~.
\end{align}
As usual for cosmological models we introduce the associated energy density and pressure for the matter content that has for the scalar field the following form:
\begin{align}
\label{eq:rhop}
\rho &= \frac{1}{2}\left( \frac{\lambda_\varphi}{A^3}\bar{\pi}_\varphi^2 + \frac{1}{\lambda_\varphi}V(\bar{\varphi}) \right) & p &=\frac{1}{2}\left( \frac{\lambda_\varphi}{A^3}\bar{\pi}_\varphi^2 - \frac{1}{\lambda_\varphi}V(\bar{\varphi}) \right).
\end{align}
Let us note here that in our notation the potential $V(\varphi)$ is twice the usual value of the potential due to on overall factor of $\frac{1}{2}$ that we chose for the scalar field action. For example, in the case of the usual quadratic inflationary potential, $V(\varphi)$ above will be $m^2 \varphi^2$ and not $\tfrac{1}{2} m^2 \varphi^2$. 

Given these we can rewrite the Hamiltonian equations of the scalar field as first order differential equations for $\rho$ and $p$:
\begin{align}
\label{eq:dotrhodotp}
\dot{\rho} &= -3\tilde{\mathcal{H}}(\rho+p) & \dot{p} &= -3\tilde{\mathcal{H}}(\rho + p) - \frac{\bar{N}}{A^{3/2}}\bar{\pi}_\varphi \frac{\mathrm{d}V}{\mathrm{d}\varphi}(\bar{\varphi}),
\end{align}
where we have used the definition of $\rho$, $p$ and the equations of motion for the background scalar field in (\ref{eq:dotbarvarphidotbarpivarphi}). 
~\\
~\\
\paragraph*{\centerline{1.b Perturbations around flat FLRW spacetimes in extended ADM phase space}}
~\\
Next, we want to consider perturbations around the spatially flat FLRW solution that was discussed in the last subsection. Since we work in extended phase space we will consider independent perturbations of all 10 metric degrees of freedom $q_{ab},N,N^a$ and their conjugate momenta $P^{ab},\Pi,\Pi_a$. In addition to the gravitational sector, we also have to introduce the perturbations of the minimally coupled scalar field. We obtain:
\begin{align}
q_{ab} &= \bar{q}_{ab} + \delta q_{ab}, & P^{ab} &= \bar{P}^{ab} + \delta P^{ab}, &
N &= \bar{N} + \delta N, & N^a &= \bar{N}^a + \delta N^a, \nonumber \\
\varphi &= \bar{\varphi} + \delta \varphi, & \pi_\varphi &= \bar{\pi}_\varphi+\delta \pi_\varphi, & \Pi &= \bar{\Pi} + \delta \Pi, & \Pi_a &= \bar{\Pi}_a+\delta \Pi_a.
\end{align}
Considering the explicit form of the flat FLRW solution discussed above these phase space variables simplify to:
\begin{align}
\label{eq:FLRWPert}
q_{ab}(\vec{x},t) &= A(t)\delta_{ab} + \delta q_{ab}(\vec{x},t), & P^{ab}(\vec{x},t) &= \tilde{P}(t)\delta^{ab} + \delta P^{ab}(\vec{x},t), \nonumber \\
N(\vec{x},t) &= \bar{N}(t) + \delta N(\vec{x},t), &  \Pi(\vec{x},t) &= \delta \Pi(\vec{x},t), \nonumber \\
N^a(\vec{x},t) &=  \delta N^a(\vec{x},t), & \Pi_a(\vec{x},t) &= \delta \Pi_a(\vec{x},t),\nonumber\\
\varphi(\vec{x},t) &= \bar{\varphi}(t) + \delta \varphi(\vec{x},t), &
\pi_\varphi(\vec{x},t) &= \bar{\pi}_\varphi(t)+\delta \pi_\varphi(\vec{x},t).
\end{align}
The linearized Einstein equations for a generic background have been derived in our companion paper \cite{Giesel:2017roz}. Moreover, these equations have been specialized to the case of the choice of a flat FLRW cosmological background spacetime. Therefore, we will only present the final results here and refer the reader to \cite{Giesel:2017roz} for a more detailed presentation. The equations of motion for the perturbation of the spatial metric and its momentum are given by:  
\begin{align}
\label{eq:deltadotqflat}
\delta \dot{q}_{ab} &= 2\tilde{\mathcal{H}}A\delta_{ab}\frac{\delta N}{\bar{N}}-2\tilde{\mathcal{H}} \left( \delta^c_a\delta^d_b-\frac{1}{2}\delta_{ab}\delta^{cd} \right) \delta q_{cd} - 4\tilde{\mathcal{H}} \frac{A}{\tilde{P}} \left( \delta_{ac}\delta_{bd}-\frac{1}{2}\delta_{ab}\delta_{cd} \right) \delta P^{cd} + 2 \delta N_{(a,b)},
\end{align}
and 
\begin{align}
\label{eq:deltadotPflat}
\delta \dot{P}^{ab} &= \frac{1}{4} \frac{1}{\sqrt{A}} \tilde{P}^2\delta^{ab} \delta N +\frac{1}{\sqrt{A}} \left( \partial^a \partial^b-\delta^{ab}\Delta \right)\delta N \nonumber \\
~ &- \bar{N}\left[ \frac{1}{A^{3/2}}\tilde{P}^2\left( \frac{5}{4}\delta^{ac}\delta^{bd}-\frac{3}{8}\delta^{ab}\delta^{cd} \right)\delta q_{cd} +\frac{1}{\sqrt{A}}\left( \delta^{ac}\delta^{bd}-\frac{1}{2}\delta^{ab}\delta^{cd} \right)\delta R^{(3)}_{cd} \right. \nonumber \\
~ &+ \left. \frac{1}{\sqrt{A}}\tilde{P}\left( \delta^a_c\delta^b_d-\frac{1}{2}\delta^{ab}\delta_{cd} \right)\delta P^{cd} \right] +\tilde{P}\left( \delta N^c_{,c} \delta^{ab}-2\delta N^{(a,b)} \right) \nonumber \\
~ &+ \frac{\kappa}{2}\bar{N}\sqrt{A} \left[ p \delta^{ab} \frac{\delta N}{\bar{N}}-\left( p \delta^{ac}\delta^{bd}+\frac{1}{2}\rho\delta^{ab}\delta^{cd} \right) \frac{\delta q_{cd}}{A} + \mathcal{P}\delta^{ab} \right]~.
\end{align}
Here we have introduced $\mathcal{P}$ and $\mathcal{E}$ as the perturbations of the  energy-density and pressure restricted to those terms that contain perturbations of the scalar field and its momentum. These are given by:
\begin{equation}
\mathcal{P} := \frac{\lambda_\varphi}{A^3}\bar{\pi}_\varphi \delta \pi_\varphi - \frac{1}{2\lambda_\varphi}\frac{\mathrm{d}V}{\mathrm{d}\varphi}(\bar{\varphi})\delta \varphi ,
\end{equation}
and
\begin{equation}
\mathcal{E} := \frac{\lambda_\varphi}{A^3}\bar{\pi}_\varphi \delta \pi_\varphi + \frac{1}{2\lambda_\varphi}\frac{\mathrm{d}V}{\mathrm{d}\varphi}(\bar{\varphi})\delta \varphi .
\end{equation}
Note, that these expressions only contain the perturbations of scalar field and its momentum, and the terms involving perturbations of the geometry are not included.

The perturbed secondary constraints turn out to be \cite{Giesel:2017roz}:
\begin{align}
\label{eq:deltaCcosmo}
\delta C &= -\frac{1}{4}\frac{\tilde{P}^2}{\sqrt{A}}\delta^{ab}\delta q_{ab} - \sqrt{A}\tilde{P}\delta_{ab}\delta P^{ab} - \frac{1}{\sqrt{A}}\left( \partial^a\partial^b - \delta^{ab}\Delta \right) \delta q_{ab} \nonumber \\
~ & + \kappa\left( -\frac{\sqrt{A}}{2}p \delta^{ab} \delta q_{ab} + \mathcal{E} \right), 
\end{align}
and
\begin{align}
\label{eq:deltaCacosmo}
\delta C_a &= -2A \delta_{ab}\delta P^{bc}_{,c} - 2\tilde{P} \left( \delta^b_a\partial^c - \frac{1}{2}\delta^{bc}\partial_a \right) \delta q_{bc} + \kappa \bar{\pi}_\varphi \delta \varphi_{,a}.
\end{align}
It is to be mentioned that the above equations agree with the work of Langlois in  \cite{Langlois} (see equations (19) and (20) in \cite{Langlois}). For a comparison, the following differences in notation must be considered: $\kappa = 2 \kappa_\text{Langlois}$, $\tfrac{1}{2}V =V_\text{Langlois}$, $\varphi$ is denoted as $\phi$, $A = \mathrm{e}^{2\alpha}$, $\bar{q}_{ab}$, $\bar{P}^{ab}$ are denoted by  $\gamma_{ij}$ and $\pi^{ij}$ respectively, $\pi_\alpha = 6A\tilde{P}$  and the respective Poisson bracket of $\gamma_{ij}$ with $\pi^{ij}$ does not involve $\kappa$. The equations of motion for lapse and shift and their momenta turn out to be:
\begin{equation}
\label{eq:deltaDotLapseShift}
\delta \dot{N} = \delta \lambda,\quad 
\delta \dot{\Pi} = -\delta C, \quad \delta \dot{N}^a=\delta \lambda^a,\quad\delta \dot{\Pi}_a = -\delta C_a.
\end{equation}
These differential equations for lapse and shift involve the perturbations of the Lagrange multipliers $\lambda$ and $\lambda^a$ associated with the primary constraints. Expressing these in terms of background quantities and perturbations we obtain:
\begin{align}
\lambda &= \bar{\lambda} + \delta \lambda = \dot{\bar{N}} + \delta \lambda \nonumber, \\
\lambda^a &= \bar{\lambda}^a + \delta \lambda^a = \delta \lambda^a ~.
\end{align}
Finally, the equations of motion for the perturbation of scalar field and its momentum are of the form: 
\begin{align}
\delta \dot{\varphi} &= \delta N \frac{\lambda_\varphi}{A^{3/2}}\bar{\pi}_\varphi + \bar{N}\frac{\lambda_\varphi}{A^{3/2}}\left( \delta \pi_\varphi - \frac{1}{2A} \bar{\pi}_\varphi \delta^{ab}\delta q_{ab} \right), \nonumber \\
\delta \dot{\pi}_\varphi &=  - \delta N \frac{A^{3/2}}{\lambda_\varphi}\frac{1}{2}\frac{\mathrm{d}V}{\mathrm{d}\varphi}(\bar{\varphi}) + \bar{N}\frac{A^{3/2}}{\lambda_\varphi}\left( \frac{1}{A}\Delta \delta \varphi - \frac{1}{4A}\frac{\mathrm{d}V}{\mathrm{d}\varphi}(\bar{\varphi})\delta^{ab}\delta q_{ab} - \frac{1}{2}\frac{\mathrm{d}^2V}{\mathrm{d}\varphi^2}(\bar{\varphi})\delta \varphi \right) \nonumber \\
~ & \hspace{1em} + \bar{\pi}_\varphi \delta N^a_{,a}.
\end{align}
This concludes our discussion on the equations of motion at the level of linear cosmological perturbation theory.
~\\
~\\
\paragraph*{\centerline{1.c Scalar-vector-tensor decomposition of the perturbations}}
~\\
One of the main advantages of a decomposition of a generic tensor into scalar, vector and tensor parts is that in linear cosmological perturbation theory the corresponding equations of motion completely decouple and thus can be analyzed independently. As a consequence, the task of finding solutions for the linearized Einstein equations presented in the next section simplifies. Again a detailed introduction of how such projectors onto the scalar, vector and tensor components can be defined can be found in our companion paper \cite{Giesel:2017roz}. In this work we will just list  the results that are needed for our analysis.

We want to define projectors for k=0 FLRW spacetimes that decompose a given symmetric tensor of rank 2 into its scalar, vector and tensor part. For this purpose we define the following differential operators:
\begin{align}
\label{eq:FLRWOper}
D^a &= \frac{1}{A}\partial^a, & \boldsymbol{\Delta} &= \frac{1}{A}\Delta, & (M^{-1})^a_b &= A\Delta^{-1}\left( \delta^a_b - \frac{1}{4}\Delta^{-1}\partial_a\partial^b \right),
\end{align}
where $\Delta^{-1}$ stands again for the Green's function of the Poisson equation,  with the associated integral kernel of $\Delta^{-1}$ denoted by $G(x,y)$, that is:
\begin{equation}
\Delta^{-1}f(x) = \int_{\bar{\Sigma}}\mathrm{d}^3y~ G(x,y)f(y).
\end{equation}
These operators allow us to define a Helmholtz decomposition of a vector into its longitudinal (scalar) and transversal parts:
\begin{align}
(\hat{P}_SV) &= A\Delta^{-1} \partial_aV^a, \nonumber \\
(\hat{P}_\perp V)^a &= V^a-\partial^a(\hat{P}_SV) . \nonumber\\
\end{align}
Further, if we introduce a projector that projects a second rank tensor onto its trace given by
\begin{equation}
(\hat{P}_\text{Tr}T) = \frac{1}{3A}\delta^{ab}T_{ab},
\end{equation}
we can formulate the following projectors:
\begin{align}
\label{eq:projectorscosmology}
(\hat{P}_{L}T)_a &:= \left( M^{-1} \right)_a^b \left[ \partial^cT_{bc}-\frac{1}{3}\partial_b(\hat{P}_\text{Tr}T) \right],\nonumber\\
(\hat{P}_\text{TT}T)_{ab} &=T_{ab} - (\hat{P}_{Tr}T)q_{ab} - 2\partial_{<a}(\hat{P}_{L}T)_{b>},\nonumber\\
(\hat{P}_{LS}T) &= \frac{3}{4}\Delta^{-2}\partial^{<a}\partial^{b>}T_{ab} ,\nonumber \\
(\hat{P}_{LT}T)_a &= \Delta^{-1}\left( \delta_a^b\partial^c -\tfrac{1}{3}\partial_a\delta^{bc} \right) T_{bc} - \frac{4}{3}\partial_a (\hat{P}_{LS}T),
\end{align}
where $L,TT,LS,LT$ are abbreviations for longitudinal, transverse-traceless, longitudinal-scalar and longitudinal-traceless respectively. Here $T_{<ab>} = T_{ab} - \tfrac{1}{3} \delta_{ab} \delta^{cd} T_{cd}. $
Given these projectors we can  decompose the metric as well as the matter perturbations shown in (\ref{eq:FLRWPert}) into their scalar, vector and tensor parts. Unfortunately, no uniform notation exists in the literature for the decomposed quantities, but we will keep our  notation close to that of  \cite{Mukhanov} and explicitly mention when we use for instance different sign conventions. We define, 
 \begin{align}
\label{eq:decomposedmetricdef}
\phi &:= \frac{\delta N}{\bar{N}}, & B &:= \frac{1}{A}(\hat{P}_S\delta \vec{N}), & S^a &:= (\hat{P}_\perp\delta \vec{N})^a, \nonumber \\
\psi &:= \frac{1}{2}(\hat{P}_\text{Tr}\delta q), & E &:= \frac{1}{A}(\hat{P}_{LS}\delta q), & F_a &:= \frac{1}{A}(\hat{P}_{LT}\delta q), & h^{TT}_{ab} &:= \frac{1}{A}(\hat{P}_{TT}\delta q)_{ab} ~.
\end{align}
Let us note that in  \cite{Mukhanov},  $\psi$ is defined with a different sign, and in \cite{Dust2} a different sign for $\phi$ is used.  $E$ is often defined such that $\Delta E = 0$ or with a  trace part \cite{Dust2}. For the latter, one uses $E_{,ab}$ instead of $E_{,<ab>}$ in  $\delta q_{ab}$. The choice of different signatures and defining shift vector with a different sign, such as in \cite{bardeen},  further complicates consistency in notation. 

Above perturbed quantities, up to the background factors,  are precisely the  ones obtained by the scalar-vector-tensor decomposition of the spatial metric perturbation $\delta q_{ab}$ and the Helmholtz decomposition of the shift vector field perturbation $\delta \vec{N}$. We can thus write the  perturbed quantities in terms of 4 scalars, 2 transversal vector fields and 1 traceless-transversal tensor field:
\begin{align}
\delta N &= \bar{N}\phi,\\
\delta N^a &= B^{,a} + S^a, ~~~~~ \mathrm{and}\nonumber \\
\delta q_{ab} &= 2A\left( \psi \delta_{ab} + E_{,<ab>} + F_{(a,b)} + \tfrac{1}{2}h^{TT}_{ab} \right) ~.
\end{align}
As we are working on phase space we have to perform a similar decomposition also for the perturbed conjugate  momenta. Analogously to the choice of variables for $\delta q_{ab}$ we define:
\begin{align}
\label{eq:decomposedmomentadef}
p_\psi &:= \frac{1}{2}\frac{1}{A\tilde{P}}(\hat{P}_\text{Tr}\delta P), & p_E &:= \frac{1}{A^2\tilde{P}} (\hat{P}_{LS}\delta P), & p_F^a &:= \frac{1}{A\tilde{P}} (\hat{P}_{LT}\delta P)^a, & p_{h^{TT}}^{ab} &:= \frac{1}{\tilde{P}}(\hat{P}_{TT}\delta P)^{ab} ~.
\end{align}
Thus, we can write $\delta P^{ab}$ in terms of different projected components:
\begin{equation}
\delta P^{ab} = 2\tilde{P}\left( p_\psi\delta^{ab} + p_E^{,<ab>} + p_F^{(a,b)} + \tfrac{1}{2}p_{h^{TT}}^{ab} \right) ~.
\end{equation}
For the temporal-temporal and temporal-spatial part of the perturbed metric we introduce the following decomposition of the associated momenta:
\begin{eqnarray}
p_\phi &:=&\frac{1}{\overline{N}}\delta\Pi,\nonumber\\
p_B &:=&\frac{1}{A}(\hat{P}_S\delta\vec{\Pi})=:\delta\hat{\Pi},\quad p_{S^a}=(\hat{P}_\perp\delta\vec{\Pi})^a=:\delta\Pi^a_\perp
\end{eqnarray}
where as before $\delta\Pi,\delta\vec{\Pi}$ denote the conjugate momenta of the perturbed lapse function $\delta N$ and the perturbed shift vector $\delta\vec{N}$ respectively.
~\\
~\\
\paragraph*{\centerline{1.d Equations of motions for the scalar, vector and tensor perturbations}}
~\\
To find the equations of motion of the scalar, vector and tensor perturbations we can use the Hamiltonian equations for $\delta{q}_{ab},\delta N^a, \delta N$ 
as well as for their conjugate momenta in (\ref{eq:deltadotqflat}),  (\ref{eq:deltadotPflat}) and (\ref{eq:deltaDotLapseShift}). For this exercise, we note that if the projectors depend explicitly on time then we need to also consider the time derivatives of the projectors.

Using $\psi = \frac{1}{6A}\delta^{ab}\delta q_{ab}$ and $E = \frac{1}{A}(\hat{P}_{LS}\delta q)$, along with (\ref{eq:deltadotqflat}), we obtain:
\begin{align}
\label{eq:dotpsiE}
\dot{\psi} &= 2\tilde{\mathcal{H}}\left( p_\psi - \frac{1}{2}\psi \right) + \tilde{\mathcal{H}}\phi + \frac{1}{3}\Delta B, \nonumber \\
\dot{E} &= -4\tilde{\mathcal{H}}(E+p_E) + B.
\end{align}
Let us note that the decomposition of perturbations allows us to write the perturbation in the spatial curvature as:
\begin{equation}
\label{eq:delRdecomposed}
\delta R^{(3)}_{ab} = - \frac{4}{3}\Delta \left( \psi -\frac{1}{3}\Delta E \right)\delta_{ab} - \left( \psi -\frac{1}{3}\Delta E \right)_{,<ab>} - \frac{1}{2}\Delta h^{TT}_{ab}.
\end{equation}
 It is also useful to introduce  the spatial energy momentum perturbation: 
\begin{equation}
\delta \tilde{T} := \frac{1}{A^{3/2}}\left( \mathcal{P} - 3(\rho+p)\psi \right).
\end{equation}

Using  the above equation together with  (\ref{eq:delRdecomposed}) and (\ref{eq:deltadotPflat}), it is straightforward to obtain the 
equations of motion for $p_\psi$ and $p_E$:
\begin{align}
\label{eq:dotGammaSigma}
\dot{p}_\psi &= \frac{1}{6}\frac{\bar{N}^2}{A\tilde{\mathcal{H}}}\Delta\left( \phi+\psi-\frac{1}{3}\Delta E \right) +\left( -\frac{1}{2}\tilde{\mathcal{H}} + \frac{\kappa}{4}\frac{\bar{N}^2}{\tilde{\mathcal{H}}}p \right)\left( p_\psi-\frac{1}{2}\psi \right) - \frac{\kappa}{8}\frac{\bar{N}^2}{\tilde{\mathcal{H}}}\delta\tilde{T} \nonumber \\
~ &-\frac{1}{2}\left( \frac{1}{2}\tilde{\mathcal{H}} + \frac{\kappa}{4}\frac{\bar{N}^2}{\tilde{\mathcal{H}}}p \right)\phi + \frac{1}{6}\Delta B,  \nonumber \\
\dot{p}_E &= -\frac{1}{4}\frac{\bar{N}^2}{A\tilde{\mathcal{H}}} \left( \phi+\psi-\frac{1}{3}\Delta E \right)+\left(\frac{5}{2}\tilde{\mathcal{H}} + \frac{\kappa}{4}\frac{\bar{N}^2}{\tilde{\mathcal{H}}}p \right) (E+p_E) -B~.
\end{align}

The equations of motion for the vector and tensor perturbations can be derived analogously by applying the corresponding projectors onto $\delta q_{ab}$ and $\delta P^{ab}$. For the vector perturbations, these turn out to be: 
\begin{equation}
\frac{\mathrm{d}}{\mathrm{d}t} \left[ \begin{matrix}
\delta^{ab}F_b \\
p_F^a
\end{matrix} \right] = \left[ \begin{matrix}
-4\tilde{\mathcal{H}} & -4\tilde{\mathcal{H}} \\
\digamma & \digamma
\end{matrix} \right] \left[ \begin{matrix}
\delta^{ab}F_b \\
p_F^a
\end{matrix} \right] + \left[ \begin{matrix}
S^a \\
-S^a 
\end{matrix} \right] ~.
\end{equation}
And, for the tensor perturbations, we obtain:
\begin{equation}
\frac{\mathrm{d}}{\mathrm{d}t} \left[ \begin{matrix}
\delta^{ac}\delta^{bd}h^{TT}_{cd} \\
p_{h^{TT}}^{ab}
\end{matrix} \right] = \left[ \begin{matrix}
-4\tilde{\mathcal{H}} & -4\tilde{\mathcal{H}} \\
\digamma+\varpi\Delta & \digamma
\end{matrix} \right] \left[ \begin{matrix}
\delta^{ac}\delta^{bd}h^{TT}_{cd} \\
p_{h^{TT}}^{ab}
\end{matrix} \right] ~.
\end{equation}
Here, we have introduced
\begin{align}\label{eq:digamma}
\digamma &:= \frac{5 {\tilde {\cal H}}}{2} +\frac{\kappa}{4}\frac{\bar{N}^2 p}{\tilde{\mathcal{H}}} ~~~~~~~~~~~~~~~~\mathrm{and}~~~~~~ & \varpi &:= -\frac{\bar{N}^2}{4A\tilde{\mathcal{H}}} ~.
\end{align}

Next, we want to discuss the equations of motion of the decomposed quantities that are associated with the perturbed lapse and shift as well as their conjugate momenta. The dynamics of the lapse and shift perturbations are related to yet undetermined functions which are the perturbations of the Lagrange-multipliers $\lambda = \bar{\lambda}+\delta \lambda$, $\lambda^a = \bar{\lambda}^a+\delta \lambda^a$. Using the decomposition of perturbation in Lagrange multiplier $\delta \lambda^a$ in terms of scalar and transversal parts:  $\delta \lambda^a = \delta \hat{\lambda}_{,a} + \delta \lambda^a_\perp$, we obtain,
\begin{align}
\label{eq:dotphidotB}
\dot{\phi} &= -\frac{\dot{\bar{N}}}{\bar{N}}\phi + \frac{\delta \lambda}{\bar{N}}, & \dot{B} &= \delta \hat{\lambda}, & \dot{S}^a &= \delta \lambda^a_\perp .
\end{align}

Using  the projected quantities, and  (\ref{eq:deltaCcosmo}) and (\ref{eq:deltaCacosmo}), the   linearized secondary constraints become: 
\begin{align}
\label{eq:deltaCdeltaCa}
\delta C &=  -\frac{3}{2}\sqrt{A}\tilde{P}^2(\psi+4 \,p_\psi) + 4\sqrt{A}\Delta\left( \psi -\frac{1}{3}\Delta E \right) + \kappa A^{3/2} \left( -3 \, p \, \psi + \mathcal{E} \right) \nonumber \\
\delta C_a &= -4A\tilde{P}\left( p_\psi - \frac{1}{2}\psi + \frac{2}{3}\Delta (E+p_E) \right)_{\hskip-0.15cm,a} - 2A\tilde{P}\Delta(F_a+p_F^b\delta_{ab}) + \kappa \bar{\pi}_\varphi\delta \varphi_{\hskip-0.05cm,a}.
\end{align}
These perturbations of the secondary constraints stabilize the perturbations of the primary constraints $\delta \Pi$, $\delta \Pi_a$:
\begin{align}
\delta \dot{\Pi} &= - \delta C, & \delta \dot{\Pi}_a &= -\delta C_a.
\end{align}

Let us note that the form of the perturbed Hamiltonian constraint $\delta C$ agrees with one in for example  \cite{Dust2} (equation (19)), where all perturbed quantities were constructed with dust clocks. However, there are differences in notation. There the authors use $\bar{N}=\sqrt{A}$, $\phi = 0$,   $\Xi$ denotes the scalar field, and  $\psi$ corresponds to $\psi + \tfrac{1}{3}\Delta E$. 
Note that due to dust clocks, the perturbed quantities  have a different interpretation because one starts from a different model, that is gravity with a minimally coupled scalar field plus the coupled Brown-Kucha\v{r} dust. 

For the reason that the perturbed spatial diffeomorphism constraint is a covector it can be decomposed into a scalar part $\delta \hat{C}$ and a transversal part $\delta C^\perp_a$ yielding:
\begin{align}
\delta \hat{C} &= -4A\tilde{P}\left( p_\psi - \frac{1}{2}\psi + \frac{2}{3}\Delta (E+p_E) \right) + \kappa \bar{\pi}_\varphi\delta \varphi, \nonumber \\
\delta C^\perp_a &= 2A\tilde{P}(F_a+p_F^b\delta_{ab}).
\end{align}
Finally, let us discuss the Hamiltonian equations of the scalar matter field perturbations. Because $\delta\varphi$ is already a scalar and $\delta\pi_\varphi$ is a scalar density no further decomposition of these quantities has to be performed. However, since also geometrical degrees of freedom are involved in the matter equations of motion, of course these equations need also to be rewritten in terms of the decomposed quantities. We get,
\begin{align}
\delta \dot{\varphi} &= \bar{N}\frac{\lambda_\varphi}{A^{3/2}}\bar{\pi}_\varphi\left( \phi -3\psi + \frac{\delta \pi_\varphi}{\bar{\pi}_\varphi} \right), \nonumber \\
\delta \dot{\pi}_\varphi &= \bar{N}\frac{A^{3/2}}{\lambda_\varphi}\left[  -\frac{1}{2}\frac{\mathrm{d}V}{\mathrm{d}\varphi}(\bar{\varphi})(\phi+3\psi) +\frac{1}{A}\Delta\delta\varphi -\frac{1}{2}\frac{\mathrm{d}^2V}{\mathrm{d}\varphi^2}(\bar{\varphi})\delta\varphi \right].
\end{align}
After having presented all necessary equations of motion that will be relevant for our analysis later, in the next session we give a brief introduction to the relational formalism and how it can be used to construct gauge invariant quantities in  general relativity.

\subsection{Relational formalism and Dirac observable in extended ADM-phase space}

In the last subsection we discussed the way linear cosmological perturbation theory can be formulated in terms of appropriate variables in extended ADM phase space such that we can analyze the scalar, vector and tensor sector independently because their corresponding equations of motion fully decouple. However, all of the scalar and vector variables introduced above share the common feature that they are not invariant under gauge transformation, that means invariance under arbitrary coordinate transformations also called diffeomorphisms or also called gauge transformations in the context of general relativity. As we analyze linear cosmological perturbation theory this invariance is required order by order and thus in our case we want to construct quantities that are invariant under linearized gauge transformations. We denote the resulting gauge invariant quantity as the gauge invariant extension of the corresponding perturbed quantity. Note that the tensor projections are already invariant under these transformations up to linear order.

In the Lagrangian framework the strategy one adopts is as follows. For the scalar as well as the vector projections one chooses a certain subset of the variables and considers the way members of these subsets transform under linearized gauge transformations. Then one adds to these chosen variables specific combinations of the remaining variables such that the final sum of the original chosen variables together with these specific combinations lead to a quantity that is invariant under linearized gauge transformations. 

Now we aim at doing the same construction in the Hamiltonian picture. A crucial concept that enters the above construction is that one considers two sets of variables both not  invariant under gauge transformations. Then one combines them in an appropriate way so as to obtain a quantity that is invariant under (linearized) gauge transformations. This construction can be naturally embedded into the so called relational formalism where gauge invariant extension of a given variable of the first of those sets is defined with respect to so called reference fields, that play the role of the second set of variables mentioned above. In order to explain how this embedding can be actually performed in the first instance we will briefly summarize how (linearized) diffeomorphisms can be implemented on the extended ADM-phase space following the seminal work of  Pons, Salisbury and Sundermeyer et al \cite{Pons1,Pons2,Pons3,Pons4}. Afterwards, we will discuss how this can be used to define a so called observable map that maps each phase space variable to its gauge invariant extension once a set of reference fields which are often also called clock fields has been chosen. 
A more detailed introduction to these topics can be found in \cite{Giesel:2017roz} and references therein. Here we are particularly interested in understanding how a given choice of clock fields is related to a certain choice of a gauge fixing condition at the Lagrangian level. Formulated in a different way we want to address the following question: What are the appropriate clock or reference fields that we must choose in order to obtain in the Hamiltonian framework gauge invariant extensions that correspond for example to the Bardeen potentials and the Mukhanov-Sasaki variable?
~\\
~\\
\paragraph*{\centerline{2.a Diffeomorphisms on extended ADM-phase space}}
~\\
Before discussing how diffeomorphism are implemented on the extended ADM-phase space, let us briefly recall how this can be done on the reduced ADM-phase space where the primary constraints $\Pi\approx 0$ and $\Pi_a\approx 0$ have already been imposed. In this case the dynamical degrees of freedom in the gravitational sector are just given by the ADM metric and its conjugate momenta $(q_{ab},P^{ab})$. Under diffeomorphism $q_{ab},P^{ab}$ will transform accordingly and in the Hamiltonian formulation these transformations can be formulated by means of  the spatial diffeomorphism constraint $C_a$ and  the Hamiltonian constraint $C$ respectively. The first one $C_a$ generates spatial diffeomorphisms within the spatial hypersurfaces one obtains from the standard ADM 3+1 decomposition of the four dimensional spacetime. The second one generates diffeomorphism orthogonal to these hypersurfaces. Note that for both generators this is only true on-shell, that is when the equations of motion as well as the constraints are satisfied. In order to introduce a notation for the action of these generators on the reduced ADM phase space, let us define the smeared versions of the constraints given by:
\begin{equation}
C[b]:=\int d^3x C(x)b(x) \quad{\rm and}\quad C_a[b^a]:=\int d^3x C_a(x)b^a(x),    
\end{equation}
where we call $b$ and $b^a$ smearing or test functions respectively. Then the action of the generators on a generic phase space function $f$ can then be written as:
\begin{equation}
\delta_{\vec{b}} f:=\{f, C_a[b^a]\}\quad{\rm and}\quad
\delta_{b} f:=\{f, C[b]\}
\end{equation}
If we restrict to the gravitational sector for the moment, then all functions $f$ on the reduced ADM-phase space can be understood as functions of the elementary variables $q_{ab},P^{ab}$. Consequently if we know the action of the generators $C[b]$ and $C_a[b^a]$ on them, we can easily compute their action on a generic $f$. One obtains on-shell for the spatial diffeomorphisms:
\begin{equation}
\delta_{\vec{b}}q_{ab}=\kappa({\cal L}_{\vec{b}}q)_{ab}\quad
{\rm and}\quad 
\delta_{\vec{b}}P^{ab}=\kappa({\cal L}_{\vec{b}}P)^{ab}
\end{equation}
and
\begin{equation}
\delta_{b}q_{ab}=\kappa({\cal L}_{\vec{b}}q)_{ab}\quad
{\rm and}\quad 
\delta_{b}P^{ab}=\kappa({\cal L}_{\vec{b}}P)^{ab},
\end{equation}
which is exactly the expected transformation behavior. In the case of the orthogonal diffeomorphisms generated by $C[b]$ the discussion is slightly more complicated, see for instance \cite{Thiemann} for a pedagogical presentation. If we consider that in the ADM 3+1 split the spacetime metric $g_{\mu\nu}$ can be expressed as $g_{\mu\nu}=q_{\mu\nu}-n_\mu n_\nu$ where $n_\mu$ is the co-normal vector of a spatial hypersurface and choose a frame in which $n_t=-b$ and $n_a=0$, corresponding to a normal vector $n^\mu$ of the form $n^t=\frac{1}{b}, n^a=-b^a/b$, then on-shell the action of $C[b]$ onto the ADM metric $q_{\mu\nu}$ can be expressed as
\begin{equation}
\delta_{b}q_{\mu\nu}=\kappa({\cal L}_{bn}q)_{\mu\nu},    
\end{equation}
where we used that in this frame $q_{tt}=0$ and $q_{at}=0$. A similar but slightly longer calculation shown for instance in \cite{Thiemann} shows that also for the ADM momenta $P^{ab}$ one gets on-shell the expected transformation behavior under orthogonal diffeomorphisms, that 
can be expressed as:
\begin{equation}
\delta_{b}P^{\mu\nu}=\kappa({\cal L}_{bn}P)^{\mu\nu}.    
\end{equation}
Note, that for the momenta on-shell involves also the constraints and hence this identity holds only on the constraint hypersurface of the Hamiltonian constraint. 

As it will be convenient for our later calculation we consider the combination of the spatial and orthogonal diffeomorphisms and define the following generator:
\begin{align}
\label{eq:G}
G_{b,\vec{b}}&= \frac{1}{\kappa} \left( C[b] + C_a[b^a] \right),
\end{align}
 Note, that if we add matter as for instance a scalar field that we will need later, we can still work with the generator $G_{b,\vec{b}}$ in which then $C$ and $C_a$ denote the total spatial and Hamiltonian constraint. The total constraints consist of the sum of the geometric and matter contributions, that is $C=C^{\rm geo}+C^{\rm matter}$ and similarly for $C_a=C^{\rm geo}_a+C^{\rm matter}_a$. At first we will restrict our discussion to the gravitational sector only and consider a generalization later when needed.

This finishes our discussion on the reduced ADM-phase space. Now, if we aim at going over the extended ADM-phase space, then we realize that since the generator $G_{b,\vec{b}}$ is a linear combination of $C$ and $C_a$ only, it trivially commutes with lapse and shift and their conjugate momenta. As a result, it certainly does not generate diffeomorphisms for the lapse and shift degrees of freedom. Consequently, if we instead work on the extended ADM-phase space where the primary constraints have not yet been reduced and thus we can treat $(N,\Pi)$ and $(N^a,\Pi_a)$ as full phase space variables, we need a modified generator, that we denote by $G'_{b,\vec{b}},$ that also generates diffeomorphisms for these set of variables. Such a generator has been derived by Garcia, Pons, Salisbury, and Shepley in Refs. \cite{Pons1,Pons2}, to which we refer the reader  for more details. A brief summary of the relevant details for the following discussion can be found in our companion paper \cite{Giesel:2017roz}. Here we will just list the results needed for the present analysis. 

In the extended phase space, the modified generator $G'_{b,\vec{b}}$ takes the following form:
\begin{equation}
\label{eq:G'}
G'_{b,\vec{b}} = \frac{1}{\kappa}\left( C[b] + C_a[b^a] + \Pi[\dot{b}+b^aN_{,a}-N^ab_{,a}] + \Pi_a[\dot{b}^a+q^{ab}(bN_{,b}-Nb_{,b})-N^ab^b_{,a}+b^aN^b_{,a}] \right).
\end{equation}
Naively, one could expect that the generator $G_{b,\vec{b}}$ needs to be extended by terms involving the primary constraints because these have a non-trivial action on lapse and shift variables. However, the particular form of the phase space dependent smearing functions for the primary constraints comes from the requirement that one wants to match the implementation of the diffeomorphisms on the extended ADM-phase space with the one in Lagrangian framework, see \cite{Pons2}. Note, that the fact that these extra terms in $G'_{b,\vec{b}}$ are proportional to the primary constraints $\Pi_\mu$ with $\mu=0,\cdots,3$ and $\Pi_0:=\Pi$, has the effect that on fields other than lapse and shift the action of $G'_{b,\vec{b}}$ reduces to the gauge transformation generated by $G_{b,\vec{b}}$ in (\ref{eq:G}).

The transformation of phase space variables under the modified generator  $G'_{b,\vec{b}}$ can be computed in a straightforward way. The configuration variables in the gravitational sector transform according to \cite{Giesel:2017roz}:
\begin{align}
\label{eq:dG'metric}
\delta_{G'_{b,\vec{b}}}N &= b_{,t}-N^ab_{,a}+b^aN_{,a}~, \nonumber \\
\delta_{G'_{b,\vec{b}}}N^a &= q^{ab}(bN_{,b}-Nb_{,b})+b^a_{,t}+b^bN^a_{,b}-N^bb^a_{,b}~, \nonumber \\
\delta_{G'_{b,\vec{b}}}q_{ab} &= \left. \dot{q}_{ab} \right|_{N=b,N^a=b^a},
\end{align}
where we have introduced the following notation
\begin{equation}
    \delta_{G'_{b,\vec{b}}}f:=\{f,G'_{b,\vec{b}}\} ~.
\end{equation}

Similarly, for the corresponding conjugate momenta we obtain:
\begin{align}
\label{eq:dG'momenta}
\delta_{G'_{b,\vec{b}}}\Pi &= (\Pi_aq^{ab}b)_{,b}+\Pi_aq^{ab}b_{,b}+(\Pi b^a)_{,a}~, \nonumber \\
\delta_{G'_{b,\vec{b}}}\Pi_a &= \Pi b_{,a}+(\Pi_ab^b)_{,b}+\Pi_bb^b_{,a}~, \nonumber \\
\delta_{G'_{b,\vec{b}}}P^{ab} &= \left. \dot{P}^{ab} \right|_{N=b,N^a=b^a} + q^{c(a}q^{b)d}\Pi_c(bN_{,d}-Nb_{,d}).
\end{align}
As can be seen from the above equation the primary constraint hypersurface $\Pi\approx 0, \Pi_a\approx 0$ is left invariant under the action of $G'_{b,\vec{b}}$. This is a necessary requirement for the generator as otherwise $G'_{b,\vec{b}}$ would map out of the physical sector. Considering the action on the momenta $P^{ab}$, we realize that $\delta_{G'_{b,\vec{b}}}P^{ab}$ contains an additional term proportional to $\Pi_a$. However, on the physical sector, where all constraints are satisfied, this term clearly vanishes. 

Given that we have a generator of diffeomorphisms on the extended ADM-phase space available, we can use it to analyze how the relevant variables for cosmological perturbations theory transform under linearized diffeomorphisms. As far as the background is considered we assume that we choose fixed coordinates. For linear perturbations  we restrict to the case of infinitesimal diffeomorphisms. This means  that the change of the tensor fields caused by diffeomorphisms is of the order of the perturbations $\epsilon$. The gauge descriptors $b$ and $\vec{b}$ will also be of the same order $\epsilon$. Now considering a generic tensor field $Q$ on the extended ADM-phase space, we obtain the following transformation behavior under linearized diffeomorphisms generated by $G'_{b,\vec{b}}$:
\begin{equation}
Q'=Q + \{ Q,G'_{b,\vec{b}} \} = \bar{Q} + \delta Q + \overline{\{ Q,G'_{b,\vec{b}} \}} + \mathcal{O}(\epsilon^2) .
\end{equation}
Here the bar on the Poisson bracket denotes that it is computed on the background spacetime. To make contact to our former notation introduced above, we define $\delta_{G'}Q := Q -\bar{Q}$ as the perturbation of the variable $Q$. Then we obtain,
\begin{equation}
\delta_{G'_{b,\vec{b}}}Q = \overline{\{ Q,G'_{b,\vec{b}} \}} \, + \, \mathcal{O}(\epsilon^2) ~.
\end{equation}
Note that the above transformation does not affect the background geometry since it does not contain any terms of the order $\epsilon^0$. That is, $\bar Q$ is unaffected by the above diffeomorphism. 
In order to find the change in the linear perturbation $\delta Q^{(1)}$, we consider only those terms which 
are of $\epsilon$ order in the above equation. However, it is to be noted that in general the action of diffeomorphism generator results in terms of order $\epsilon^2$ and higher. Thus, the change in first order perturbations can be written as,
\begin{equation}
(\delta_{G'_{b,\vec{b}}}\delta Q)^{(1)} = \overline{\{ Q,G'_{b,\vec{b}} \}} ~.
\end{equation}
In the following we drop the superscript above the parenthesis for brevity. 
Since we need the transformation behavior of the earlier introduced projected quantities it is convenient to decompose $\vec{b}$ into its scalar and transversal part: $\vec{b} = \hat{b}^{,a}+b^a_\perp$. Looking at the results in equations (\ref{eq:dG'metric}) and (\ref{eq:dG'momenta}) we can rewrite these taking the decomposition of the spatial descriptor $b^a$ into account. This leads to the following form of the transformations:
\begin{align}
\label{eq:dG'metricFRWandMomenta}
\delta_{G'_{b,\vec{b}}}N &= b_{,t} , ~~~~~
\delta_{G'_{b,\vec{b}}}N^a = -\frac{\bar{N}}{A}b^{,a}+\hat{b}^{,a}_{,t}+b^a_{\perp,t} , \nonumber \\
\delta_{G'_{b,\vec{b}}}q_{ab} &= 2A \left[ \left( \frac{\tilde{\mathcal{H}}}{\bar{N}}b+\frac{1}{3}\Delta \hat{b} \right)\delta_{ab} + \hat{b}_{,<ab>}+b^c_{\perp,(a}\delta_{b)c} \right] ,\nonumber \\ \delta_{G'_{b,\vec{b}}}\Pi &= \bar{\Pi}\Delta \hat{b} = 0 , ~~~~~
\delta_{G'_{b,\vec{b}}}\Pi_a = \bar{\Pi}b_{,a} = 0, \nonumber \\
\delta_{G'_{b,\vec{b}}}P^{ab} &= 2\tilde{P} \left[ \left( \left( -\frac{1}{4}\frac{\tilde{\mathcal{H}}}{\bar{N}}-\frac{\kappa}{8}\frac{\bar{N}}{\tilde{\mathcal{H}}}p \right)b +\frac{1}{6}\Delta \left( \frac{\bar{N}}{A\tilde{\mathcal{H}}}b+\hat{b}\right) \right)\delta^{ab} - \left( \frac{\bar{N}}{4A\tilde{\mathcal{H}}}b+\hat{b}\right)^{,<ab>}-b_{\perp}^{(a,b)} \right].
\end{align}
Now, we consider as our matter content a minimally coupled scalar field, already mentioned earlier for the spatially flat FLRW background solution. Adapting the form of the Hamiltonian and spatial diffeomorphism constraint accordingly as discussed above, we further obtain the following transformation behavior of the matter degrees of freedom under linearized diffeomorphisms:
\begin{align}
\label{eq:trafoscalarfield}
\delta_{G'_{b,\vec{b}}}\varphi &= \frac{\lambda_\varphi}{A^{3/2}}\bar{\pi}_\varphi b, & \delta_{G'_{b,\vec{b}}}\pi_\varphi &= \bar{\pi}_\varphi\Delta\hat{b}-\frac{1}{2}\frac{A^{3/2}}{\lambda_\varphi}\frac{\mathrm{d}V}{\mathrm{d}\varphi}(\bar{\varphi})~b.
\end{align}
As the final step we use the results in (\ref{eq:dG'metricFRWandMomenta}) and (\ref{eq:trafoscalarfield}) to derive the way projected quantities transform under linearized diffeomorphisms. For this purpose we take advantage of the fact that all of the projectors introduced in (\ref{eq:FLRWOper}) only depend on the background quantities and therefore can be pulled out of the Poisson bracket defined on the linearized extended ADM-phase space. This means that we have for a generic phase space variable $Q$, the following relation:
\begin{equation}
\hat{P}\delta_{G'_{b,\vec{b}}}Q = \delta_{G'_{b,\vec{b}}}\hat{P}Q ~.
\end{equation}
Using the above equation and the action of  scalar, vector and tensor projectors on (\ref{eq:dG'metricFRWandMomenta}) we find that the scalar configuration variables transform as follows:
\begin{align}
\label{eq:trafoscalarsmetric}
\phi &\to \phi +\frac{1}{\bar{N}}b_{,t}, ~&~ B &\to B-\frac{\bar{N}}{A}b+\hat{b}_{,t}, \nonumber \\
\psi &\to \psi+\frac{\tilde{\mathcal{H}}}{\bar{N}}b+\frac{1}{3}\Delta\hat{b}, ~&~ E &\to E+\hat{b} ~.
\end{align}
Their conjugate momenta have the following transformation behavior under linearized diffeomorphisms:
\begin{align}
\label{eq:traforscalarmomenta}
p_\phi &\to p_\phi, ~&~ p_B & \to p_B,\nonumber\\
p_\psi &\to p_\psi-\left( \frac{1}{4}\frac{\tilde{\mathcal{H}}}{\bar{N}}+\frac{\kappa}{8}\frac{\bar{N}}{\tilde{\mathcal{H}}}p \right)b+\frac{1}{6}\Delta \left( \frac{\bar{N}}{A\tilde{\mathcal{H}}}b+\hat{b}\right), ~&~ p_E &\to p_E-\frac{\bar{N}}{4A\tilde{\mathcal{H}}}b-\hat{b},
\end{align}
where we used in the first line the result in (\ref{eq:dG'metricFRWandMomenta}), namely that $\delta_{G'_{b,\vec{b}}}\Pi=0$ and $\delta_{G'_{b,\vec{b}}}\Pi^a=0$ because the primary constraints are satisfied for the background solution.

Further, the vector variables transform under linearized diffeomorphisms in the following way:
\begin{align}
S^a &\to S^a+b^a_{\perp,t}, ~&~ p_{S^a}&\to  p_{S^a},\nonumber\\
F_a &\to F_a+b_{\perp a}, ~&~ p_F^a &\to p_F^a-b^a_{\perp} ~.
\end{align}
The tensor sector encoded in $h^{TT}_{ab}$ and $p_{TT}^{ab}$ consists of already gauge invariant variables and hence no further analysis is needed here. In the Lagrangian framework  usually an infinitesimal diffeomorphism is parametrized as $x^\mu\to x^\mu+\epsilon^\mu$ where $x^\mu$ denotes spacetime coordinates. In order to compare our results here with the literature (e.g. \cite{Mukhanov}) we have expressed $\epsilon^\mu$ in terms of the descriptors $b$ and $b^a$. We have $\epsilon^\mu = b n^\mu + X^\mu_a b^a$ and for an adaptive frame we can use $X^\mu_a = \delta^\mu_a$ and $n^\mu = N^{-1}(1,-N^a)$.
~\\
~\\
\paragraph*{\centerline{2.b Construction of Dirac observables within the relational formalism}}
~\\
In our companion paper \cite{Giesel:2017roz} we discussed in detail how Dirac observables can be constructed in the context of the relational formalism. The techniques developed in this formalism by several authors \cite{RovelliObservable,RovelliPartial,Dittrich,Dittrich2,Thiemann2,Pons3,Pons4} allow us to construct observables, which are gauge invariant extensions, for generic phase space functions. The general starting point is a Hamiltonian system with constraints for which general relativity formulated in ADM variables is a prominent example. A detailed discussion of 
 the construction of observables for constrained systems and in particular its  application to general relativity was presented in the review \cite{Giesel:2017roz}.   Here we only summarize the main points for our analysis and refer to \cite{Giesel:2017roz} as well as the original literature   \cite{RovelliObservable,RovelliPartial,Dittrich,Dittrich2,Thiemann2,Pons3,Pons4} for further details.

The idea of the relational formalism is to construct observables for a constrained system by means of formulating values that a chosen set of field variables can take in a relational manner. By this we mean that in the framework of general relativity it is not a  gauge invariant statement to say that the metric takes a certain value at a spacetime coordinate $x^\mu$. However, what can be formulated in a gauge invariant way is the values that the metric takes if other so called reference or clock fields take a certain value. A familiar example is finding the volume of a definite finite region in space. If we compute the volume and then apply a coordinate transformation, the actual number the volume takes changes and thus this number is not invariant under diffeomorphisms. But if we are able to define the volume of that region relative to another field, for instance we define the region as the part of space where some matter density is non-vanishing, then we can formulate a diffeomorphism invariant expression for the volume of this region.

Hence, the main idea behind this formalism is to provide a framework in which observables for gauge variant phase space functions can be constructed by defining their values as well as their dynamics relative to other reference/clock fields that are themselves gauge variant fields on phase space. For general relativity such reference fields correspond to  the choice of physical spatial and temporal coordinates. 

It was shown earlier that once a set of clock fields has been chosen, one can define an observable map that maps a given phase space function onto its gauge invariant extension \cite{Dittrich,Dittrich2,Thiemann2,Pons3,Pons4}. Of course the actual form of this map crucially depends on the chosen clock fields as well as the constraints the system under consideration possesses. We now summarize how such an observable map can be constructed for the extended ADM-phase space.

Let us start by recalling that on the extended ADM-phase space we have for each spacetime point eight first class constraints: four primary and four secondary ones, that we write in the following compact notation as:
\begin{equation}
\Pi_\mu=(\Pi,\Pi_a)\quad{\rm and}\quad C_\mu=(C,C_a)\quad\mu=0,\cdots,3 ~.
\end{equation}
As discussed above, the quantity $G'_{b,\vec{b}}$ in (\ref{eq:G'}) generates diffeomophisms for all geometric as as well as matter degrees of freedom on extended phase space. Our aim is to construct quantities, called observables, denoted by $O$ that are invariant under diffeomorphisms, that is general coordinate transformations. At the canonical level this condition can be formulated by requiring that such quantities need to at least weakly Poisson commute with the generator $G'_{b,\vec{b}}$, that is $\{O,G'_{b,\vec{b}}\}\approx 0$. Here weakly means equality on the constraints hypersurface. 

Next we discuss the role of the clock fields in this formalism. The general strategy one embarks on is that for each constraint in the system one introduces a clock that has to satisfy certain conditions that we will discuss below. Now in the case of general relativity in the reduced ADM-phase space we have for each spacetime point four constraints $C_\mu$. Hence, we have $4\times\infty$ many constraints and a possible choice for clocks would be four scalar fields because this have exactly $4\times\infty$ many degrees of freedom. In order that the chosen fields can be used as clock fields they have to satisfy the following condition:
\begin{equation}
\label{eq:detTC}
\det\left(\{ T^\mu(x),C_\nu(y) \}\right) \neq 0 \quad \forall \,\, \mu,\nu = 0,\cdots,3 ~.
\end{equation}
Here $x$ and $y$ denote local coordinates on the spatial manifold $\Sigma$. The above condition on clocks needs to hold in some local neighborhood of $x$ and $y$. Whether or not the clocks can be used globally depends on whether the above condition is globally satisfied. 

If the clocks satisfy the condition (\ref{eq:detTC}), it ensures that the matrix $\mathcal{A}^\mu_\nu(x,y)$ defined as:
\begin{equation}
\mathcal{A}^\mu_\nu(x,y) := \{ T^\mu(x),C_\nu(y) \},
\end{equation}
is invertible and clocks locally parameterize the gauge orbits. This allows us to use the inverse of $\mathcal{A}$ denoted by $\mathcal{B} := \mathcal{A}^{-1}$ in order to construct an equivalent set of first class constraints given by:
\begin{equation}
\tilde{C}_\mu(x) := \int\mathrm{d}^3y~\mathcal{B}^\nu_\mu(y,x)C_\nu(y).
\end{equation}
It turns out that the clock fields are weakly canonically conjugate to this equivalent set of constraints. That is,  $\{T^\mu(x)\,\tilde{C}_\nu(y)\}\approx\delta^\mu_\nu \delta(x,y)$. Further, the mutually Poisson bracket between the constraints of the equivalent set $\tilde{C}_\mu$ yields terms that are at least quadratic in the constraints. The above procedure of 
choosing constraints $\tilde C_\mu$ is called {\it{weak abelianization}} because the associated Hamiltonian vector fields $\chi_\mu := \{ \cdot,\tilde{C}_\mu \}$ are weakly abelian. 

Given a set of clock fields $T^\mu$ satisfying the assumptions above, we can define a set of gauge fixing constraints:
\begin{equation}
G^\mu=\tau^\mu-T^\mu,
\end{equation}
where $\tau^\mu$ is a generic function of spacetime coordinates and can be interpreted as the value that the clock $T^\mu$ takes. Since $\tau^\mu$ is phase space independent also, $-G^\mu$ is weakly canonically conjugate to $\tilde{C}_\mu$ if $T^\mu$ is.

Now it has been proven in \cite{Dittrich} that the following formal power series
\begin{eqnarray}
\label{eq:obsformula3}
\mathcal{O}_{f,T}[\tau] &=& f + \sum\limits_{n=1}^{\infty} \frac{1}{n!} \int\mathrm{d}^3x_1 ... \int\mathrm{d}^3x_n~ G^{\mu_1}(x_1) ... G^{\mu_n}(x_n) \{...\{ f,\tilde{C}_{\mu_1}(x_1) \}, ...\tilde{C}_{\mu_n}(x_n) \},\nonumber\\
\end{eqnarray}
is gauge invariant, where we have suppressed the $\mu$-label for the clocks $T^\mu$ and their values $\tau^\mu$ to keep our notation compact and simple. Gauge invariance means, that at least weakly the observable commutes with all constraints $C_\mu$ relevant in the reduced ADM-phase space, that means  $\{\mathcal{O}_{f,T}[\tau],C_\mu\}\approx 0$. The map $f\to \mathcal{O}_{f,T}[\tau]$ returns the value of a generic phase space function $f$ at those values where the clock fields $T^\mu$ take the values $\tau^\mu$. This can be also seen by rewriting the formula for observables in a slightly different but equivalent way in terms of the gauge flow induced by the constraints denoted by $\alpha_{\tilde{G}_{b,\vec{b}}}$:
\begin{equation}
\label{eq:obsformula2}
\mathcal{O}_{f,T}[\tau] \approx \left. \alpha_{\tilde{G}_{b,\vec{b}}}(f) \right|_{b=G^0,b^a=G^a}.
\end{equation}
An important point, also relevant for the above formal power series is the following. First the gauge flow acting on $f$ is computed for general phase space independent descriptors $b,\vec{b}$ and only afterwards these descriptors are identified with the phase space dependent gauge fixing conditions. The observable formula in (\ref{eq:obsformula3}) is sufficient for all  phase space functions other than lapse and shift, and moreover if the clock fields also do not depend on lapse and shift either. 

Now let us briefly discuss how this formula can be generalized in case we want to have gauge invariant extensions of variables that involve lapse and shift degrees of freedom as shown in \cite{Pons1,Pons2,Pons3,Pons4}. Still we assume that the clock fields do not depend on lapse and shift variables which is sufficient for the cases discussed in this article. In case we drop also this assumption one can still define an observable map but the weak abelianization gets more complicated in general. 

In the reduced ADM-phase space where lapse and shift are Lagrange multipliers, the stability of a gauge fixing condition $G^\mu$, that is $\dot{G}^\mu\approx 0$, involves lapse and shift. Consequently, the stability requirement of $G^\mu$ just fixes the Lagrange multipliers $N$ and $N^a$. In the case of the extended ADM-phase space, where $N,N^a$ are elementary phase space variables, the situation is different. The lapse and shift are dynamical variables in the extended phase space, and as a result the stability condition for gauge fixing results in secondary gauge fixing constraints. To be precise, we get
\begin{align}
G^{(2)\mu}(x) &:= \dot{G}^{\mu}(x)=[\dot{\tau}^\nu - \dot{T}^\mu](x) \nonumber \\
~ &= [\partial_t \tau^\nu - \partial_tT^\mu](x) - \kappa^{-1}\int_\Sigma \mathrm{d}^3y~ \mathcal{A}^\mu_\nu(x,y) N^\nu(y).
\end{align}
A consequence of the secondary gauge fixing constraints is that weak abelianization is required for secondary as well as 
the primary constraints, modifying the observable formula. In particular,  the descriptors in the diffeomorphism generator $G'_{b,\vec{b}}$ are replaced by $G^\mu$, and the time derivatives of descriptors by $G^{(2)\mu}$. As a side remark note that given the eight primary and secondary constraints, one could naively think that we will need eight instead of four clock fields in the extended phase space. However, since the generator $G'_{b,\vec{b}}$ is not an independent linear combination of these eight constraints and in particular involves four descriptors. Hence, a choice of only four clock fields is natural.

In the extended phase space, the procedure to obtain the weakly abelianized constraints goes as follows. We need to 
compute the matrix elements    $\mathfrak{A}^I_J :=  -\{ \mathfrak{G}^I,\mathfrak{C}_J \} $ between gauge fixing 
constraints $\mathfrak{G}^I:=(G^\mu,G^{(2)\mu})$, and primary and secondary constraints $\mathfrak{C}_I := (C_\mu,\Pi_\mu)$ for $I=1,..,8.$. The negative sign in the definition of the matrix elements is because of the relative sign difference between the gauge fixing constraints and the clocks. Using our notation $\mathcal{A}^\mu_\nu = \{ T^\mu,C_\nu \}$, and the identity $\dot{T}^\mu=\partial_tT^\mu + \int\mathrm{d}^3x~ \mathcal{A}^\mu_\nu(\cdot,x) N^\nu(x)$, we find 
\begin{equation}
\mathfrak{A}^I_J = -\left[ \begin{matrix}
\mathcal{A}^\mu_\nu & 0 \\
\{ \dot{T}^\mu,C_\nu \} & \mathcal{A}^\mu_\nu
\end{matrix} \right] ~.
\end{equation}
Its inverse matrix denoted by $\mathfrak{B}:=\mathcal{A}^{-1}$ can be in the following way \cite{Pons3}:
\begin{equation}
\mathfrak{B}^I_J = (\mathfrak{A}^{-1})^I_J = \left[ \begin{matrix}
\mathcal{B}^\mu_\nu & 0 \\
S^\mu_\nu & \mathcal{B}^\mu_\nu
\end{matrix} \right],
\end{equation}
where
\begin{equation}
S^\mu_\nu(x,y) = -\int\mathrm{d}^3z\int\mathrm{d}^3v~ \mathcal{B}^\mu_\rho(x,z)\mathcal{B}^\sigma_\nu(v,y)\{ \dot{T}^\rho(z),C_\sigma(v) \}.
\end{equation}
Given $\mathfrak{B}^I_J$ we can construct the set of weakly abelianized constraints as:
\begin{equation}
\tilde{\mathfrak{C}}_I(x) = \int\mathrm{d}^3y~\mathfrak{B}^J_I(y,x)\mathfrak{C}_J(y) ~.
\end{equation}
Let us define the equivalent abelian set of constraints by $\tilde{\mathfrak{C}}_I(x) =: (\tilde{\tilde{C}}_\mu,\tilde{\Pi}_\mu)$. Then $\tilde{\Pi}_\mu$ and $\tilde{\tilde{C}}_\mu$ expressed in terms of the original constraints are given by:
\begin{align}
\tilde{\Pi}_\mu(x) &= \int\mathrm{d}^3y~\mathcal{B}^\nu_\mu(y,x)\Pi_\nu(y) ,  \nonumber \\
\tilde{\tilde{C}}_\mu(x) &= \int\mathrm{d}^3y~\mathcal{B}^\nu_\mu(y,x) \left[ C_\nu(y) - \int\mathrm{d}^3z\int\mathrm{d}^3v~ \mathcal{B}^\sigma_\rho(v,z)\{ \dot{T}^\rho(z),C_\nu(y) \}\Pi_\sigma(v) \right].
\end{align}
Also in the case of the extended phase space we aim at writing down a formal power series formula for the observable map. This final formula crucially simplifies if we take a result proven in  \cite{Pons3} into account, namely that up to second order in the gauge generators, we can replace the rather complicated generator $G'_{b,\vec{b}}$ in (\ref{eq:G'}) by a simpler gauge generator. Explicitly, we have \cite{Pons3}:
\begin{equation}
G'_{b,\vec{b}} +\mathcal{O}(C^2) = \tilde{G}'_{\xi,\vec{\xi}} =: \kappa^{-1}\left( \tilde{\Pi}_\mu[\dot{\xi}^\mu] + \tilde{\tilde{C}}_\mu[\xi^\mu] \right), 
\end{equation}
where $\xi^\mu = \int\mathrm{d}^3x \mathcal{A}^\mu_\nu(\cdot,x)b^\nu(x)$. As a result we can construct observables 
using  $\tilde{G}'_{\xi,\vec{\xi}}$:
\begin{equation}
\label{eq:obsGR1}
\mathcal{O}_{f,T}[\tau] \approx \left. \alpha_{ \tilde{G}'_{\xi,\vec{\xi}} }(f) \right|_{\xi^\mu=G^\mu,\dot{\xi}^\mu=G^{(2)\mu}}.
\end{equation}
As before we can rewrite this formula in a formal power series that for the extended phase space has the following form:
\begin{equation}
\label{eq:obsGR2}
\mathcal{O}_{f,T}[\tau] = f+\sum\limits_{n=1}^{\infty} \frac{1}{n!} \int\mathrm{d}^3y_1...\int\mathrm{d}^3y_n~ \mathfrak{G}^{I_1}(y_1)...\mathfrak{G}^{I_n}(y_n) \{...\{ f,\tilde{\mathfrak{C}}_{I_1}(y_1) ~. \},...\tilde{\mathfrak{C}}_{I_n}(y_n) \} ~.
\end{equation}
Here we used the definitions of $\mathfrak{G}^I = (G^\mu,G^{(2)\mu})$ and $\tilde{\mathfrak{C}}_I(x) = (\tilde{\tilde{C}}_\mu,\tilde{\Pi}_\mu)$. This observable formula is the required generalization needed for the extended ADM-phase space to have the same techniques available for constructing gauge invariant quantities as in the reduced ADM-phase space. Since we aim at applying these techniques in the context of cosmological perturbation theory, in the next subsection we will briefly discuss how observables can be computed perturbatively.
~\\
~\\
\paragraph*{\centerline{2.c Construction of observables in linear perturbation theory}}
~\\

For the reason that we aim at constructing observables associated with the scalar, vector and tensor perturbations, the question arises how the observable map discussed in the last subsection needs to be modified if we consider instead of a generic phase space function $f$ its corresponding projection $\hat{P}f$. We discuss this in the case that our $f$ will be chosen among the linear perturbations of the geometry or matter sector. Note that the action of the projectors $\hat{P}$ only affects the function $f$ in the observable formula. This can be seen from the observable formula in (\ref{eq:obsGR2}). The projectors $\hat{P}$ involve background quantities as well as derivative operators with respect to the variables the function $f$ depends on. However, the constraints as well as the gauge fixing conditions are evaluated at a different, independent variable and thus the action of $\hat{P}$ on them becomes trivial. Furthermore, the iterated Poisson bracket involved in the observable formula is evaluated with respect to the phase space of the linear perturbations. As discussed earlier for perturbations around a flat FLRW background the relevant projectors depend only on differential operators and background quantities. In other words as far as the phase space of the perturbations is considered those projectors are phase space independent. Given this, we have for all projectors $\hat{P}$ considered in our further computations that the following relation holds:
\begin{align}
\mathcal{O}_{\hat{P}f,T}[\tau] &= \hat{P} \mathcal{O}_{f,T}[\tau].
\end{align}
In the following we want to use the observable map in the framework of linear cosmological perturbation theory and we aim at showing that the conventional gauge invariant quantities such as the Mukhanov-Sasaki variable or the Bardeen potentials can be obtained naturally from the application of the observable map with an appropriate choice of clock variables. In order to do so, we need to discuss how perturbations of these observables can be formulated. At first we consider only phase space functions that are independent of lapse and shift. In that case we can use the observable formula  shown in (\ref{eq:obsformula3}). Its first order perturbation leads to:
\begin{align}
\delta \mathcal{O}_{f,T}[\tau] &= \delta f + \int\mathrm{d}^3y~ \delta G^\mu(y)\overline{\mathcal{O}_{ \{ f,\tilde{C}_\mu(y) \},T }[\tau] } \nonumber \\
~ &+ \sum\limits_{n=1}^{\infty} \frac{1}{n!} \int\mathrm{d}^3y_1...\int\mathrm{d}^3y_n~ \bar{G}^{\mu}(y_1) ... \bar{G}^{\mu_n}(y_n) \{ ... \{ f,\tilde{C}_{\mu_1}(y_1) \}, ...,\tilde{C}_{\mu_n}(y_n) \}^{(1)}.
\end{align}
As discussed above, the functions $\tau^\mu$ involved therein should be in the range of the clocks $T^\mu$, that is there exists a gauge such that $T^\mu=\tau^\mu$ leading to the gauge fixing conditions $G^\mu=\tau^\mu-T^\mu$. For the background solution we assume that these gauge fixing conditions are satisfied, that is $\bar{G}^\mu=0$ and thus $\bar{T}^\mu=\bar{\tau}^\mu$. As a consequence, the formula for the linear perturbations of the observables 
can be written as:
\begin{align}
\label{eq:pertobsGen}
\delta \mathcal{O}_{f,T}[\tau] &= \delta f +\int\mathrm{d}^3y\delta G^\mu(y)\overline{\{ f,\tilde{C}_\mu(y) \}}\approx \delta f + \int\mathrm{d}^3y\int\mathrm{d}^3z~ \delta G^\mu(y)\bar{\mathcal{B}}^\nu_\mu(z,y)\overline{\{ f,C_\nu(z) \}}.
\end{align}
Note that as expected in the gauge $\delta G^\mu=0$ the gauge variant quantities $\delta f$ coincide with their corresponding observables $\delta \mathcal{O}_{f,T}[\tau]$.

Generalizing to the case that $f$ can also depend on lapse and shift, we need to consider perturbations of the observable formula in (\ref{eq:obsGR2}). Assuming again that $\bar{G}^\mu=0$ and $\bar{G}^{(2)\mu}=0$, we get,
\begin{align}
\label{eq:pertobs2}
\delta \mathcal{O}_{f,T}[\tau] &= \delta f + \int\mathrm{d}^3y \left[ \delta \dot{G}^\mu(y)\overline{\{ f,\tilde{\Pi}_\mu(y) \}} + \delta T^\mu(y)\overline{\{ f,\tilde{\tilde{C}}_\mu(y) \}} \right] \nonumber \\
~ &\approx \delta f + \int\mathrm{d}^3y \int\mathrm{d}^3z \bar{\mathcal{B}}^\nu_\mu(z,y) \left[ \delta \dot{G}^\mu(y)\overline{\{ f,\Pi_\nu(z) \}} - \delta G^\mu(y) \left( \overline{\{ f,C_\nu(z) \}} \right. \right. \nonumber \\
~ &\hspace{8em} + \left. \left. \int\mathrm{d}^3w \int\mathrm{d}^3v~ \bar{\mathcal{B}}^\rho_\sigma(w,v) \overline{\{ \dot{T}^\sigma(v),C_\nu(z) \}}~ \overline{\{ f, \Pi_\rho(w) \}} \right) \right].
\end{align}
In section \ref{Sec:ConstrObs}, we will extensively use these observable formulae to obtain gauge invariant variables natural to various gauge choices. For these gauges, we will have to choose $\delta\tau^\mu=0$, that is use $\delta G^\mu = -\delta{T}^\mu \approx 0$. The stability of this gauge fixing constraint and of subsequent conditions will play a central role in our analysis.

~\\
~\\
\paragraph*{\centerline{2.d Poisson brackets of linearized cosmological  perturbations}}
~\\
Both of the formulae for the linearized observables derived above show that we need to compute the Poisson bracket of the quantity $f$, whose observable we want to construct, with the constraints and then evaluated it on the background solution, which in  our case is the flat FLRW spacetime. As discussed in detail in \cite{Giesel:2017roz} (in appendix A), these Poisson brackets can be related to certain Poisson brackets on the linearized phase space. Explicitly, we have for two generic phase space functions $f,g$:
\begin{align}
\label{eq:backPoissonaslinearPoisson}
\overline{\{ f(x),g(y) \}} &= \{ \delta f(x),\delta g(y) \}_\delta~,
\end{align}
where on the right-hand side the Poisson bracket $\{\cdot,\cdot\}_\delta$ denotes the Poisson bracket of the linearized phase space. 
~\\

To compute various observables, we need several Poisson brackets of the linearized cosmological perturbations with the linearized constraints. Since the tensor perturbations are already gauge invariant we here focus on the Poisson brackets involving scalar and vector perturbations only. 

To calculate Poisson brackets among the decomposed perturbations, we use the fact that one can pull the phase space independent projectors $\hat{P}_I$ out of the Poisson brackets. This leads to
\begin{align}
\{ \hat{P}_I(\delta q)(x), \hat{P}_J(\delta P)(y) \} &= \hat{P}_I^{ab}(x)\hat{P}_{Jcd}(y)\{ \delta q_{ab}(x),\delta P^{cd}(y) \} =\kappa \hat{P}_I^{(ab)}(x)\hat{P}_{Jab}(y)\delta (x,y)
\end{align}
where $I,J$ label  different projectors introduced above. For the elementary phase space variables in the extended ADM phase space we get:
\begin{align}
\label{eq:scavetealgebra}
\{\phi(x),p_{\phi}(y)\} &=\kappa \delta(x,y), \nonumber \\
\{B(x),p_B(y)\} &=-\kappa G(x,y), \nonumber\\
\{S^a(x),p_{S^b}(y)\}&=\kappa\delta^a_b \delta(x,y) +2\kappa G(x,y),     \nonumber\\
\{ \psi(x),p_\psi(y) \} &= \kappa\frac{1}{12A\tilde{P}}\delta(x,y), \nonumber \\
\{ E(x),p_E(y) \} &= \kappa\frac{3}{8A\tilde{P}} \int\mathrm{d}^3zG(x,z)G(z,y), \nonumber \\
\{ F_a(x),p^b_F(y) \} &= \kappa\frac{1}{2A\tilde{P}}\left[ -\delta_a^bG(x,y) + \int\mathrm{d}^3z G(x,z)\frac{\partial^2G(z,y)}{\partial z^a \partial z_b} \right].
\end{align}
All of the remaining Poisson brackets in the scalar-vector sector vanish. Considering the form of the secondary constraints in terms of the projected variables in (\ref{eq:deltaCdeltaCa}) we can calculate the necessary Poisson brackets of the perturbations and the linearized constraints (see also \cite{Giesel:2017roz} for more details). The results are given by:
\begin{align}\label{eq:dfdC1}
\frac{1}{\kappa} \{ \psi(x),\delta C(y) \} &= \frac{\tilde{\mathcal{H}}}{\bar{N}}\delta(x,y), & \frac{1}{\kappa} \{ \psi(x),\delta C_a(y) \} 
&= \frac{1}{3} \frac{\partial}{\partial x^a} \delta(x,y), \nonumber \\
\frac{1}{\kappa} \{ E(x),\delta C(y) \} &= 0, & \frac{1}{\kappa} \{ E(x),\delta C_a(y) \} &= \frac{\partial G}{\partial x^a}(x,y),
\end{align}
and
\begin{align}\label{eq:dfdC2}
\frac{1}{\kappa} \{ p_\psi(x),\delta C(y) \} &= \left( -\frac{\tilde{\mathcal{H}}}{4\bar{N}}+\frac{\bar{N}}{6A\tilde{\mathcal{H}}}\Delta 
-\frac{\kappa}{8}\frac{\bar{N}}{A^{3/2}\tilde{\mathcal{H}}}F^-_\varphi \right) \delta(x,y), & \frac{1}{\kappa} \{ p_\psi(x),\delta C_a(y) \} 
&= \frac{1}{6}\frac{\partial}{\partial x^a}\delta(x,y), \nonumber \\
\frac{1}{\kappa} \{ p_E(x),\delta C(y) \} &= -\frac{\bar{N}}{4A\tilde{\mathcal{H}}}\delta(x,y), & \frac{1}{\kappa} \{ p_E(x),\delta 
C_a(y) \} &= -\frac{\partial G}{\partial x^a}(x,y).
\end{align}
For the vector perturbations we find the non-vanishing Poisson brackets as:
\begin{align}
\label{eq:dfdC3}
\frac{1}{\kappa} \{ F_a(x),\delta C(y) \} &= 0, & \frac{1}{\kappa} \{ F_a(x),\delta C_b(y) \} &= \left( \delta_{ab}\delta(x,y)- 
\frac{\partial^2G(x,y)}{\partial y^a \partial y^b} \right), \nonumber \\
\frac{1}{\kappa} \{ p^a_F(x),\delta C(y) \} &= 0, & \frac{1}{\kappa} \{ p^a_F(x),\delta C_b(y) \} &= -\left( \delta^a_b\delta(x,y)- 
\frac{\partial^2G(x,y)}{\partial y_a \partial y^b} \right) ~.
\end{align}
We will use these results frequently in the next section where the explicit form of the observables is derived and discussed.

\section{Gauge choices and corresponding clocks}
\label{Sec:ConstrObs}

The goal of this section is to construct geometrical clocks and the associated observables for metric and momentum perturbations at the linear order for various gauges used in cosmological perturbation theory. The observables turn out to be gauge invariant variables specific to the choice of a particular gauge condition. The main idea in this 
construction is the following. The hypersurface defined by the linearized gauge fixing constraint $\delta G^\mu = 0$ involves   the perturbations $\delta T^\mu$, that is $\delta G^\mu=\delta\tau^\mu-\delta T^\mu$. Hence, a choice of perturbed clocks $\delta T^\mu$ can be directly related to a family of a gauge fixings parametrized by $\delta\tau^\mu$. In particular, the first order observable $\delta \mathcal{O}_{f,T}[\tau]$ equals $\delta f$ on $\delta G^\mu = 0$. Each gauge condition used in cosmological perturbation theory is 
tied to a specific choice of clocks which in our framework naturally leads to a set of observables or gauge invariant variables. Using our framework, we can systematically find 
 clock fields that yield the Bardeen potentials, Mukhanov-Sasaki variable etc. 

In the context of the relational formalism the choice of a clock corresponds to the choice of a reference field that on  one hand is used to construct gauge invariant quantities and on the other hand defines a notion of physical time or spatial coordinates respectively. By this we mean that the evolution of such gauge invariant quantities is not defined with respect to coordinate time, but with respect to the values that the temporal reference field takes, that is $\tau^0$ in our notation. As discussed in \cite{Pons4, Dust1}, for a certain type of coordinate gauge fixing conditions, one can show that the gauge invariant evolution of the observables and the gauge-fixed evolution can be mapped into each other under an appropriate identification and both formulations yield equivalent results. Exactly, by means of this property we are able to rederive the gauge invariant observables as well as their equations of motion common in linearized cosmological perturbation theory in the context of the relational formalism. 

Now, in case we apply perturbation theory, we have gauge fixing conditions for the background solution $\bar{G}^\mu=\bar{\tau}^\mu-\bar{T}^\mu$ and of course these conditions need to be consistent with the equations of motion of the background clocks $\bar{T}^\mu$. The same is true at the linear order where a choice of $\delta G^\mu=\delta\tau^\mu-\delta T^\mu$ can be only considered if such a choice does not contradict the equations of motion of $\delta T^\mu$.

In order to reproduce the gauges common in cosmological perturbation theory, we will often choose components of the perturbed metric as clocks whose corresponding components vanish on the background such as for instance linear perturbations of the shift vector. Thus, in order to ensure consistency with the background solutions for some gauges we have to choose $\bar{\tau}^\mu=0$. 
However, in general this would still allow us to work with a generic value of $\delta\tau^\mu$ and would yield a family of gauge fixings represented by $\delta G^\mu=\delta\tau^\mu-\delta T^\mu$. Depending on the gauge under consideration, in certain cases we will also need to choose $\delta\tau^\mu=0$ to reproduce the gauge invariant observables conventionally constructed for cosmological perturbation theory. An example of this is the case of the longitudinal gauge, for which in the Lagrangian framework the longitudinal part of the spatial metric perturbation $E$ and the perturbation in shift vector $B$  are set to zero. As explained before, the observable formula, when applied to the clocks $T^\mu$, maps them onto the values $\tau^\mu$. Carrying this over to the perturbative case, we can access gauges like the longitudinal gauge with clocks chosen from the linear perturbations of the metric and its momenta simply by setting $\delta\tau^\mu=0$. In such a  case we have $\delta G^\mu=-\delta T^\mu$ and the construction formula for the observables further simplifies to:
\begin{align}
\label{eq:pertobs1}
\delta \mathcal{O}_{f,T}[\tau]\Big|_{\tau=0} &= \delta f - \int\mathrm{d}^3y\delta T^\mu(y)\overline{\{ f,\tilde{C}_\mu(y) \}}\nonumber \\
&\approx \delta f - \int\mathrm{d}^3y\int\mathrm{d}^3z~ \delta T^\mu(y)\bar{\mathcal{B}}^\nu_\mu(z,y)\overline{\{ f,C_\nu(z) \}}.
\end{align}
Note that as expected in the gauge $\delta G^\mu=0$, that is $\delta T^\mu=0$ if $\delta\tau^\mu=0$ has been chosen, the gauge variant quantities $\delta f$ coincide with their corresponding observables $\delta \mathcal{O}_{f,T}[\tau]$.

As has been shown in \cite{Dust1,Dust2}, if one can identify non-linear clocks, that is clocks for full general relativity, then one can first construct manifestly gauge invariant quantities and afterwards consider the linearized perturbations around the FLRW solution. In this case, the non-linear temporal clock determines the temporal clock at any order. As a result, one has a well-defined notion of physical time for the background and for arbitrary orders in perturbation theory. From this perspective, we  would interpret a clock for which we have $\bar{\tau}^\mu=0$, to be not natural because with such a choice we are not able to define a notion of physical time for the background. As a consequence, it might be impossible to find non-linear clocks that reproduce such clocks in the background as well as at the linear order. Moreover, it may seem that if we need to choose $\bar{\tau}^\mu=0$ as well as $\delta \tau^\mu=0$ in order to reproduce a common gauge in cosmology, the interpretation of the evolution of the gauge invariant observables in the context of the relational formalism is problematic. However, it turns out that even in these cases the observable map opens the possibility to obtain these gauge invariant quantities very systematically and a consistent and judicious notion of evolution can be formulated. As far as the derivation of their evolution equations is considered the relational formalism is very useful because it technically simplifies to explicitly derive these evolution equations. We will discuss these important points in a separate investigation in more detail.
For this article, we focus on different gauges and discuss the precise relationship between different gauge fixing conditions in linear cosmological perturbation theory, clocks and observables. For some gauges, a generalization is possible allowing us to identify the clock in the relational formalism with the physical time. These will be discussed in section \ref{Sec:GenGauge}. 

We will consider  the following five common gauges, which are discussed in cosmological perturbation theory (see e.g. \cite{bardeen,Mukhanov,kodama-sasaki}). Though we will focus our discussion on scalar perturbations, the clocks and the observables for the 
transversal vector components are also found in the process. These gauges are:
\begin{enumerate}
\item \textbf{Longitudinal gauge:} $E=B=0$,
\item \textbf{Spatially flat gauge:} $\psi=E=0$,
\item \textbf{Uniform field gauge:} $\delta \varphi = E = 0$,
\item \textbf{Synchronous gauge:} $\phi=B=0$,
\item \textbf{Comoving gauge:} $\delta \varphi = B = 0$ ~.
\end{enumerate}

Above gauges can be divided into two broad categories. The first three of these share a common feature that the longitudinal part of the spatial metric perturbation vanishes. This is the isotropic threading of spacetime. There is no residual freedom in 
the spatial coordinates. In contrast, the latter two gauges correspond to a time slicing such that the shift perturbations vanish. There is a residual freedom associated with the shift in spatial coordinates in these cases. It turns out that the construction of the geometrical clocks and observables is much more straightforward and less involved in the first three gauges than the last two.

In the following we begin with the longitudinal gauge where we demonstrate the entire method  to appropriate clocks and construct the respective observables in  detail. The calculations for the other gauges in subsequent subsections follow the same strategy, albeit for some subtleties and technical issues related in particular to 
the vanishing of the shift perturbation. For this reason, discussion of other gauges, except for latter issues, is shorter in presentation than the longitudinal gauge.

\subsection{Longitudinal Gauge}

The longitudinal or the Newtonian gauge amounts to an isotropic threading and a shear free slicing. Using its definition in the Lagrangian picture,  we impose the following in the constraint framework. Note that we will use the weak equality symbol $\approx$ to denote equalities on the gauge fixing constraint hypersurface. This gauge is identified as:
\begin{align}\label{eq:l-gauge}
B \approx 0, ~~~~~~~~~~~~~~ E \approx 0  .
\end{align}
In the Lagrangian formulation the corresponding velocities have also to be set equal to zero for stability reasons. In the Hamiltonian picture, the stability analysis of the gauge fixing condition will yield a further condition among the perturbed quantities involving scalar perturbations of the momenta. The stability conditions are,
\be
\dot{B} \approx  0, ~~~~~~~~~~~~~~   \dot{E} \approx 0 ~.
\ee
Using the Hamilton's equations  (\ref{eq:dotpsiE}) and (\ref{eq:dotphidotB}), we can write 
\begin{equation}\label{eq:l-dotB}
\dot{B} = \delta \hat{\lambda} \approx 0
\end{equation}
and 
\be\label{eq:l-dotE}
\dot{E} = - 4\tilde{\mathcal{H}}(E+p_E) + B \approx - 4\tilde{\mathcal{H}}p_E \approx 0~.
\ee
 (\ref{eq:l-dotB}) does not require further stability conditions because the perturbation  $\delta \hat{\lambda}$ gets fixed. The same is not true for (\ref{eq:l-dotE}). We further need to impose $\dot p_E \approx 0$. 
The stability of $p_E \approx 0$, using (\ref{eq:dotGammaSigma}) yields:
\begin{equation}
\phi \approx -\psi ~.
\end{equation}
Using (\ref{eq:dotphidotB}), the above equation results in fixing the perturbation of the Lagrange multiplier as 
\be\label{eq:deltalambda-lg}
\delta \lambda \approx \dot{\bar{N}} \phi +2\bar{N}\tilde{\mathcal{H}}(\psi - p_\psi) ~.
\ee
No further conditions are required for the stability of the longitudinal gauge. 

In order to consider this gauge in the framework of geometrical clocks, we need to find the gauge descriptors consistent with the longitudinal gauge. At the linear order in perturbations, we need the first order transformation properties of perturbations under the action of the gauge operator $G'_{b,\vec{b}}$. Using these properties of metric perturbations and their momenta, given by (\ref{eq:trafoscalarsmetric}) and (\ref{eq:traforscalarmomenta}), the following relations must hold:
\begin{align}\label{eq:descriptorslongitudinalgauge}
b &\overset{!}{=} \frac{4\tilde{\mathcal{H}}A}{\bar{N}}(E+p_E), & \hat{b} &\overset{!}{=} -E ~.
\end{align}
Using these relations we can obtain the longitudinal gauge by an appropriate choice of gauge descriptors, even if one started 
with an arbitrary gauge. Note that the above relations do not imply that the gauge descriptors are functions of the phase space variables. The above identification is to be made {\it{after}} the gauge transformation with phase space independent descriptors has been calculated.

\subsubsection{Geometrical clocks}
The linearized gauge fixing constraints $\delta G^\mu=\delta\tau^\mu-\delta T^\mu \approx 0$ involve the perturbed clocks $\delta T^\mu$. For the background we have chosen $\bar{T}^\mu \overset{!}{=} \overline{\tau}^\mu$ and the latter choice needs of course to be consistent with the background solution. The range of values allowed for $\delta{\tau}^\mu$ of course depends on the specific choice of clocks $\bar{T}^\mu$ and their linear perturbations respectively. For the longitudinal gauge the components of $E$ and $B$ are set to zero in the Lagrangian framework. As explained before the observable formula when applied to the clocks $T^\mu$ maps them onto the values $\tau^\mu$. Carrying this over to the perturbative case, we can access gauges like the longitudinal gauge with clocks chosen from the linear perturbations of the metric and its momenta simply by setting $\overline{\tau}^\mu=0$ and $\delta\tau^\mu=0$. The first choice is necessary because the background metric components associated with $E$ and $B$ just vanish. For the choice regarding the perturbations we get $\delta G^\mu=-\delta T^\mu$ and hence the linearized gauge fixing conditions $\delta G^\mu$ constitute the hypersurface defined by $\delta T^\mu\approx 0$. Thus, we aim at choosing a set of perturbed clocks $\delta T^\mu$ such that $\delta T^\mu \approx 0$ is equivalent to the longitudinal gauge. 
This means, we have to find perturbed clocks $\delta T^0$ and $\delta \hat{T}$, where the latter is the scalar projection of $\delta T^a$, such that the following conditions:
\begin{align}
\delta T^0 &\approx 0, & \delta \hat{T} &\approx 0, & \delta \dot{T}^0 &\approx 0, & \delta \dot{\hat{T}} &\approx 0,
\end{align}
are equivalent to the longitudinal gauge on the phase space, identified by
\begin{align}
B &\approx 0, & E &\approx 0, & p_E &\approx 0, & \phi &\approx - \psi ~.
\end{align}
To find the appropriate clocks consistent with longitudinal gauge, we recall that for a phase space function $f$ assumed to be independent of lapse and shift, the first order observable formula reads \cite{Giesel:2017roz}:
\begin{equation}\label{eq:obsform}
\delta \mathcal{O}_{f,T}[\tau] = \delta f + \left. \overline{\{ f,\tilde{C}_\mu[b^\mu] \}} \right|_{b^\mu = -\delta T^\mu}~,
\end{equation}
where $\tilde C_\mu$ denote weakly abelianized constraints:
\begin{equation}
\tilde{C}_\mu(x) = \int\mathrm{d}^3y \mathcal{B}^\nu_\mu(y,x)C_\nu(y) ~.
\end{equation}
Here, as noted earlier, the matrix $\mathcal{B}$ is the inverse of matrix ${\cal A}$ defined as 
\be
\mathcal{A}^\mu_\nu(x,y) := \{ T^\mu(x),C_\nu(y) \} ~. 
\ee
And, the 
second term in the observable formula (\ref{eq:obsform}) corresponds to the change of 
$f$ under an infinitesimal diffeomorphism with descriptors $b^\mu = -\delta T^\mu$ evaluated after the Poisson bracket has been calculated. 

Using the identification in (\ref{eq:descriptorslongitudinalgauge}) we are led to the following perturbed clocks for the longitudinal gauge:
\begin{align}
\delta T^0 &\overset{!}{=} -\frac{4\tilde{\mathcal{H}}A}{\bar{N}}(E+p_E) = 2\sqrt{A}\tilde{P}(E+p_E), & \delta \hat{T} \overset{!}{=} E~,
\end{align}
where $\tilde P = P_A/3$ is defined via (\ref{eq:BgrdMomenta}), with  $P_A$ as  the conjugate 
momentum of $A$. 

Note that the gauge 
descriptors 
$\hat b$ and $b$ do not determine $\delta T_\perp^a$. However, the latter can be fixed knowing $\delta \hat T$. As $\delta \hat{T}$ is the 
longitudinal scalar part of $\delta q_{ab}$ we choose $\delta T_\perp^a$ to be the longitudinal transversal part thereof such that:
\begin{equation}
\delta T^a = \bar{q}^{ab}(\hat{P}_L\delta q)_b = \delta^{ab}( E_{,b}+F_b ) ~.
\end{equation}

The consistency of above clocks can be verified using $\delta T^\mu \approx 0$ and  $\delta \dot{T}^\mu \approx 0$. It is easily seen that 
\begin{align}
\delta \hat{T} \approx 0 &\Rightarrow E \approx 0, &    &  &  \delta T^0 \approx 0 &\Rightarrow p_E \approx 0, \nonumber \\
\delta \dot{\hat{T}} \approx 0 &\Rightarrow B \approx 0 &  & \mathrm{and} &  \delta \dot{T}^0 \approx 0 &\Rightarrow \psi = -\phi ~.
\end{align}
Stabilization of above conditions fix the Lagrange multipliers $\delta \lambda$ and  $\delta \hat{\lambda}$ consistent with  the longitudinal gauge and result in  (\ref{eq:l-dotB}) and (\ref{eq:deltalambda-lg}).

On the other hand, 
the Lagrange multiplier $\delta \lambda^a_\perp$ gets fixed as follows. The condition $\delta \dot{T}^a_\perp = \delta^{ab}\dot{F}_b \approx 0$ yields $p_F^a \approx  \tfrac{1}{4\tilde{\mathcal{H}}}S^a$. Demanding its further stabilization yields $\delta \lambda^a_\perp + 3 \tilde{\mathcal{H}} S^a \approx 0$. We note that this condition arises just by using the equations of motion for $p_F^a$ and $\dot S^a = \delta \lambda^a_\perp$, and does not involve conditions from other clocks or gauge descriptors corresponding to the scalar perturbations of the longitudinal gauge. This is not surprising once we recall that the clock $\delta \dot{T}^a_\perp$ is introduced without using the gauge descriptors $\hat b$ and $b$ which identify the longitudinal gauge. It will be useful to note that above constraint on $\delta \lambda^a_\perp$ and $S^a$ arises when ever the clock $\delta \dot{T}^a_\perp$  is the longitudinal scalar part of the perturbation in the spatial metric. Thus, this will be true for the spatially flat and uniform field gauges as well.

An important contribution to the observable map are the abelianized constraints that involve the inverse of the matrix ${\mathcal A}^\mu_\nu$. As explained in section \ref{Sec:ReviewObs}, the requirement that the inverse of the matrix exists is a condition on the clock fields that we can choose. At the level of linear perturbation theory we choose clock fields defined on the linearized phase space and hence the condition on the existence of the inverse of the matrix ${\mathcal A}^\mu_\nu$ has to be fulfilled on that phase space. Since a Poisson bracket between two linear functions on the linearized phase space yields a resulting expression that depends on the background quantities only, we will denote the matrix relevant for our further analysis with $\bar{\mathcal{A}}^\mu_\nu$ (see also section \ref{Sec:ReviewObs} for more details\footnote{Note, that for some calculations, as for instance derivation of the equations of motion and the Dirac bracket of the observables, 
one also needs the perturbation $\delta \mathcal{A}^\mu_\nu$ of this matrix. However, this perturbation will not be relevant for the work discussed in this paper.}). Using the Poisson brackets of perturbations of metric components and their momenta, with perturbations of the Hamiltonian and spatial diffeomorphism constraints, it is straightforward to verify that 
\begin{equation}
\bar{\mathcal{A}}^\mu_\nu(x,y) = \{ \delta T^\mu(x),\delta C_\nu(y) \} = \kappa \delta^\mu_\nu \delta(x,y) ~.\,
\end{equation}
As a result, at the level of the linearized phase space the matrix ${\mathcal A}^\mu_\nu$ can be trivially inverted. 
However, there is a subtlety with the clocks in the longitudinal gauge. It turns out that the above choice of clocks is non-commuting. One 
can check that though 
\begin{equation}\label{eq:lonclockcomm00}
\{ \delta T^0(x),\delta T^0(y) \} = 0
\end{equation}
but
\begin{equation}\label{eq:lonclockcomm0a}
\{ \delta T^0(x),\delta T^a(y) \} = -\frac{3}{4} \frac{\kappa}{\sqrt{A}}\int \mathrm{d}^3z \, G(x,z) \frac{\partial G(z,y)}{\partial y_a} ~,
\end{equation}
where $G(x,y)$ is is the integral kernel of the Green's function of the Poisson equation:
\begin{equation}
\Delta^{-1}f(x) = \int_{\bar{\Sigma}}\mathrm{d}^3y~ G(x,y)f(y) ~.
\end{equation}
The non-commuting clocks suggest that the linearized observable algebra will be non-standard and may have a complicated form, an aspect relevant once the quantization of such an observable algebra is wished to be achieved. Nevertheless, we will see below that the Bardeen potentials follow very naturally from the longitudinal clocks as gauge invariant extensions of $\psi$ and $\phi$.

\subsubsection{Observables} 
Now we use the clocks corresponding to longitudinal gauge to construct first order observables of the scalar, vector and tensor perturbations. We use the notation
\begin{equation}
\mathcal{O}_{\delta f, T}^{(1)}[\tau] = \delta \mathcal{O}_{f,T}[\tau]
\end{equation}
for the first order gauge invariant extension of a perturbation $\delta f$. As discussed above for the longitudinal gauge we need to choose $\overline{\tau}^\mu=0$ and $\delta\tau^\mu=0$. Therefore, we will either write this out explicitly or will suppress the  $\tau^\mu$ dependency of the observables in what follows for simplicity where possible. Let us first calculate the observables for perturbations $\delta f$ other than lapse and shift. In this case we can use the following perturbed observable formula \cite{Giesel:2017roz}:
\begin{align}
\delta \mathcal{O}_{f,T}[\tau]\Big|_{\tau=0} &= \delta f - \int\mathrm{d}^3y\delta T^\mu(y)\overline{\{ f,\tilde{C}_\mu(y) \}} \nonumber \\
~ &\approx \delta f - \int\mathrm{d}^3y\int\mathrm{d}^3z~ \delta T^\mu(y)\bar{\mathcal{B}}^\nu_\mu(z,y)\overline{\{ f,C_\nu(z) \}}
\end{align}
with 
\begin{equation}\label{eq:beq}
\bar{\mathcal{B}}^\mu_\nu(x,y) = \kappa^{-1}\delta^\mu_\nu\delta(x,y) 
\end{equation}
to derive the following formula for observables constructed with the longitudinal clocks:
\begin{equation}
\mathcal{O}_{\delta f,T}^{(1)} \approx \delta f + \frac{4\tilde{\mathcal{H}}A}{\kappa \bar{N}} \int\mathrm{d}^3y (E+\Sigma)(y)\overline{\{ f,C(y) \}} - \frac{1}{\kappa}\int\mathrm{d}^3y ~\delta^{ab}(E_{,b}+F_b)(y)\overline{\{ f,C_a(y) \}} ~.
\end{equation}
Note, that in the above observable formula we neglected terms proportional to the secondary constraints. In the following expressions, we 
will write an equal sign $=$ instead of the weak equality $\approx$ to distinguish between the gauge fixing constraint hypersurface and the 
hypersurface where the linearized secondary constraints are fulfilled. 
Using the expressions for the Poisson brackets of the perturbations 
with the linearized secondary constraints (\ref{eq:dfdC1}) and (\ref{eq:dfdC2}) that have been derived in \cite{Giesel:2017roz}, it is straightforward to calculate the first order 
observables of the scalar-vector-tensor perturbations. The observable corresponding to $\psi$ turns out to be one of the 
Bardeen's potential:
\begin{align}\label{Psi}
\mathcal{O}_{\psi(x),T}^{(1)} &= \psi(x) +\frac{4\tilde{\mathcal{H}}A}{\bar{N}}\int\mathrm{d}^3y(E+p_E)(y)\frac{\tilde{\mathcal{H}}}{\bar{N}}\delta(x,y)-\int\mathrm{d}^3y~\delta^{ab}(E_{,b}+F_b)(y)\frac{1}{3}\frac{\partial}{\partial x_a}\delta(x,y) \nonumber \\
~ &= \psi(x)+\frac{4\tilde{\mathcal{H}}^2A}{\bar{N}^2}(E+p_E)(x)-\frac{1}{3}\Delta E(x) \nonumber \\
~ &=: \Psi(x) ~.
\end{align}
Similarly, we can compute:
\begin{align}\label{eq:Upsilon}
\mathcal{O}_{p_\psi(x),T}^{(1)} &= p_\psi(x) +\frac{4\tilde{\mathcal{H}}A}{\bar{N}}\int\mathrm{d}^3y(E+p_E)(y)\left( -\frac{1}{4}\frac{\tilde{\mathcal{H}}}{\bar{N}}+\frac{1}{6}\frac{\bar{N}}{\tilde{\mathcal{H}}A}\Delta-\frac{\kappa}{8}\frac{\bar{N}}{\tilde{\mathcal{H}}} p \right)\delta(x,y) \nonumber \\
~ &- \int\mathrm{d}^3y ~\delta^{ab}(E_{,b}+F_b)(y)\frac{1}{6}\frac{\partial}{\partial x^a}\delta(x,y) \nonumber \\
~ &= \left[p_\psi-\frac{1}{6}\Delta E + \frac{2}{3}\Delta (E+p_E) - \left( \frac{\tilde{\mathcal{H}}^2A}{\bar{N}^2}+\frac{\kappa A}{2} p \right)(E+p_E)\right](x) \nonumber \\
~ &=: \Upsilon(x)
\end{align}
which is the gauge invariant extension of the momentum $p_\psi$ usually not considered in the Lagrangian framework but which also appears in the equation of motion for $\Psi$ in the canonical framework of gauge invariant cosmological perturbation theory (see \cite{Giesel:2017roz}). If we apply the linearized observable map to the clock degrees of freedom that are for the longitudinal gauge encoded in $E$ and $p_E$ these quantities are mapped to zero for the reason that we have chosen $\delta\tau^\mu=0$. That this is indeed the case can be seen below:
\begin{align}
\mathcal{O}_{E(x),T}^{(1)} &= E(x) -\int\mathrm{d}^3y~\delta^{ab}(E_{,b}+F_b)(y)\frac{\partial G}{\partial x^a}(x,y) \nonumber \\
~ &= E(x) -\int\mathrm{d}^3y E(y)\Delta G(x,y) \nonumber \\
~ &= 0, 
\end{align}
and
\begin{align}
\mathcal{O}_{p_E(x),T}^{(1)} &= p_E(x)
+\frac{4\tilde{\mathcal{H}}A}{\bar{N}}\int\mathrm{d}^3y(E+p_E)(y)\left( -\frac{\bar{N}}{4\tilde{\mathcal{H}}A} \right)\delta(x,y) \nonumber \\
~ &+\int\mathrm{d}^3y ~\delta^{ab}(E_{,b}+F_b)(y)\frac{\partial G}{\partial x^a}(x,y) \nonumber \\
~ &= 0 ~.
\end{align}
Here we have used $\partial^aF_a=0$ as well as the symmetry of the Green's function and its partial derivatives. For the transversal vector perturbations one can derive using the background Poisson brackets in (\ref{eq:dfdC3}), the following:
\begin{align}
\mathcal{O}_{F_a(x),T}^{(1)} &= F_a(x)  - \int\mathrm{d}^3y(E_{,b}+F_b)(y)\left( \delta^b_a\delta(x,y) - \frac{\partial^2G(x,y)}{\partial y^a\partial y^b} \right) \nonumber \\
~ &= 0
\end{align}
and
\begin{align}\label{eq:nua}
\mathcal{O}_{p_F^a(x),T}^{(1)} &= p_F^a(x) + \int\mathrm{d}^3y(E_{,b}+F_b)(y)\left( \delta^{ab}\delta(x,y) - \frac{\partial^2G(x,y)}{\partial y^a\partial y^b} \right) \nonumber \\
~ &= p_F^a(x)+\delta^{ab}F_b(x) =: \nu^a(x) ~.
\end{align}
Note that the fact that the gauge invariant extension of $F_a$ just vanishes is again consistent with $\delta T^a=\delta^{ab}(E_{,b}+F_b)$ being clock variables. 
For the scalar field perturbation and its momentum, we obtain,
\begin{align}
\mathcal{O}_{\delta \varphi(x),T}^{(1)} &= \delta \varphi(x) + \frac{4\tilde{\mathcal{H}}A}{\kappa \bar{N}} \int\mathrm{d}^3y(E+p_E)(y)\overline{\{ \varphi(x),C(y) \}} \nonumber \\
~ & \hspace{3em} - \frac{1}{\kappa}\int\mathrm{d}^3y\delta^{ab}(E_{,b}+F_b)(y)\overline{\{ \varphi(x),C_a(y) \}} \nonumber \\
~ &= \delta \varphi(x) + \frac{4\tilde{\mathcal{H}}\lambda_\varphi}{\kappa \sqrt{A}}\bar{\pi}_\varphi (E+p_E) =: \delta \varphi^{(gi)}(x) ~.
\end{align}
Similarly for $\delta \pi_\varphi$ we get:
\begin{align}
\mathcal{O}_{\delta \pi_\varphi(x),T}^{(1)} &= \delta \pi_\varphi(x)- \frac{2\tilde{\mathcal{H}}A^{5/2}}{\bar{N}\lambda_\varphi}\frac{\mathrm{d}V}{\mathrm{d}\varphi}(\bar{\varphi})(E+p_E)(x) - \bar{\pi}_\varphi \Delta E(x) =: \delta \pi_\varphi^{(gi)}(x) ~.
\end{align}\\

Finally, let us compute the observables corresponding to lapse and shift perturbations. For these we have to use the more involved perturbed observable formula \cite{Giesel:2017roz}:
\begin{align}\label{eq:pertobs2a}
\mathcal{O}_{\delta N^\mu(x),T}^{(1)} &= \delta N^\mu(x) - \int\mathrm{d}^3y\delta \dot{T}^\nu(y)\overline{\{N^\mu(x), \tilde{\Pi}_\nu(y)\}} - \int\mathrm{d}^3y\delta T^\nu(y)\overline{\{N^\mu(x),\tilde{\tilde{C}}_\nu(y)\}} \nonumber \\
~ &= \delta N^\mu(x) - \int\mathrm{d}^3y\delta \dot{T}^\nu(y)\kappa \bar{\mathcal{B}}_\nu^\mu(x,y) - \int\mathrm{d}^3y\delta T^\nu(y)\overline{\{N^\mu(x),\tilde{\tilde{C}}_\nu(y)\}}~,
\end{align}
where we once again neglected terms proportional to the primary and secondary constraints. Here 
\be
\tilde{\Pi}_\mu(x) = \int\mathrm{d}^3y~\mathcal{B}^\nu_\mu(y,x)\Pi_\nu(y)
\ee
and
\be
\tilde{\tilde{C}}_\mu(x) = \int\mathrm{d}^3y~\mathcal{B}^\nu_\mu(y,x) \left[ C_\nu(y) - \int\mathrm{d}^3z\int\mathrm{d}^3v~ \mathcal{B}^\sigma_\rho(v,z)\{ \dot{T}^\rho(z),C_\nu(y) \}\Pi_\sigma(v) \right] ~.
\ee
Using the latter we can compute 
\begin{align}\label{eq:NtiltilC}
\overline{\{N^\mu(x),\tilde{\tilde{C}}_\nu(y)\}} &= 
-\kappa\int\mathrm{d}^3z\int\mathrm{d}^3v \bar{\mathcal{B}}^\mu_\sigma(x,v)\bar{\mathcal{B}}^\rho_\nu(z,y) \overline{\{ \dot{T}^\sigma(v),C_\rho(z) \}} \nonumber \\
~ &= -\kappa\int\mathrm{d}^3z\int\mathrm{d}^3v \bar{\mathcal{B}}^\mu_\sigma(x,v)\bar{\mathcal{B}}^\rho_\nu(z,y) \{ \delta 
\dot{T}^\sigma(v),\delta C_\rho(z) \} ~.
\end{align}
The time derivatives in the above formulas are evaluated on shell, i.e. they have to be replaced by the respective Hamilton's equations of motion. These are given by,
\begin{align}
\delta \dot{T}^0 &= \frac{\mathrm{d}}{\mathrm{d}t}[2\sqrt{A}\tilde{P}(E+p_E)] = \bar{N}\left( \phi+\psi-\frac{1}{3}\Delta E \right) + \frac{4\tilde{\mathcal{H}}^2A}{\bar{N}}(E+p_E) ,\nonumber \\
\delta \dot{T}^a &= \dot{E}^{,a}+\delta^{ab}\dot{F}_b = - 4\tilde{\mathcal{H}}\left[ (E+p_E)^{,a} + \nu^a \right] + \delta N^a ~.
\end{align}
Using these results, it is straightforward to obtain

\begin{align}
\overline{\{N(x),\tilde{\tilde{C}}(y)\}} &= 0, 
& \overline{\{N(x),\tilde{\tilde{C}}_a(y)\}} &= 0, \nonumber \\
\overline{\{N^a(x),\tilde{\tilde{C}}(y)\}} &= -\frac{\bar{N}}{A} \frac{\partial}{\partial x^a} \delta(x,y), & \overline{\{N^a(x),\tilde{\tilde{C}}_b(y)\}} &= 0 ~.
\end{align}
This yields the following expressions for the observables corresponding to the lapse and shift perturbations:
\begin{equation}
\mathcal{O}_{\delta N,T}^{(1)} = -\bar{N}\left( \psi - \frac{1}{3}\Delta E + \frac{4\tilde{\mathcal{H}}^2A}{\bar{N}^2}(E+p_E) \right) = - \bar{N} \Psi,
\end{equation}
and
\begin{equation}
\mathcal{O}_{\delta N^a,T}^{(1)} = 4\tilde{\mathcal{H}}\nu^a ~.
\end{equation}
Using the definitions $\delta N = \bar{N}\phi$ and $\delta N^a = B^{,a}+S^a$ we find,
\begin{align}
\mathcal{O}_{\phi,T}^{(1)} &=: \Phi =: \mathcal{O}_{-\psi,T} = -\Psi ~, & \mathcal{O}_{B,T}^{(1)} &= 0 ~, &  \mathcal{O}_{S^a,T}^{(1)} &= 4\tilde{\mathcal{H}}\nu^a ~.
\end{align}
For their conjugate momenta we obtain,
\begin{equation}
\mathcal{O}_{p_\phi,T}^{(1)} = \frac{1}{\bar{N}}\delta\Pi, \quad \mathcal{O}_{p_B,T}^{(1)} = \hat{\delta\Pi} ~,\quad  \mathcal{O}_{p_{S^a},T}^{(1)} = \delta\Pi^a_\perp ~,
\end{equation}
where we used that the projected primary constraints Poisson commute with the generator $G'$.
The result for the observables associated to $S^a$ given by $\mathcal{O}_{S^a,T}^{(1)}$ is also consistent with our former result for the observable $\mathcal{O}_{p_F^a,T}^{(1)}$ if we take the relation $p_F^a=\frac{1}{4\tilde{\cal H}}S^a$ into account. 
As expected, we find that the first order observable associated to $\phi$ is the negative of the first order observable of $\psi$. 
This reflects the well known result that the two Bardeen potentials $\Psi$ and $\Phi$ coincide up to a minus sign for a  scalar field as matter content, see for instance \cite{Giesel:2017roz,Mukhanov}.
\\

In summary, we find that the geometrical clocks corresponding to the longitudinal gauge  lead to the  gauge invariant quantities $\Psi$, $\Upsilon$, $\delta \varphi^{(gi)}$ and $\delta \pi_\varphi^{(gi)}$ for the scalar degrees of freedom and in the case of the vector degrees of freedom we get an observable proportional to $\nu^a$ and a corresponding 
momentum observable.  In addition, 
taking into account tensor perturbations, we can similarly write an observable corresponding to $h_{ab}^{TT}$, and a corresponding momentum observable. For the configuration variables, we get in total seven gauge invariant variables: 3 scalars ($\Phi$, $\Psi$, $\mathcal{O}_{\delta \varphi,T}^{(1)}$), 2 vector components $ (\mathcal{O}_{S^a,T}^{(1)}) $, and 
2 tensor components $ (\mathcal{O}_{h^{TT}_{ab},T}^{(1)}) $. However, only three of these should be independent which amount to three physical degrees of freedom in the gravity plus scalar field configuration space. 

This can be understood by noting that in our analysis in the extended phase space, imposing gauge fixing constraint in terms of clocks, $\delta G^\mu = \delta T^\mu \approx 0$ reduces 4 degrees of freedom. The stability of clocks further reduce degrees of freedom by four. Thus, 8 degrees of freedom are reduced by geometrical clocks and their stability. As a result, in the extended phase space out of 10 configuration degrees of freedom in the metric perturbations and 1 scalar degree of freedom, only 3 remain. Further stabilization of clocks result in fixing of the four Lagrange multipliers $\delta \lambda^\mu$. It should be noted that this is a general argument which applies to all the gauges, if no residual freedom is left after gauge fixing. Let us contrast the situation in the conventional picture in the reduced ADM-phase space. There the gauge fixing constraints $\delta G^\mu \approx 0$ reduce 4 degrees of freedom. Their stability fixes the 4 Lagrange multipliers, which are the lapse function and the shift vector. In the reduced ADM-phase space, with lapse and shift treated as Lagrange multipliers, one starts with 6 metric degrees of freedom and 1 scalar field. After imposing the gauge constraints, one again obtains 3 physical degrees of freedom.

A detailed analysis of the degrees of freedom for different gauge fixing constraints and which of the observables correspond to these degrees of freedom will be done else where. Here let us briefly comment on the  case of the longitudinal gauge. The question is which of the 7 gauge invariant variables corresponding to the configuration space are independent when one considers the longitudinal gauge. One can show that after expressing 10 Einstein's equations in terms of above 7 gauge invariant variables, only 3 of the Einstein's equations are independent. The independent gauge invariant variables turn out to be $\Psi$ and $\mathcal{O}_{h^{TT}_{ab},T}^{(1)}$. Thus, we find that using stability of geometrical clocks we recover the correct 
counting of degrees of freedom via the observable formulae.  In the next subsection, we will see that in the case of the spatially flat gauge, the independent degrees of freedom are captured by the 
Mukhanov-Sasaki variable and $\mathcal{O}_{h^{TT}_{ab},T}^{(1)}$.

The example of longitudinal gauge shows that in our framework the resulting gauge invariant quantities are obtained in a straightforward and systematic way.  
The conventional formalism for perturbations which results in identification of Bardeen potentials framework can therefore be naturally embedded into the observable formalism in the case of the longitudinal clocks. Let us summarize all the observables:

\begin{align}
\mathcal{O}_{\phi,T}^{(1)} &= -\Psi, & \mathcal{O}_{B,T}^{(1)} &= 0, & \mathcal{O}_{S^a,T}^{(1)} &= 4\tilde{\mathcal{H}}\nu^a, \nonumber \\
\mathcal{O}_{\psi,T}^{(1)} &= \Psi, & \mathcal{O}_{E,T}^{(1)} &= 0, & \mathcal{O}_{F_a,T}^{(1)} &= 0, \nonumber \\
\mathcal{O}_{p_\psi,T}^{(1)} &= \Upsilon, & \mathcal{O}_{p_E,T}^{(1)} &= 0, & \mathcal{O}_{p_F^a,T}^{(1)} &= \nu^a, \nonumber \\
\mathcal{O}_{\delta \varphi,T}^{(1)} &= \delta \varphi^{(gi)}, & \mathcal{O}_{\delta \pi_\varphi,T}^{(1)} &= \delta \pi_\varphi^{(gi)} .
\end{align}
For the momenta involving the decomposed primary constraints we get:
\begin{equation}
\mathcal{O}_{p_\phi,T}^{(1)} = \frac{1}{\bar{N}}\delta\Pi, \quad \mathcal{O}_{p_B,T}^{(1)} = \hat{\delta\Pi} ~,\quad  \mathcal{O}_{p_{S^a},T}^{(1)} = \delta\Pi^a_\perp ~.
\end{equation}
As expected all elementary variable that are involved in the clock fields, for the longitudinal gauge these are $E,p_E,B,F_a$ are mapped onto zero by the observable map which is completely consistent with the requirement $\delta T^0\approx 0$ and $\delta T^a\approx 0$. 

\subsection{Spatially Flat Gauge}
We now present the case of the spatially flat gauge. The gauge corresponds to an isotropic threading with a flat slicing.  In this gauge the perturbation in intrinsic curvature due to scalar perturbations vanish. It is defined as,
\begin{align}
\psi &\approx 0, & E &\approx 0 ~.
\end{align} 
It is easily seen that (\ref{eq:delRdecomposed}) then implies $\delta R^{(3)}_{ab} = 0$ in the absence of tensor perturbations.  Using Hamilton's equations for $\psi$ and $E$, the stability of spatially flat gauge conditions amounts to 
\be
\dot{\psi} \approx \tilde{\mathcal{H}}\phi + 2\tilde{\mathcal{H}}p_\psi + \frac{1}{3}\Delta B \overset{!}{\approx} 0 ~~~ \mathrm{and} ~~~ \dot{E} \approx -4\tilde{\mathcal{H}}\Sigma + B \overset{!}{\approx} 0, 
\ee
which imply
\be \label{eq:spatiallyflat2}
 \phi \approx -2 p_\psi - \frac{4}{3}\Delta p_E, ~~~~ \mathrm{and} ~~~  B \approx 4\tilde{\mathcal{H}} p_E ~.
\ee
The stability check of the above two equations do not yield any new constraints, but fix  the perturbations of the Lagrange multipliers $\delta \lambda$ and $\delta \hat{\lambda}$. 

Following the analogous strategy used for the longitudinal gauge to identify the gauge descriptors, we find that the spatially flat 
gauge can be implemented by choosing the following descriptors in the gauge generator $G'_{b,\vec{b}}$:
\begin{align}
b \overset{!}{=} - \frac{\bar{N}}{\tilde{\mathcal{H}}}\left( \psi - \frac{1}{3}\Delta E \right) ~~~~~ \mathrm{and} ~~~~~  \hat{b} &\overset{!}{=} -E ~.
\end{align}

\subsubsection{Geometrical clocks}
To choose the perturbed clocks $\delta T^\mu$ corresponding to the spatially flat gauge, let us recall that the gauge constraint hypersurface is given via $\delta G^\mu = \delta \tau^\mu - \delta T^\mu \approx 0$. For the background metric, the metric components defining the spatially flat gauge are zero, which implies that $\bar \tau^\mu = \bar T^\mu = 0$. Setting $\delta \tau^\mu = 0$, we can identify the  perturbed clocks such that $\delta T^\mu \approx 0$ corresponds to the spatially flat gauge fixing constraint. 
Similar to the case of longitudinal gauge we choose the perturbed clocks equal to the negative of the above expressions for the gauge descriptors. We obtain,
\begin{align}
\delta T^0 &\overset{!}{=} \frac{\bar{N}}{\tilde{\mathcal{H}}}\left( \psi - \frac{1}{3}\Delta E \right), & \delta T^a &\overset{!}{=} \delta^{ab}(E_{,b}+F_b) ~.
\end{align}
Here $\delta T^a$ is composed of their derivative of $\delta \hat T$ determined by $\hat b$, and $\delta T^a_\perp$ chosen as the longitudinal transversal part of the perturbation in spatial metric.    
This choice again yields $\bar{\mathcal{A}}^\mu_\nu(x,y) = \kappa \delta^\mu_\nu\delta(x,y)$. Since the gauge fixing constraints $\delta T^0 \approx 0$ and $\delta \hat{T} \approx 0$ are equivalent to the 
conditions $\psi \approx 0$ and $E \approx 0$, the stabilization of perturbed clocks, $\delta T^0$ and $\delta \hat T$ yield (\ref{eq:spatiallyflat2}). The stability conditions of these clocks also fix the Lagrange multipliers $\delta \lambda$ and $\delta \hat \lambda$.  The stabilization of $\delta T^a_\perp = \delta^{ab} F_b \approx 0$ results in $p_F^a \approx  \tfrac{1}{4\tilde{\mathcal{H}}}S^a$. And, the stability of the latter gives  $\delta \lambda^a_\perp + 3 \tilde{\mathcal{H}} S^a \approx 0$, as in the case of the longitudinal gauge.

Note that the clocks corresponding to the spatially flat gauge only involve metric perturbations and do not depend on the momentum perturbations. As a result, this set of clocks commute in contrast to the non-commuting clocks for the longitudinal gauge. 
 Further, the algebra of observables corresponding to the non clock degrees of freedom will be the respectively simpler standard algebra of configuration and momentum degrees of freedom. 

\subsubsection{Observables} 
The first order observables of the perturbations can be constructed in the same way as the case of longitudinal gauge using the Poisson brackets of metric  and momentum perturbations with those of the linearized constraints and the time derivatives 
of the perturbations in clocks. In these calculations it is useful to note that (\ref{eq:beq}) holds for the spatially flat gauge too. After some straight forward computations we find the following form of the first order observables for the 
clocks corresponding to spatially flat gauge:
\begin{align}
\mathcal{O}_{\phi,T}^{(1)} &= - 2\Upsilon - \left(\frac{1}{2} + \frac{\kappa}{\tilde P^2} A p\right) \Psi, & \mathcal{O}_{B,T}^{(1)} &= \frac{\bar{N}^2}{\tilde{\mathcal{H}}A}\Psi, & \mathcal{O}_{S^a,T}^{(1)} &= 4\tilde{\mathcal{H}}\nu^a, \nonumber \\
\mathcal{O}_{\psi,T}^{(1)} &= 0, & \mathcal{O}_{E,T}^{(1)} &= 0, & \mathcal{O}_{F_a,T}^{(1)} &= 0, \nonumber \\
\mathcal{O}_{p_\psi,T}^{(1)} &= \Upsilon +  \alpha \Psi, & \mathcal{O}_{p_E,T}^{(1)} &= \frac{1}{\tilde{P}^2}\Psi, & \mathcal{O}_{p_F^a,T}^{(1)} &= \nu^a, \nonumber \\
\mathcal{O}_{\delta \varphi,T}^{(1)} &= v, & \mathcal{O}_{\delta \pi_\varphi,T}^{(1)} &= \pi_v ~.
\end{align}
Here
\be\label{eq:alpha}
\alpha := \left(\frac{1}{4} + \frac{\kappa}{2 \tilde P^2} A p\right) -\frac{2}{3}\frac{1}{\tilde{P}^2}\Delta,
\ee
$v$ is the Mukhanov-Sasaki variable 
\be\label{eq:Mukhanov-Sasaki}
v :=\delta \varphi -\frac{\lambda_\varphi}{A^{3/2}\tilde{\mathcal{H}}}\bar{N}\bar{\pi}_\varphi \left( \psi -\frac{1}{3}\Delta E\right)
\ee 
and $\pi_v$ is the corresponding gauge invariant momentum observable:
\begin{align}\label{eq:piv}
\pi_v &:= \delta \pi_\varphi - \bar{\pi}_\varphi\Delta E + \frac{1}{2}\frac{A^{3/2}}{\lambda_\varphi}\frac{\mathrm{d}V}{\mathrm{d}\varphi}(\bar{\varphi}) \frac{\bar{N}}{\tilde{\mathcal{H}}}\left( \psi -\frac{1}{3}\Delta E \right) ~.
\end{align}
For the momenta involving the degrees of freedom of the primary constraints we obtain:
\begin{equation}
\mathcal{O}_{p_\phi,T}^{(1)} = \frac{1}{\bar{N}}\delta\Pi, \quad \mathcal{O}_{p_B,T}^{(1)} = \hat{\delta\Pi} ~,\quad  \mathcal{O}_{p_{S^a},T}^{(1)} = \delta\Pi^a_\perp ~.
\end{equation}
It turns out that the gauge invariant observables which naturally emerges for the choice of clocks consistent with spatially flat gauge are not  the Bardeen potentials. As discussed in section \ref{Sec:Intro}, the natural gauge for the Bardeen potentials is the longitudinal gauge and if any other gauge is chosen, none of the Dirac observables is equal to the Bardeen potentials.  Instead the Mukhanov-Sasaki variable $v$ 
 occurs as the first order gauge invariant extension of the scalar field perturbation $\delta \varphi$. This is expected, because the Mukhanov-Sasaki variable $v$ coincides with the scalar field perturbation in the spatially flat gauge. 
 Finally, let us note that as in the case of longitudinal gauge, including the observables for tensor perturbations, there are 7 (configuration) gauge invariant variables for the metric and scalar field perturbations. Of these, only three are independent 
 which include the Mukhanov-Sasaki variable $v$ and  $\mathcal{O}_{h^{TT}_{ab},T}^{(1)}$. Thus, the degrees of freedom in metric and scalar field sector turn out to be three. As discussed in the case of the longitudinal gauge, this can be shown in the extended phase space using stability of geometrical clocks in our framework.

\subsection{Uniform Field Gauge}
In the presence of the scalar field, as in our analysis, we can choose the gauge such that perturbations in the scalar field vanish. The scalar field is thus homogeneous in this case. In the observable formalism the scalar field perturbation will serve as a clock, with respect to which evolution of other metric and matter variables can be studied.   The metric perturbation $\psi$ captures the curvature perturbation, which itself turns out to be gauge invariant \cite{hwang-noh}.  In order to fix the gauge, another condition is required, such as the isotropic threading which requires vanishing of the longitudinal scalar part of the perturbation in the spatial metric. The uniform field gauge requires:

\begin{align}
\delta \varphi &\approx 0 & E &\approx 0 .
\end{align}
The stability of the gauge conditions yields:
\begin{align}
\phi &\approx 3\psi - \frac{\delta \pi_\varphi}{\bar{\pi}_\varphi} & B &\approx 4\tilde{\mathcal{H}} p_E ~.
\end{align}
As before, the stability requirements fix the Lagrange multipliers $\delta \lambda$ and $\delta \hat{\lambda}$.

Using the transformation properties of $\delta \varphi$ (\ref{eq:trafoscalarfield}) and metric perturbations (\ref{eq:trafoscalarsmetric}) under the action of the gauge generator $G'_{b,\vec{b}}$, we obtain the following gauge descriptors for the uniform field gauge:
\be\label{eq:b-uniform}
b \overset{!}{=} - \frac{A^{3/2} \delta \varphi}{\lambda_\varphi \bar \pi_\varphi} ~~~\mathrm{and} ~~~~  \hat b \overset{!}{=} - E ~.
\ee

\subsubsection{Geometrical Clocks} 
As before, we choose perturbed clocks such that we can reproduce the common results in linearized cosmological perturbation theory. For the background gauge fixing condition we are allowed to choose a non-trivial and time-dependent $\bar{\tau}^0$ because the temporal descriptor involve the scalar field linearly. For the spatial part we still need to choose $\bar{\tau}^a=0$ since the associated linearized descriptor is completed determined by $E$ whose associated background quantity vanishes. 
Now at the linearized level it turns out that a choice of $\delta\tau^\mu\approx 0$ leading to the gauge fixing conditions $\delta G^\mu=-\delta T^\mu\approx 0$ will rediscover the results obtained in cosmology. However, as discussed in section   \ref{Sec:GenGauge} for the uniform field gauge we can relax that assumption and also consider generalized gauge fixing conditions with $\delta\tau^\mu\not=0$. With the choice of  setting $\delta \tau^\mu = 0$, we can identify using the gauge descriptors the following perturbed clocks for the uniform field gauge:

\begin{align}
\delta T^0 &\overset{!}{=} \frac{A^{3/2}}{\lambda_\varphi\bar{\pi}_\varphi}\delta \varphi,  & \delta T^a &\overset{!}{=} \delta^{ab}(E_{,b}+F_b) ~.
\end{align}
The clocks satisfy $\bar{\mathcal{A}}^\mu_\nu(x,y) = \kappa \delta^\mu_\nu\delta(x,y)$ which simplifies the calculations of the observables. It is straightforward to check that as in the case of the spatially flat gauge,  these clocks do commute. It is easy to verify that the stability of the above clocks yield conditions which are consistent with the uniform field gauge and the related stability conditions. Note, that $\delta T^a$ is the same for all choices of clocks that were yet considered for the isotropic threading of spacetime.  Hence, stability of $\delta T^a_\perp \approx 0$ yields $p_f^a \approx \frac{1}{4 \tilde{\cal H}} S^a$ as in the case of the longitudinal and spatially flat gauges.

\subsubsection{Observables}
The observables can be constructed using the linearized observable formulas in (\ref{eq:pertobs1}) and (\ref{eq:pertobs2a}). Inserting the uniform field gauge clocks into the respective formulas, the following first order observables can be derived:
\begin{align}
\mathcal{O}_{\phi,T}^{(1)} &= -(3\varkappa + \varsigma)v - \frac{\pi_v}{\bar{\pi}_\varphi}, & \mathcal{O}_{B,T}^{(1)} &= \frac{\bar{N}^2}{\tilde{\mathcal{H}}A}(\Psi + \varkappa v), & \mathcal{O}_{S^a,T}^{(1)} &= 4\tilde{\mathcal{H}}\nu^a, \nonumber \\
\mathcal{O}_{\psi,T}^{(1)} &= -\frac{1}{\varkappa}v, & \mathcal{O}_{E,T}^{(1)} &= 0, & \mathcal{O}_{F_a,T}^{(1)} &= 0, \nonumber \\
\mathcal{O}_{p_\psi,T}^{(1)} &= \Upsilon + \alpha (\Psi + \varkappa v), & \mathcal{O}_{p_E,T}^{(1)} &= \frac{1}{\tilde{P}^2}(\Psi + \varkappa v), & \mathcal{O}_{p^a_F,T}^{(1)} &= \nu^a, \nonumber \\
\mathcal{O}_{\delta \varphi,T}^{(1)} &= 0, & \mathcal{O}_{\delta \pi_\varphi,T}^{(1)} &= \pi_v + \bar{\pi}_\varphi \varsigma v ~.
\end{align}
Here we have defined,
\be\label{eq:varkappa}
\varkappa := \frac{A^{3/2}}{\lambda_\varphi \bar{\pi}_\varphi}\frac{\tilde{\mathcal{H}}}{\bar{N}} 
\ee
and 
\be\label{eq:varsigma}
\varsigma := \frac{1}{2}\frac{A^3}{\lambda_\varphi^2 \bar{\pi}_\varphi^2}\frac{\mathrm{d}V}{\mathrm{d}\varphi}(\bar{\varphi}) ~.
\ee
Apart from the above observables, we also obtain the following related to primary constraints,
\begin{equation}
\mathcal{O}_{p_\phi,T}^{(1)} = \frac{1}{\bar{N}}\delta\Pi, \quad \mathcal{O}_{p_B,T}^{(1)} = \hat{\delta\Pi} ~,\quad  \mathcal{O}_{p_{S^a},T}^{(1)} = \delta\Pi^a_\perp ~.
\end{equation}

Note, that the combination $\Psi+\varkappa v$ appearing in the linearized observables can be related to $\delta \varphi^{(gi)}$:
\begin{align}
\delta \varphi^{(gi)} &= \frac{1}{\varkappa}(\Psi+\varkappa v), & \delta \pi_\varphi^{(gi)} &= \pi_v - \frac{\varsigma}{\varkappa}\bar{\pi}_\varphi \Psi ~.
\end{align}
Further, the first order observable for $\psi$ is proportional to the Mukhanov-Sasaki variable $v$. This is, however, not surprising because instead of interpreting $v$ as a gauge invariant extension of the scalar field involving the geometric perturbations $\psi$ and $E$ we can analogously interpret it as a gauge invariant extension of $\psi$ involving $\delta \varphi$ and $E$. Similar to the longitudinal and spatially flat gauge, we will have three in independent physical degrees of freedom, one in the scalar and two in the tensor sector.

 \subsection{Synchronous Gauge}
 In the cosmological perturbation theory, synchronous gauge has been studied extensively, see for e.g. \cite{Weinberg}. The underlying idea is to  use the gauge freedom of the theory to set the temporal-temporal and temporal-spatial components of the metric perturbation $\delta g_{tt}$, $\delta g_{ti}$ equal to zero. In terms of the ADM variables this gauge is equivalent to choosing vanishing lapse and shift perturbations, that is $\delta N^\mu =0$. For the scalar perturbations this gauge requires:
\begin{align}
\phi &\approx 0, & B &\approx 0 ~.
\end{align}
The stability of these gauge conditions directly translate to the conditions on the perturbations of the Lagrange multipliers, via (\ref{eq:dotphidotB}). No other conditions on any other perturbations arise. 
Using (\ref{eq:trafoscalarsmetric}), we find the gauge descriptors for the synchronous gauge as: 
\begin{align}
b(x,t) &\overset{!}{=} -\int\limits^{t}\mathrm{d}t'\bar{N}(t')\phi(x,t') - c_1(x), \nonumber \\
\hat{b}(x,t) &\overset{!}{=} -\int\limits^{t}\mathrm{d}t'\frac{\bar{N}}{A}(t')\left[ \int\limits_{}^{t'}\mathrm{d}t''\bar{N}(t'')\phi(x,t'')+c_1(x) \right] - \int\limits^{t}\mathrm{d}t'B(x,t') -c_2(x) ~.
\end{align}
In comparison to the longitudinal and spatially flat gauges, the gauge descriptors for the synchronous gauge have more non-trivial expressions.  First, they involve time integrals which are quite non-trivial to deal with in passage to the Hamiltonian formulation. Second, there are two arbitrary functions $c_1(x)$ and $c_2(x)$ which are constant in time and hence can be any constants of motion. 
 These arbitrary `constants' are the non-physical gauge modes in the synchronous gauge (see for e.g. \cite{Mukhanov}). Finally, the gauge descriptors involve lapse and shift perturbations. Thus, the resulting clocks will depend on the latter.

\subsubsection{Geometrical clocks}
Let us consider the clocks for the synchronous gauge using the identification of the gauge descriptors. For the background metric, $B$ vanishes and in agreement with this we choose $\bar{\tau}^a=0$. For the temporal gauge fixing condition we obtain $\bar{T}^0=\bar{N}$ because we have $N=\bar{N}+\delta N=\bar{N}+\bar{N}\phi$. Thus we realize that we can choose a non-vanishing time dependent $\tau^0$ for the synchronous gauge and therefore also in this gauge we can potentially define a notion of physical time for the background solution. Similar to the uniform field gauge also here in principle a generalized perturbed gauge fixing condition with $\delta\tau^\mu\not=0$ can be formulated as also discussed in section  \ref{Sec:GenGauge}. However, for being able to reproduce the results of the Lagrangian framework, we chose again $\delta \tau^\mu=0$ and write the linearized gauge fixing constraint as $\delta G^\mu = - \delta T^\mu \approx 0$. As a result, the perturbed clocks turn out to be: 
\begin{align}\label{eq:synch-clock}
\delta T &\overset{!}{=} \hat{I}\bar{N}\phi + c_1, \nonumber \\
\delta \hat{T} &\overset{!}{=} \hat{I}\frac{\bar{N}}{A}\left[ \hat{I}\bar{N}\phi+c_1 \right] +\hat{I}B +c_2 ~.
\end{align}
Since the descriptors involve a time integral the same will be true for the corresponding clocks. For this purpose we denoted this time integral as $\hat{I}$ which is defined such that it satisfies
\begin{equation}
\frac{\mathrm{d}}{\mathrm{d}t}\left(\hat{I}f\right)(t):=\frac{\mathrm{d}}{\mathrm{d}t}\int\limits^t \mathrm{d}t'f(t')=f(t) 
\end{equation}
where $f$ is an arbitrary phase space function and time derivatives of $f$ are expressed via the Poisson bracket of $f$ and the perturbed Hamiltonian $\delta H^{(2)}$ on the linearized phase space. Note, that $\hat{I}$ and Poisson brackets do not commute in general and also $\hat{I}$ will in general be non unique and may only exist on a subset of the phase space.

For the reason that $\delta \hat T$ is the scalar part of the shift perturbation, we choose $\delta T^a_\perp$ as the transverse part of the latter:
\be
\delta T^a_\perp \overset{!}{=} \hat{I} S^a +c_3 ~, 
\ee
where like $c_1$ and $c_2$, $c_3$ is a function of spatial coordinates which is a constant in time. 

Using the on-shell relations of the perturbations in the metric and the Lagrange multiplier we obtain:
\begin{align}
[\hat{I}\bar{N}\phi](x,t) &= \int\limits^{t}\mathrm{d}t'\int\limits^{t'}\mathrm{d}t''\delta \lambda(x,t''),  & [\hat{I}B](x,t) &= \int\limits^{t}\mathrm{d}t'\int\limits^{t'}\mathrm{d}t''\delta \hat{\lambda}(x,t'') ~
\end{align}
and
\be
[\hat{I}S^a](x,t) = \int\limits^{t}\mathrm{d}t'\int\limits^{t'}\mathrm{d}t''\delta {\lambda^a_\perp}(x,t'') ~.
\ee
For the stability of these clocks, we need 
\be
\delta \dot{T}^0 = \bar{N}\phi \approx 0, ~~~~~~~~~ \delta \dot{\hat{T}} = \frac{\bar{N}}{A}\left[ \hat{I}\bar{N}\phi+c_1 \right] +B \approx 0 ~,
\ee
and 
\be
\delta \dot{{T^a_\perp}} = S^a \approx 0 ~.
\ee
The first condition yields $\phi \approx 0$. Noting that $\delta T^0 \approx 0$ implies $c_1 \approx 0$, from the second condition we obtain $B \approx 0$. And, the third condition yields $S^a \approx 0$.   The stability of the clocks immediately yield:  $\delta \lambda \approx 0$, $\delta \hat \lambda \approx 0$ and $\delta \lambda^a_\perp \approx 0$. 
Let us note that these are not the most general clocks for the synchronous gauge. The reason is that we have obtained them by fixing $c_1 \approx 0$, $c_2 \approx 0$ and $c_3 \approx 0$, the latter two getting fixed using  $\delta \hat{T} \approx 0$ and $\delta T^a_\perp \approx 0$. Other choices of $c_1$, $c_2$ and $c_3$ are possible by defining,
$\sigma^\mu := \tau^\mu - \bar{T}^\mu $, as a result of which we obtain $\delta T^\mu \approx \sigma^\mu$. Following the above analysis, one is then led to relations between components of $\sigma^\mu$ and $c_1$, $c_2$ and $c_3$. The constant functions $c_1$ and $c_2$ are thus determined by choice of $\sigma^\mu$. In the following we will consider the choices where constants $c_1, c_2$ and $c_3$ are all vanishing.

\subsubsection{Observables}
The clocks we have found for the synchronous gauge, by construction satisfy:
\begin{align}
\overline{\{ T,G'_{b,\vec{b}} \}}, &= b & \overline{\{ \hat{T},G'_{b,\vec{b}} \}} = \hat{b}~.
\end{align}
Using these the observable formula can be solved for the descriptors quite easily. The first order observable formula becomes:
\begin{equation}\label{gen-obsformula}
\mathcal{O}^{(1)}_{f,T}(x) = \delta f(x) +\left. \overline{\{ f(x),G'_{b,\vec{b}} \}} \right|_{b^\mu=-\delta T^\mu} ~.
\end{equation}
Thus the first order observable of a scalar perturbation is just the expression of its infinitesimal transformation behavior with $b \to -\delta T$ and $\hat{b} \to -\delta \hat{T}$. This results in the following first order observables:

\begin{align}
\mathcal{O}^{(1)}_{\phi,T} &= 0, ~&~ \mathcal{O}^{(1)}_{B,T} &= 0, \nonumber \\
\mathcal{O}^{(1)}_{\psi,T} &= \psi- \frac{\tilde{\mathcal{H}}}{\bar{N}}\hat{I}\bar{N}\phi-\frac{1}{3}\Delta \hat{I}\left( \frac{\bar{N}}{A}\hat{I}\bar{N}\phi +B \right), ~&~ \mathcal{O}^{(1)}_{E,T} &= E - \hat{I}\left(\frac{\bar{N}}{A}\hat{I}\bar{N}\phi +B \right), \nonumber \\
\mathcal{O}^{(1)}_{p_E,T} &= p_E + \frac{\bar{N}}{4A\tilde{\mathcal{H}}}\hat{I}\bar{N}\phi + \hat{I}\left(\frac{\bar{N}}{A}\hat{I}\bar{N}\phi +B \right), \nonumber
\end{align}
\vskip-0.5cm
\ba
\mathcal{O}^{(1)}_{p_\psi,T} &=& p_\psi + \left( \frac{\tilde{\mathcal{H}}}{4\bar{N}}+\frac{\kappa}{8}\frac{\bar{N} p}{\tilde{\mathcal{H}}} \right)\hat{I}\bar{N}\phi -  \frac{1}{6}\Delta\left( \frac{\bar{N}}{A\tilde{\mathcal{H}}}\hat{I}\bar{N}\phi + \hat{I}\left(\frac{\bar{N}}{A}\hat{I}\bar{N}\phi +B \right) \right), \nonumber \\
\mathcal{O}^{(1)}_{F_a,T} &=& F_a - \hat I S_a,  ~~~~~~~~~~~~ \mathcal{O}^{(1)}_{{p^a_F},T} = p_F^a + \hat I S_a, ~~~~~~~~~~~~ \mathcal{O}^{(1)}_{S^a,T} = 0, \nonumber \\
\mathcal{O}_{\delta \varphi,T}^{(1)} &=& \delta \varphi - \frac{\lambda_\phi}{A^{3/2}} \bar \pi_\phi \hat I \bar N \phi, \nonumber \\
  \mathcal{O}_{\delta \pi_\varphi,T}^{(1)} &=&  \pi_{\varphi} - \bar \pi_\varphi \Delta \hat I \left(\frac{\bar N}{A} \hat I \bar N \phi + B\right) + \frac{1}{2} \frac{A^{3/2}}{\lambda_\varphi} \frac{\mathrm{d}V}{\mathrm{d}\varphi} \hat I \bar N \phi ~.
\ea
And for the observables corresponding to momenta of lapse and shift perturbations we get,
\begin{equation}
\mathcal{O}_{p_\phi,T}^{(1)} = \frac{1}{\bar{N}}\delta\Pi, \quad \mathcal{O}_{p_B,T}^{(1)} = \hat{\delta\Pi} ~,\quad  \mathcal{O}_{p_{S^a},T}^{(1)} = \delta\Pi^a_\perp ~.
\end{equation}
Though we have found the observables corresponding to the synchronous gauge, we must note that the operator $\hat I$ might not be unique. This is to be contrasted with the results for the previous gauges where no such ambiguity exists.

\subsection{Comoving gauge}
In this gauge the slicing is chosen such that the scalar field perturbations vanish. In the presence of fluids this translates to comoving slicing in which the time slices are orthogonal to the fluid velocity. In particular, the fluid velocity perturbation must match the perturbation in the shift. For the case of the scalar field, we obtain, 
\be
\delta \varphi \approx 0, ~~~~~ B \approx 0 ~.
\ee
As in the case of the uniform field gauge, the scalar field is homogeneous. Though the time slicing is fixed, there is a residual freedom in the choice of origin of spatial coordinates. As we will see, this will get reflected in the presence of an arbitrary constant in the geometrical clocks corresponding to scalar perturbations. Stability of the comoving gauge conditions require:
\be
\delta \dot \varphi \approx 0, ~~ \mathrm{and} ~~ \dot B \approx 0 ~.
\ee
The second condition fixes the perturbation, $\delta \hat \lambda \approx 0$. Whereas the first condition, using Hamilton's equations,  results in 
\be
\phi - 3 \psi - \frac{\delta \pi_\varphi}{\bar \pi_\varphi} \approx 0 ~.
\ee
The stability of the above condition results in fixing the perturbation $\delta \lambda$. Using the transformation properties of $\delta \varphi$ and $B$ we can find the gauge descriptors as before. The difference in contrast to previous gauges is that 
the gauge descriptor $b$ takes a simple form as in the longitudinal gauge, whereas the gauge descriptor $\hat b$ involves time integrals as in the synchronous gauge. These are given by
\be\label{eq:b-comoving}
b(x,t) \overset{!}{=} -\frac{A^{3/2} \delta \varphi}{\lambda_\varphi \bar \pi_\varphi} ,
\ee
and 
\be
\hat b(x,t) \overset{!}{=} - \int \mathrm{d} t' \frac{\bar N(t') A^{1/2}(t')}{\lambda_\varphi \bar \pi_\varphi(t')}  - \int \mathrm{d} t'' B(x,t'') - c_4(x) ~.
\ee
As in the case of the synchronous gauge, the latter gauge descriptor involves an integration in time which creates some ambiguity in the Hamiltonian formulation. The presence of the function $c_4(x)$ represents a residual gauge 
freedom corresponding to a shift of the spatial coordinates.

\subsubsection{Geometrical Clocks}
As in the other gauges, we want to to choose the clocks such that we can reproduce the gauge fixing constraints used in cosmology.
Considering the background quantities in the comoving gauge, we realize that we need to choose $\bar{\tau}^a=0$
since the metric component corresponding to the shift perturbation is zero. However, as far as the temporal background clock is considered we have the freedom to choose $\bar{\tau}^0\not=0$ and again this in principle allows to define a notion of physical time for the background solution. At the linearized level we choose $\delta\tau^a=0$ and also here a non-vanishing $\delta\tau^0$ is consistent with the linearized equations of motion. But, likewise to the other gauges already discussed, for reproducing the exact gauge fixing conditions used in cosmological perturbation theory, we consider the specific choice of $\delta\tau^0=0$. The generalized gauge fixing condition with $\delta\tau^0\not=0$ will be analyzed in  more detail in section \ref{Sec:GenGauge}. With $\delta \tau^\mu$ set to vanish, the perturbed clocks using the expressions of $b$ and $\hat b$ are:
\be
\delta T^0 \overset{!}{=} \frac{A^{3/2} \delta \varphi}{\lambda_\varphi \bar \pi_\varphi}, ~~~~ \delta \hat T \overset{!}{=} \hat I \frac{\bar N A^{1/2} \delta \varphi}{\lambda_\varphi \bar \pi_\varphi}  + \hat I B + c_4 ~.
\ee
Since $\delta \hat T$ is proportional to the shift perturbation, it leads us to 
identify
\be
\delta T^a_\perp \overset{!}{=} \hat I S^a + c_5 ~,
\ee
where $c_5$, like $c_4$, is an arbitrary constant in time which depends on spatial coordinates.  Note that in contrast to the uniform field gauge, where also $\delta \varphi \approx 0$, $\delta \hat T$ contains shift perturbation. This is what results in the presence of $c_4$ for the scalar perturbation which reflects the residual freedom mentioned earlier. The stability of $\delta T^0 \approx 0$ results in the fixing of $\delta \lambda$. On the other hand, stability of $\delta \hat T$ and 
$\delta T^a_\perp$ fix $\delta \lambda^a$, consistent with the gauge conditions for the comoving gauge.

\subsubsection{Observables}
To find the observables we use the general formulas as noted earlier, (\ref{eq:pertobs1}) and (\ref{eq:pertobs2a}), and also (\ref{gen-obsformula}). Using the clocks corresponding to the comoving gauge, we obtain:
\ba
\mathcal{O}_{\phi,T}^{(1)} &=& -(3\varkappa + \varsigma)v - \frac{\pi_v}{\bar{\pi}_\varphi}, ~~~~~~~~~~~~~~~~~ \mathcal{O}_{B,T}^{(1)} = 0, ~~~~~~~~~~~~~~~~~ \mathcal{O}_{S^a,T}^{(1)} = 0, \nonumber \\
\mathcal{O}_{\psi,T}^{(1)} &=& \psi - \frac{\tilde {\cal H} A^{3/2} \delta \varphi}{\bar N \lambda_\varphi \bar \pi_\varphi} -  \frac{\Delta}{3} \hat I (\beta \delta \varphi + B), ~~~~~~~~~~~~ \mathcal{O}_{E,T}^{(1)} = E - \hat I(\beta \delta \varphi + B), \nonumber \\
\mathcal{O}_{F_a,T}^{(1)} &=& F_a - \hat I S_a, ~~~~~~~~~~~~~~ \mathcal{O}_{p_E,T}^{(1)} = p_E + \frac{\beta \delta \varphi}{4 {\tilde{\cal H}}} + \hat I (\beta \delta \varphi + B), \nonumber \\
\mathcal{O}_{p_\psi,T}^{(1)}  &=&  p_\psi + \frac{A^{3/2} \delta \varphi}{4 \lambda_\varphi \bar \pi_\varphi} \left(\frac{\tilde {\cal H}}{\bar N} + \frac{\kappa \bar N p}{2 \tilde{\cal H}}\right) - \frac{\Delta}{6}\left(\frac{\beta}{\tilde {\cal H}} \delta \varphi + \hat I (\beta \delta \varphi + B)\right),  \nonumber \\
\mathcal{O}_{p_F^a,T}^{(1)} &=& p_F^a + \hat I S^a, ~~~~~~~~~~~~~~~~~~~ \mathcal{O}_{\delta \varphi,T}^{(1)} = 0, \nonumber \\
\mathcal{O}_{\delta \pi_\varphi,T}^{(1)} &=& \delta \pi_\varphi - \frac{\dot \pi_\varphi}{\dot \varphi} \delta \varphi - \bar \pi_\varphi \Delta \hat I \left(\bar \pi_\varphi \beta \delta \varphi + B\right), ~
\ea
where 
\be\label{eq:beta}
\beta := \frac{\bar N A^{1/2}}{\lambda_\varphi \bar \pi_\varphi} ~.
\ee
And as for all other gauges, we also have 
\begin{equation}
\mathcal{O}_{p_\phi,T}^{(1)} = \frac{1}{\bar{N}}\delta\Pi, \quad \mathcal{O}_{p_B,T}^{(1)} = \hat{\delta\Pi} ~,\quad  \mathcal{O}_{p_{S^a},T}^{(1)} = \delta\Pi^a_\perp ~.
\end{equation}
Unlike the case of the uniform field gauge, some of the observables consist of the $\hat I$ operator. These observables, as in the case of the synchronous gauge, have a certain non-uniqueness associated with the $\hat I$.

\subsection{Generalized gauge fixing constraints and modified gauges}
\label{Sec:GenGauge}
As mentioned earlier for the uniform field, the synchronous and the comoving gauge, the equation of motions are consistent with choosing temporal functions $\bar{\tau}^0$ and $\delta\tau^0$ that do not vanish. This corresponds to gauge fixing conditions $\bar{G}^\mu=\bar{\tau}^\mu-\bar{T}^\mu$ and $\delta G^\mu=\delta\tau^\mu-\delta T^\mu$. We showed in the previous subsections that a choice of $\delta\tau^0=0$, that is $\delta G^\mu=-\delta T^\mu\approx 0$, was necessary to reproduce the common gauges and associated gauge invariant quantities like for instance the Bardeen potentials and the Mukhanov-Sasaki variable.
However, if we want to interpret our chosen clocks as true physical clocks in the context of the relational formalism, then a vanishing temporal parameter $\delta\tau^0$ seems to be problematic since in general the observable map of a function $f$ returns the value of $f$ at those values where the clocks take the values $\tau^\mu$. As far as the spatial clocks are considered we can set $\delta\tau^a$ to zero because the physical evolution of the observables is defined with respect to the temporal clock only.
More generally, if we choose $\tau^a=\tau^a(x^j)$ such that it is time independent, this corresponds to an induced slicing for which the shift vector vanishes on the constraint hypersurface where the gauge-fixing conditions are satisfied. 
In this subsection we want to analyze generalized gauges for the uniform field, the synchronous and the comoving gauge. For this purpose we choose the simple generalization of a temporal function $\tau^0(t)\not=0$ corresponding in linearized cosmological perturbation theory to $\bar{\tau}^0(t)\not=0$ and $\delta\tau^0(t)=t$. In the linearized theory the latter  choice corresponds to $\delta G^0=t-\delta T^0\approx 0$ allowing us to define a notion of physical time for the observables constructed in these gauges. That we choose a linear dependence on $t$ for $\delta\tau^0$ is the most simple choice for a coordinate gauge fixing constraints as discussed for instance in \cite{Dust1,Pons4}. As usual in the relational formalism with such generalized gauge fixing constraints we would construct a $\tau^0$-dependent family of observables and hence more general observables than conventionally used in the context of cosmological perturbation theory. By introducing  a non-vanishing function $\delta\tau^0(t)$ we also modify the gauge and hence gauge fix the clock fields differently. As a consequence also the stability conditions of the clocks become modified and involve additional terms. Nevertheless, these modified conditions still merge into our former results presented above if we again consider the choice of $\delta\tau^0=0$.

In the following we focus on the way uniform field, synchronous and comoving gauges are modified when the gauge fixing constraint is $\delta G^0 = t- \delta T^0 \approx 0$. As a result, the stability requirement of the temporal gauge fixing constraint leads to modified conditions on the geometric and matter perturbations respectively. However, as far as the spatial gauge fixing condition is considered our former results still apply because we still choose $\delta\tau^a=0$, that is $\delta G^a=-\delta T^a$. Note that for the uniform field and the comoving gauge, the perturbed temporal clock is the same and hence we do not need to discuss these two cases separately since the results are identical. The latter gauges will be discussed after the case of the synchronous gauge that we will start with.

Let us consider a generalization of the gauge fixing constraint which led to the synchronous gauge in the above analysis. The perturbed temporal clock in the synchronous gauge was determined by the perturbation in the lapse. Requiring  $\delta G^0 = t- \delta T^0 \approx 0$, then the generalized gauge descriptor $b$ is given by 
\begin{equation}
b(x,t) \overset{!}{=} t -\int\limits^{t}\mathrm{d}t'\bar{N}(t')\phi(x,t') - c_1(x) ~,
\end{equation}
which leads to the following perturbed temporal clock $\delta T^0 = \hat{I}\bar{N}\phi $. Reinserting this back into the gauge fixing constraint yields,
\begin{equation}
    \delta T^0 \overset{!}{=} \hat{I}\bar{N}\phi  \nonumber \approx t ~.
\end{equation}
Here as in the case of the synchronous gauge we have chosen the constant $c_1$ to be vanishing. The spatial clocks and corresponding gauge descriptors remain unchanged.  
 The stability of the above gauge fixing constraint gives:
\begin{equation}
    \delta \dot T^0 {=} \bar{N}\phi  \nonumber \approx 1 ~.
\end{equation}
Using (\ref{eq:trafoscalarsmetric}), we easily see that the stability of above equation yields $\delta \lambda = 0$ exactly as in the case of the perturbation in temporal clock for the synchronous gauge. Recall that one of the conditions for the synchronous gauge, $\phi \approx 0$,  is equivalent to the stability of the gauge fixing: $\delta T^0 \approx 0$. The stability of the generalized gauges requires that $\bar N \phi \approx 1$. Thus, if the perturbed temporal clock 
$\delta T^0$ is required to be linear in time and the perturbed spatial clock takes vanishing value then the synchronous gauge condition for the perturbations modifies to: 
\begin{equation}
    \phi \approx \frac{1}{\bar N}~, ~~~~~~~~~~~ B \approx 0 ~.
\end{equation}\\

 Similarly, we can consider the generalization of the gauge fixing constraint leading to the uniform field and the comoving gauges. For both the cases, the perturbation in the temporal clock is the same. The gauge descriptor corresponding to $\delta G^0 = t - \delta T^0 \approx 0$ changes from (\ref{eq:b-uniform}) (or (\ref{eq:b-comoving})) to 
\begin{equation}
b  \overset{!}{=} t - \frac{A^{3/2} \delta \varphi}{\lambda_\varphi \bar \pi_\varphi} ~.
\end{equation}
The associated perturbed temporal clock is chosen to have the form $\delta T^0 \overset{!}{=} A^{3/2} \delta \varphi/\lambda_\varphi \bar \pi_\varphi$ leading to the following gauge fixing constraints
 \be
 \delta T^0 = \frac{A^{3/2} \delta \varphi}{\lambda_\varphi \bar \pi_\varphi} \approx t ~.
  \ee
The stability of the above constraint yields a differential equation relating time derivatives of background quantities with $\delta \dot \varphi$: 
\be
\delta \dot T^0 = \left(\frac{A^{3/2} \delta \varphi}{\lambda_\varphi \bar \pi_\varphi}\right)^{\boldsymbol{\cdot}} \delta \varphi ~ + ~ \frac{A^{3/2} \delta \varphi}{\lambda_\varphi \bar \pi_\varphi} \delta \dot \varphi \approx 1
\ee
The solution of the above equation yields, $ \delta \varphi \approx (t + t_{\delta \varphi}^\prime) ~\lambda_\varphi \bar \pi_\varphi/A^{3/2}$, where $t_{\delta \varphi}^\prime$ is a constant of integration which we set to zero. We then need to consider the stability of the above equation with respect to the equations of motion for perturbed variables. This  results in determining the value of $\delta \lambda$ in terms of background and perturbation variables, as it was the case for the uniform field and comoving gauges. Thus we obtain a consistent perturbed temporal clock that can be gauge fixed to be linear in time. The associated modified uniform field gauge in this case is given by:
\begin{equation}
     \delta \varphi \approx t  ~\frac{\lambda_\varphi \bar \pi_\varphi}{A^{3/2}} ~, ~~~~~~~ E \approx 0
\end{equation}
Similarly a modified comoving gauge reads:
\begin{equation}
     \delta \varphi \approx t  ~\frac{\lambda_\varphi \bar \pi_\varphi}{A^{3/2}} ~, ~~~~~~~ B \approx 0 ~.
\end{equation}
This concludes our discussion of the generalized gauge fixing constraints that lead to modified uniform field, synchronous and comoving gauges for which in the relational formalism a notion of physical time can be defined via these geometrical clocks. Likewise to the case of the unmodified gauges, we could apply the observable map now and obtain a family of Dirac observables parametrized by $t$, where this parameter is interpreted as physical time. For the choice of $t=0$ these observables coincide with the observables constructed in the former subsections.
\\

Finally,  let us summarize all the results obtained so far in this section for various gauges in our formalism of reference clocks  in tables \ref{tab:clockchoices1} and \ref{tab:clockchoices2}. In table \ref{tab:clockchoices1} we summarize the results for longitudinal, spatially flat and uniform field gauges all which involve isotropic threading. Table \ref{tab:clockchoices2} summarizes the synchronous and comoving gauges which have vanishing perturbations of the shift. Table \ref{tab:symbols} summarizes various symbols and key equations in tables \ref{tab:clockchoices1} and \ref{tab:clockchoices2}. 

\begin{table}[tbh!]
\begin{center}
\renewcommand{\arraystretch}{1.5}
\begin{tabular}{c||c|c|c}
Variable & Longitudinal & Spatially flat & Uniform field \\
\hline \hline
$\delta T^0$ & $2\tilde{P}\sqrt{A}(E+p_E)$ & $\frac{\bar{N}}{\tilde{\mathcal{H}}}(\psi-\frac{1}{3}\Delta E)$ & $\varkappa \frac{\bar{N}}{\tilde{\mathcal{H}}}\delta \varphi$ \\
$\delta T^a$ & $\delta^{ab}(E_{,b}+F_b)$ & $\delta^{ab}(E_{,b}+F_b)$ & $\delta^{ab}(E_{,b}+F_b)$ \\
\hline
$\phi$ & $-\Psi$ &  $- 2\Upsilon - \left(\frac{1}{2} + \frac{\kappa}{\tilde P^2} A p\right) \Psi $   & $-(3\varkappa + \varsigma)v-\frac{\pi_v}{\bar{\pi}_\varphi}$ \\
$B$ & $0$ & $\frac{\bar{N}^2}{A\tilde{\mathcal{H}}}\Psi$ & $\frac{\bar{N}^2}{A\tilde{\mathcal{H}}}(\Psi + \varkappa v)$ \\
$S^a$ & $4\tilde{\mathcal{H}}\nu^a$ & $4\tilde{\mathcal{H}}\nu^a$ & $4\tilde{\mathcal{H}}\nu^a$ \\
$\psi$ & $\Psi$ & $0$ & $-\frac{1}{\varkappa}v$ \\
$E$ & $0$ & $0$ & $0$ \\
$F_a$ & $0$ & $0$ & $0$ \\
$p_\psi$ & $\Upsilon$ & $\Upsilon + \alpha \Psi$ & $\Upsilon + \alpha(\Psi + \varkappa v)$ \\
$p_E$ & $0$ & $\frac{1}{\tilde{P}^2}\Psi$ & $\frac{1}{\tilde{P}^2}(\Psi + \varkappa v)$ \\
$p^a_F$ & $\nu^a$ & $\nu^a$ & $\nu^a$ \\
$\delta \varphi$ & $v + \frac{1}{\varkappa}\Psi$ & $v$ & $0$ \\
$\delta \pi_\varphi$ & $\pi_v- \bar{\pi}_\varphi\frac{\varsigma}{\varkappa}\Psi$ & $\pi_v$ & $\pi_v+\bar{\pi}_\varphi \varsigma v$ \\
\end{tabular}
\caption{Summary of geometrical clocks and linearized observables corresponding to various metric perturbations and their momenta for the gauges where the longitudinal part of the spatial metric perturbation vanishes. Various symbols are summarized in table \ref{tab:symbols}. \label{tab:clockchoices1}} 
\end{center}
\end{table}

\begin{table}[h!]
\begin{center}
\renewcommand{\arraystretch}{1.5}
\begin{tabular}{c||c|c}
Variable & Synchronous & Comoving  \\
\hline \hline
$\delta T^0$ & $\hat{I}\bar{N}\phi$ &  $\frac{A^{3/2} \delta \varphi}{\lambda_\varphi \bar \pi_\varphi}$\\
$\delta \hat T$ & $\hat{I}\frac{\bar{N}}{A}\left[ \hat{I}\bar{N}\phi\right] +\hat{I}B $& $\hat I \frac{\bar N A^{1/2} \delta \varphi}{\lambda_\varphi \bar \pi_\varphi}  + \hat I B$\\
$\delta T^a_\perp$ &  $\hat{I} S^a$ &$\hat{I} S^a$  \\
\hline
$\phi$ & $0$ &  $-(3\varkappa + \varsigma)v - \frac{\pi_v}{\bar{\pi}_\varphi}  $   \\
& &  \\
$B$ & $0$ & $0$ \\
& &  \\
$S^a$ & $0$ & $0$ \\
& &  \\
$\psi$ & $\psi- \frac{\tilde{\mathcal{H}}}{\bar{N}}\hat{I}\bar{N}\phi-\frac{1}{3}\Delta \hat{I}  \left(\frac{\bar{N}}{A}\hat{I}\bar{N}\phi +B \right) $ & $\psi - \frac{\tilde {\cal H} A^{3/2} \delta \varphi}{\bar N \lambda_\varphi \bar \pi_\varphi} -  \frac{\Delta}{3} \hat I (\beta \delta \varphi + B)$  \\
& &  \\
$E$ & $E - \hat{I} \left(\frac{\bar{N}}{A}\hat{I}\bar{N}\phi +B \right)  $ & $ E - \hat I(\beta \delta \varphi + B)$ \\
& &  \\
$F_a$ & $F_a - \hat I S_a$ & $F_a - \hat I S_a$ \\ 
& &  \\
$p_\psi$ & \makecell{$p_\psi + \frac{1}{4}\left( \frac{\tilde{\mathcal{H}}}{\bar{N}}+\frac{\kappa}{2}\frac{\bar{N} p}{\tilde{\mathcal{H}}} \right)\hat{I}\bar{N}\phi$\\$ -  \frac{1}{6}\Delta\left( \frac{\bar{N}}{A\tilde{\mathcal{H}}}\hat{I}\bar{N}\phi + \hat{I}  \left(\frac{\bar{N}}{A}\hat{I}\bar{N}\phi +B \right) \right)$} & \makecell{$  p_\psi + \frac{A^{3/2} \delta \varphi}{4 \lambda_\varphi \bar \pi_\varphi} \left(\frac{\tilde {\cal H}}{\bar N} + \frac{\kappa \bar N p}{2 \tilde{\cal H}}\right)$\\$ - \frac{\Delta}{6}\left(\frac{\beta}{\tilde {\cal H}} + \hat I (\beta \delta \varphi + B)\right) $} \\
& &  \\
$p_E$ & \makecell{$p_E + \frac{\bar{N}}{4A\tilde{\mathcal{H}}}\hat{I}\bar{N}\phi$\\ $+ \hat{I}  \left(\frac{\bar{N}}{A}\hat{I}\bar{N}\phi +B \right) $} & $p_E + \frac{\beta \delta \varphi}{4 {\tilde{\cal H}}} + \hat I (\beta \delta \varphi + B)$  \\
& &  \\
$p^a_F$ & $p_F^a + \hat I S_a$ & $p_F^a + \hat I S^a$  \\
& &  \\
$\delta \varphi$ & $\delta \varphi - \frac{\lambda_\phi}{A^{3/2}} \bar \pi_\phi \hat I \bar N \phi$ & $0$  \\
& &  \\
$\delta \pi_\varphi$ & \makecell{$ \delta \pi_{\varphi} - \bar \pi_\varphi \Delta \hat I \left(\frac{\bar{N}}{A}\hat{I}\bar{N}\phi +B \right) $ \\+ $\frac{1}{2} \frac{A^{3/2}}{\lambda_\varphi} \frac{\mathrm{d}V}{\mathrm{d}\varphi} \hat I \bar N \phi $} & \makecell{$\delta \pi_\varphi - \frac{\dot \pi_\varphi}{\dot \varphi} \delta \varphi$\\ $- \bar \pi_\varphi \Delta \hat I \left(\bar \pi_\varphi \beta \delta \varphi + B\right)$}  \\
\end{tabular}
\caption{Summary of geometrical clocks and first order observables for the gauge choices corresponding to vanishing perturbation in shift. Unlike table \ref{tab:clockchoices1}, we have split the perturbation in clocks corresponding to shift since the clock components are different. For definition of symbols, see table \ref{tab:symbols}. \label{tab:clockchoices2}}
\end{center}
\end{table}

\begin{table}[h!]
 \begin{center}
  \renewcommand{\arraystretch}{1.5}
\begin{tabular}{c||c|c}
Symbol & Relation to background and perturbation variables & Equation \\
\hline \hline
$\alpha$ & $\frac{1}{4} + \frac{\kappa}{2 \tilde P^2} A p -\frac{2}{3}\frac{1}{\tilde{P}^2}\Delta$ & \eqref{eq:alpha} \\
$\beta$ & $\frac{\bar N A^{1/2}}{\lambda_\varphi \bar \pi_\varphi}$ & \eqref{eq:beta} \\
$\varkappa$  & $\frac{A^{3/2}}{\lambda_\varphi \bar{\pi}_\varphi}\frac{\tilde{\mathcal{H}}}{\bar{N}}$ & \eqref{eq:varkappa}  \\
$\nu_a$ & $p_F^a(x)+\delta^{ab}F_b(x)$ & \eqref{eq:nua} \\
$\Psi$ &    $\psi(x)+\frac{4\tilde{\mathcal{H}}^2A}{\bar{N}^2}(E+p_E)(x)-\frac{1}{3}\Delta E(x)  $    & \eqref{Psi} \\
$\pi_v$ & $\delta \pi_\varphi - \bar{\pi}_\varphi\Delta E + \frac{1}{2}\frac{A^{3/2}}{\lambda_\varphi}\frac{\mathrm{d}V}{\mathrm{d}\varphi}(\bar{\varphi}) \frac{\bar{N}}{\tilde{\mathcal{H}}}\left( \psi -\frac{1}{3}\Delta E \right)$ & \eqref{eq:piv} \\
$\varsigma$ & $\frac{1}{2}\frac{A^3}{\lambda_\varphi^2 \bar{\pi}_\varphi^2}\frac{\mathrm{d}V}{\mathrm{d}\varphi}(\bar{\varphi})$ &  \eqref{eq:varsigma}  \\
$\Upsilon$ & $ p_\psi +\frac{\Delta E}{2} + \frac{2}{3}\Delta p_E - \left( \frac{\tilde{\mathcal{H}}^2A}{\bar{N}^2}+\frac{\kappa A}{2} p \right)(E+p_E)$ & \eqref{eq:Upsilon} \\
$v$ & $\delta \varphi -\frac{\lambda_\varphi}{A^{3/2}\tilde{\mathcal{H}}}\bar{N}\bar{\pi}_\varphi \left( \psi -\frac{1}{3}\Delta E\right)$ & \eqref{eq:Mukhanov-Sasaki} \\
\end{tabular}
\caption{Definitions of various symbols used in tables \ref{tab:clockchoices1} and \ref{tab:clockchoices2}. \label{tab:symbols}}   
\end{center}
 \end{table}

\section{Conclusions and Outlook}
\label{Sec:Concl}

 The main objective of our manuscript was to apply the relational formalism in the extended phase space to linearized cosmological perturbation theory and to understand the relationship between the choice of clocks, gauge fixing conditions and the associated gauge invariant quantities. Our manuscript, which is a companion article to the review \cite{Giesel:2017roz}, extends the results already present in the literature for the reason that the consideration of the extended phase space opens a window to a larger class of gauge fixing conditions that can not be dealt with if only the reduced ADM-phase space is considered where lapse and shift are treated as Lagrange multipliers. Let us note that in a seminal work, Langlois formulated the canonical description using ADM variables and derived the analogue of the Mukhanov-Sasaki variable in phase space \cite{Langlois}. However, in this study lapse and shift were treated as Lagrange multipliers and therefore the analysis had in built restrictions. As an example, it is impossible to obtain the Bardeen potentials in Langlois' analysis because one of the Bardeen potentials is tied to perturbations in the lapse. As we have shown this restriction can be avoided by formulating linear canonical perturbation theory in the extended phase space. 
 This generalization that is strongly based on earlier work by Pons, Salisbury, Sundermeyer et al \cite{Pons1,Pons2,Pons3,Pons4}
 was started in the review \cite{Giesel:2017roz}. There, as a preparation for the work in this article, the phase space analogues of gauge invariant quantities such as for instance the Bardeen potential and the Mukhanov-Sasaki variable were constructed. These results were used in our work as the explicit form of these gauge invariant quantities defined on the extended ADM-phase space were taken as the guiding principle for the choice of clocks.  
 
 A difference of our analysis to the conventional approach in cosmological perturbation theory lies in the context of the relational formalism. The latter has been earlier applied to study linearized cosmological perturbations \cite{Dittrich-Tambornino1,Dittrich-Tambornino2,Dust1,Dust2}, but the applications have been limited to the reduced ADM-phase space or its corresponding extension in terms of Ashtekar variables respectively. Our manuscript provides the first application of the relational formalism to cosmological perturbation theory in the extended ADM-phase space. As a result, clocks and Dirac observables can be understood even for lapse and shift variables which are treated dynamically on the same footing as all remaining phase space variables.

In the relational formalism every gauge fixing conditions is determined by a choice of clocks. Therefore in this manuscript, we have taken the  approach where we have chosen linearized clocks which yield the commonly used gauge conditions in the linear cosmological perturbation theory. Five gauge fixing constraints were considered: the longitudinal gauge, spatially flat gauge, the uniform field gauge, the synchronous gauge and the comoving gauge. In the first three cases, the gauge freedom is completely fixed, whereas in the latter two cases there is residual freedom in the shift in the spatial coordinates. For each of these gauge choices, we identified the clocks constructed from the metric and in some cases metric and matter perturbations at the linear order. The associated observables to these clocks correspond to the gauge invariant variables which are naturally tied in their physical interpretation  to the same gauge fixing conditions.
 
 For the longitudinal gauge fixing constraint, the geometrical clocks we chose result in the Bardeen potentials as the associated and  independent Dirac observables understood as the gauge invariant extensions of the lapse and the trace of the spatial metric perturbation. For the spatially flat gauge, the Mukhanov-Sasaki variable is the naturally associated Dirac observable corresponding to the scalar field perturbation. Similarly, for the other gauges we find a natural set of gauge invariant  quantities using geometric clocks and construct their associated Dirac observables.

The connection between the gauge fixing conditions and gauge invariant variables is well known in the conventional treatment of cosmological perturbation theory. Our analysis bring out this relationship  from the perspective of the relational formalism in the canonical perturbation theory. As emphasized earlier in section  \ref{Sec:Intro}, each gauge invariant quantity in cosmological perturbation theory has a direct relationship with metric or matter perturbations only in a specific gauge. This becomes transparent in our procedure in terms of Dirac observables. As an example, even though the Bardeen potentials are gauge invariant quantities, irrespective of the chosen gauge, they only appear as the {\it{natural}} Dirac observables when the clocks are chosen to be consistent with the longitudinal gauge -- their natural gauge as far as their physical interpretation is addressed. In any other gauge, they do not appear as natural Dirac observables and thus their physical interpretation is lost. The same is true for the relationship between other gauges and gauge invariant quantities. We note that this conclusion is expected from the conventional treatment of cosmological perturbation theory where the naturalness of gauge invariant quantities vis-\`a-vis gauge fixing conditions is well known, but has been borne out for the first time in the language of Dirac observables.  Thus, our canonical analysis make apparent the subtle and self-consistent relationship between the choice of clocks, gauge fixing conditions and Dirac observables.

It is useful to point out a subtlety in comparing our work based on the extended ADM formulation to extract gauge invariant quantities  with the conventional  framework based on the Lagrangian approach. Let us recall following the early work in cosmological perturbation theory by Sachs \cite{Sachs:1964}, and Stewart and Walker \cite{Stewart:1974uz} (further developed by Bruni, Dunsby, Ellis and Sonego \cite{Bruni:1992dg,Sonego:1997np,Bruni:1999et}) that one starts by introducing of two spacetime manifolds $M$ and $M_0$ where $M$ defines the physical spacetime and $M_0$ the background spacetime. The latter can be understood as a fiducial manfiold in this construction. The perturbations are then defined via a point identification map between $M_0$ and $M$, and the choice of a specific point identification map can be understood as a choice of gauge. The gauge invariant quantities are then those whose values do not depend on the point identification map and this kind of gauge invariance was called gauge of second kind by Sachs due to the fact that it occurs in addition to the usual coordinate freedom present in general relativity which we will refer to as gauge invariance of the first kind.

Now following the relational formalism the idea is the following. First let us neglect the aspects of perturbation theory. Then one would like to choose reference fields (clocks) that enter the observable map which allow to solve the constraints present in the canonical theory of general relativity. That is we construct gauge invariant quantities with respect to the gauge invariance of the first kind. As a result one obtains a reduced phase space with only physical degrees of freedom and a physical Hamiltonian that generates the dynamics of the observables. This dynamics is unconstrained unlike the
Lagrangian approach where the constraints of general relativity are still present among the 10 Einstein's equations. 

If we aim at formulating  a perturbative setting of this reduced canonical theory, the second kind of gauge transformation has to be considered as well. In this case the reference fields (clocks) serve a twofold purposes because on the one hand they are used to construct Dirac observables and on the other hand in the above language also as a point identification map that eliminates the second kind of gauge freedom.

Thus, to compare the complexity of the two approaches it is necessary to introduce the demand of gauge invariance of the first kind in the conventional Lagrangian based approach to cosmological perturbation theory. An aspect where this issue 
becomes immediately relevant is for instance when one derives the gauge invariant dynamics
of for instance the Bardeen potential and the Mukhanov-Sasaki variable in linear perturbation theory. In contrast to the conventional way, this dynamics can be obtained in a very straightforward and efficient way in our formalism \cite{gsw}. This occurs thanks to the implementation of the gauge invariance of the first kind in our formalism which allows one to derive the dynamics of these gauge invariant
quantities purely at the gauge invariant level without the need to go
back to the gauge variant form of the Einstein equations and derive
from it the associated dynamics of the observables.

An important question in cosmological perturbation theory is what are the gauge invariant quantities when we go beyond linear order. In the conventional analysis, this is a non-trivial problem as one has to construct these quantities order by order in perturbation theory, with the results obtained  at lower order not being generalizable to higher orders. This task becomes easier in the relational formalism. The reason for this is that one can construct manifestly gauge invariant quantities already at the non-linear level, that is full general relativity, as it has for instance be done in \cite{Dust1,Dust2,Giesel-Ma}. This yields the gauge invariant Einstein's equations at the full non-linear level. Perturbations of these equations involve by construction only quantities that are manifestly gauge invariant and hence are  invariant under coordinate transformations up to arbitrary high orders. Thus, even in linear perturbation theory around a flat FLRW background we consider the linearity as far as the perturbations on phase space are considered, but have manifestly gauge invariance for the diffeomorphisms. As has been shown in the case of dust matter model \cite{Dust2}, if one truncates this manifestly gauge invariant quantities at linear order, one can reproduce the results in linearized cosmological perturbation theory, where one linearizes the Einstein equation first and afterwards construct linearized gauge invariant quantities. 

We should note that non-linear gauge invariant quantities have earlier been studied in the covariant formulation \cite{Ellis:1989jt} and further developed in \cite{Langlois:2005ii, Langlois:2006vv}. The latter method is based on constructing gauge invariant quantities by introducing flow lines associated with fundamental observers. Using
the Stewart-Walker lemma such quantities are automatically gauge invariant if the corresponding quantities vanish in the background spacetime. One obtains full non-linear gauge invariant quantities and as shown by Langlois and Vernizzi one can recover the usual gauge invariant quantities at linear and second order \cite{Langlois:2005ii, Langlois:2006vv}.  As in the case of dust model \cite{Dust2}, one can first construct non-linear gauge invariant quantities and then apply perturbation theory which results in a formulation of perturbation theory at the gauge invariant level.  The key difference of this approach in comparison to our analysis is that the formalism by Langlois and Vernizzi considers only the gauge invariance of the second kind but not the first
kind as can been seen for instance from the fact that the constraints are still part of the theory. Furthermore, the introduction of flow lines associated with fundamental observers can be understood in the
relational formalism as the choice of an idealized observer, that is one that causes no backreaction. In our 
formalism observers are the reference fields (clocks) which are dynamically coupled to the system. Thus, backreaction will be  inherently included in our approach. This is evident via the imprint of the chosen clocks in the equations of motion of the observables usually via their energy and momentum densities. In order to compare the two approaches and their complexities in studying second order perturbation theory, one would need to include gauge invariance of the first kind in the approach of Langlois and Vernizzi.

We showed here that the common gauge invariant variables in linear cosmological perturbation theory can be systematically constructed if we apply the relational formalism and the observable map to the extended phase space of linear cosmological perturbation theory. An open question that arises from our results is whether we can find non-linear geometrical clocks that reduce at the linear order to those we have identified here. Such clocks have been constructed in relational formalism for the the case of dust matter in Ref. \cite{Dust1}, though they have been used  to study first order cosmological perturbation theory \cite{Dust2}. To generalize our current analysis to second order perturbation theory, one will need to find non-linear clocks using purely metric perturbations or its combinations with scalar field perturbations. For this purpose we had to generalize for instance the work in Ref. \cite{Dust1} to the extended ADM phase space and carefully analyze the stability of our chosen clocks that becomes more complicated when
geometrical than matter clocks are chosen. The identification of these
clocks at linear order in our present analysis give us vital insights on the nature of non-linear
geometrical clocks which will be investigated in a future work. Such an analysis would extend our results beyond the linear order and would also in higher order have a systematic and straightforward way to construct gauge invariant quantities with a clear physical interpretation of such quantities. 

If we want to go beyond the classical theory and consider the quantization of the associated reduced phase spaces that follow from a certain choice of geometrical clocks, then it might be the case that models where one chooses purely matter clocks as for instance in \cite{Thiemann3,Dust1,Giesel:2007wn,Giesel:2012rb,Giesel:2016gxq,Domagala:2010bm, Husain:2011tm} are of advantage. The reason for this is that in general the Poisson algebra of the Dirac observables that need to be considered in a reduced phase space quantization is more complicated than the standard kinematical algebra. Consequently, to find representations of the algebra, that is finding the associated quantum theory, might be very difficult. Along with this comes the fact that the decomposed scalar, vector- and tensor gauge variant quantities in general satisfy also a more complicated algebra than the original ADM variables due to the projectors that are used to define such decomposed quantities. The matter field reference models in \cite{Thiemann3,Dust1,Giesel:2007wn,Giesel:2012rb,Giesel:2016gxq,Domagala:2010bm, Husain:2011tm} are all designed in such a way that the Poisson algebra of the Dirac observables is as simple as the kinematical algebra. However, a real conclusion on this point can only be drawn after we have identified some candidates for non-linear clocks, which will be one of our future projects. 

In the present manuscript we have focused our discussion on the classical perturbation theory in the canonical setting. 
However, one important application of our analysis lies in the canonical quantization program, such as Wheeler-deWitt quantization or loop quantum gravity. For these approaches, our framework provides a natural setting to investigate quantum gravitational effects on cosmological perturbations. For this purpose, the first step is to formulate a canonical formulation of classical cosmological perturbation theory at the linear order. Our analysis accomplishes this step using the extended phase space using ADM variables. The next step will be to incorporate quantum gravitational constraints in this setting and repeat the analysis with appropriate clocks at the quantum level. An interesting question will be to examine the role of clocks in the cosmological perturbation theory at the quantum level. 


\section*{Acknowledgements}
P.S. is grateful to the Institute for Quantum Gravity at Friedrich-Alexander-Universit\"{a}t Erlangen-N\"{u}rnberg for its generous support and a warm hospitality at various stages of this work. 
P.S. is supported by NSF grant PHY-1454832.


\end{document}